\pdfoutput=1
\documentclass[aps,rmp,reprint,amsmath,amssymb,longbibliography,onecolumn]{revtex4-1}
\usepackage{xcolor}
\usepackage{wrapfig}
\usepackage{graphicx}
\usepackage{nicefrac}
\usepackage{array,booktabs}
\usepackage{multirow}
\usepackage{url}
\usepackage{enumitem}

\usepackage{hyperref}
 \hypersetup{
     colorlinks=true,
     citecolor=black, 
     filecolor=black,
     linkcolor=[RGB]{55,85,114}, %[RGB]{271,27,91}, %figures pink-red
     urlcolor=[RGB]{55,85,114}, %figures blue
}

\definecolor{blue}{rgb}{0,0,1}

\newcommand{\blue}[1]{\textcolor{black}{#1}}

\begin{document}

%\title{Deformation and flow of amorphous solids: insights from elastoplastic models} %%on RMP
\title{Deformation and flow of amorphous solids: An updated review of mesoscale elastoplastic models} %%on arXiv

\author{Alexandre Nicolas}
\affiliation{LPTMS, CNRS, Univ. Paris-Sud, Universit\'e Paris-Saclay, 91405 Orsay, France.}

\author{Ezequiel E. Ferrero}
\affiliation{Centro At\'omico Bariloche, 8400 San Carlos de Bariloche, R\'io Negro, Argentina}

\author{Kirsten Martens}
\affiliation{Univ. Grenoble Alpes, CNRS, LIPhy, 38000 Grenoble, France}

\author{Jean-Louis Barrat}
\affiliation{Univ. Grenoble Alpes, CNRS, LIPhy, 38000 Grenoble, France}

\begin{abstract}

The deformation and flow of disordered solids, such as metallic glasses and concentrated emulsions, involves swift localized rearrangements of particles that induce a long-range deformation field. To describe these heterogeneous processes, elastoplastic models handle the material as a collection of `mesoscopic' blocks alternating between an elastic behavior and plastic relaxation, when they are loaded above a threshold. Plastic relaxation events redistribute stresses in the system in a very anisotropic way. We review not only the physical insight provided by these models into practical issues such as strain localization, creep and steady-state rheology, but also the fundamental questions that they address with respect to criticality at the yielding point and the statistics of avalanches of plastic events. Furthermore, we discuss connections with concurrent mean-field approaches and with related problems such as the plasticity of crystals and the depinning of an elastic line.

\end{abstract}

\maketitle

\tableofcontents{}

\section*{Frequently used notations}

\begin{tabular}{cc}
 $\Sigma$ & Macroscopic shear stress \\
 $\Sigma_y$ & Macroscopic yield stress \\
 $\sigma$ & Local shear stress \\
 $\sigma_y$ & Local yield stress \\
 $\mu$ & Shear modulus \\
 $\gamma$ & Shear strain \\
 $\dot\gamma$ & Shear rate \\
 EPM & Elastoplastic model \\
 MD & Molecular dynamics \\
 rhs (lhs) & right-hand side (left-hand side) \\
 ST & Shear transformation \\
\end{tabular}

\newpage{}

\section*{Introduction}

\begin{center}
\begin{figure}
\begin{centering}
\includegraphics[width=0.85\textwidth]{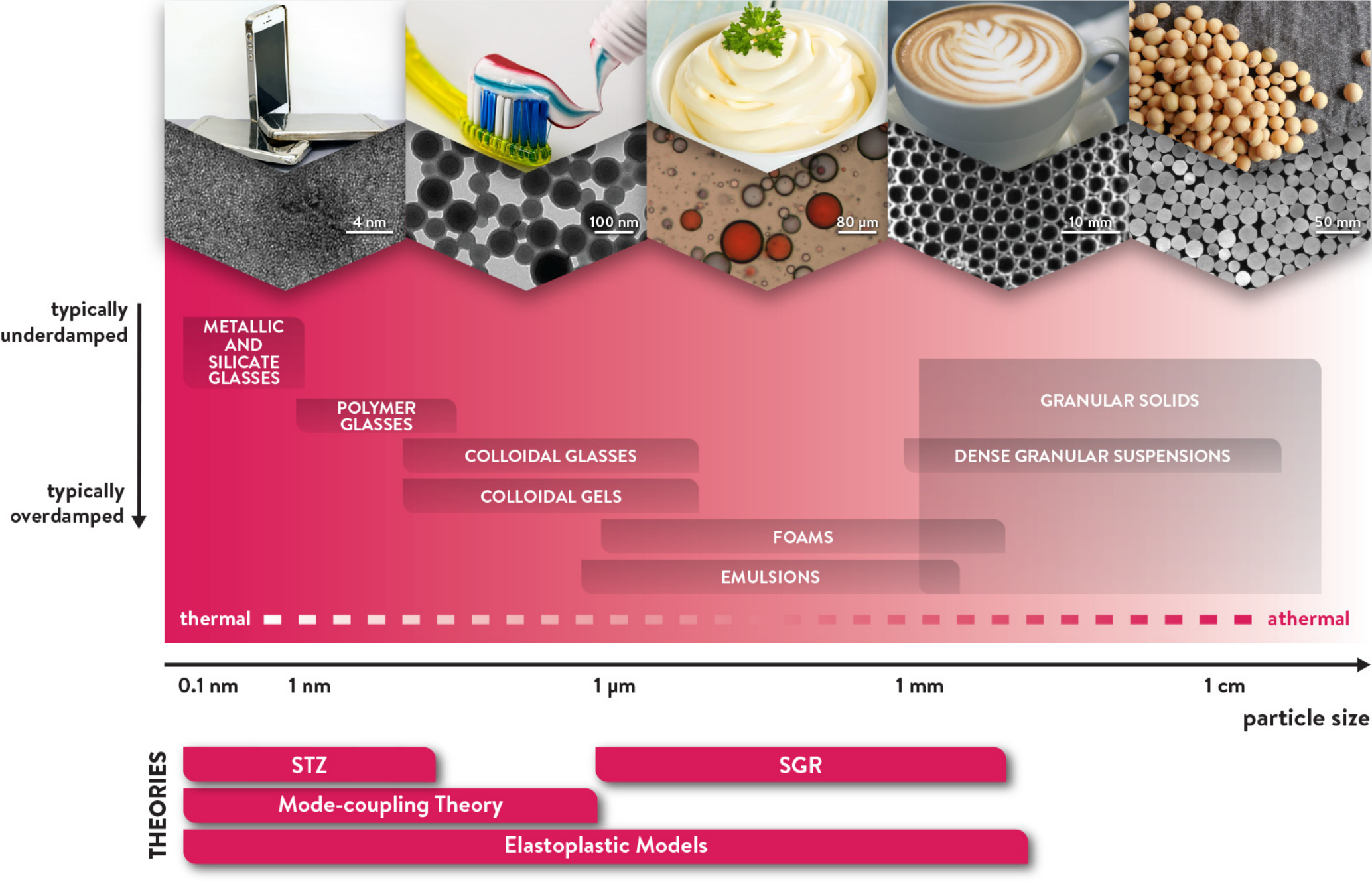}
\par\end{centering}

\caption{\label{fig:amorphoussolids}Overview of amorphous solids. \emph{From
left to right, top row}: cellular phone case made of metallic glass
(1); toothpaste (2); mayonnaise (3); coffee foam (4); soya beans (5).
\emph{Second row}: a transmission electron microscopy (TEM) image
of a fractured bulk metallic glass ($\mathrm{Cu_{50}Zr_{45}Ti_{5}}$)
by X. Tong et. al (Shanghai University, China); TEM image of blend (PLLA/PS)
nanoparticles obtained by miniemulsion polymerization, from
L. Becker Peres et al. (UFSC, Brazil); emulsion of water droplets
in silicon oil observed with an optical microscope by N. Bremond (ESPCI
Paris); a soap foam filmed in the lab by M. van Hecke (Leiden University,
Netherlands); thin nylon cylinders of different diameters pictured
with a camera, from T. Miller et al. (University of Sydney, Australia).
The white scale bars are approximate. \emph{Just below}, a chart of
different amorphous materials, classified by the size and the damping
regime of their elementary particles. \emph{At the bottom}: some popular
modeling approaches, arranged according to the length scales of the
materials for which they were originally developed. STZ stands for
the shear transformation zone theory of \citet{Langer2008}, and SGR
for the soft glassy rheology theory of \citet{sollich1997rheology}.}
\end{figure}

\par\end{center}

19th-century French Chef Marie-Antoine \citet{careme1842cuisinier}
claims that `mayonnaise' comes from the French verb `\emph{manier}'
(`to handle'), because of the continuous whipping that is required
to make the mixture of egg yolk, oil, and vinegar thicken. This etymology
may be erroneous, but what is certain is that the vigorous whipping
of these liquid ingredients can produce a viscous substance, an emulsion
consisting of oil droplets dispersed in a water-based phase. At high
volume fraction of oil, mayonnaise even acquires some resistance to
changes of shape, like a solid; it no longer yields to small forces,
such as its own weight. Similar materials, sharing solid and liquid
properties, pervade our kitchens and fridges: Chantilly cream,
heaps of soya grains or rice are but a couple of examples. They also
abound on our bathroom shelves (shaving foam, tooth paste, hair gel),
and in the outside world (sand heaps, clay, wet concrete), see Fig.~\ref{fig:amorphoussolids}
for further examples. All these materials will deform, and may flow,
if they are pushed hard enough, but will preserve their shape otherwise.
Generically known as amorphous (or disordered) solids, they 
\blue{seem to} have no more in common than what the etymology implies:
their structure is
disordered, that is to say, deprived of regular pattern at ``any''
scale, as liquids, but they are nonetheless solid. So heterogeneous
a categorization may make one frown, but has proven useful in framing
a unified theoretical description \citep{Barrat2007deformation}.
In fact, the absence of long range order or of a perceptible microstructure
makes the steady-state flow of amorphous solids simpler, and much
less dependent on the preparation and previous deformation history,
than that of their crystalline counterparts. A flowing amorphous material
is therefore a relatively simple realization of a state of matter
driven far from equilibrium by an external action, a topic of current
interest in statistical physics. 

A matter of clear industrial interest, the prediction of the mechanical
response of such materials under loading
is a challenge for Mechanical Engineering, too. This problem naturally
brings in its wake many questions of fundamental physics. Obviously,
it is not exactly solvable, since
it involves the coupled mechanics equations of the $N\gg1$ elementary
constituents of the macroscopic material; this is a many-body problem
with intrinsic disorder and very few symmetries. Two paths can be
considered as alternatives: (i) searching for empirical laws in the
laboratory, and/or (ii) proposing approximate, coarse-grained mathematical
models for the materials. The present review is a pedagogical journey
along the second path.

\begin{wrapfigure}{r}{0.45\textwidth}%
\begin{centering}
\includegraphics[clip,width=0.445\textwidth]{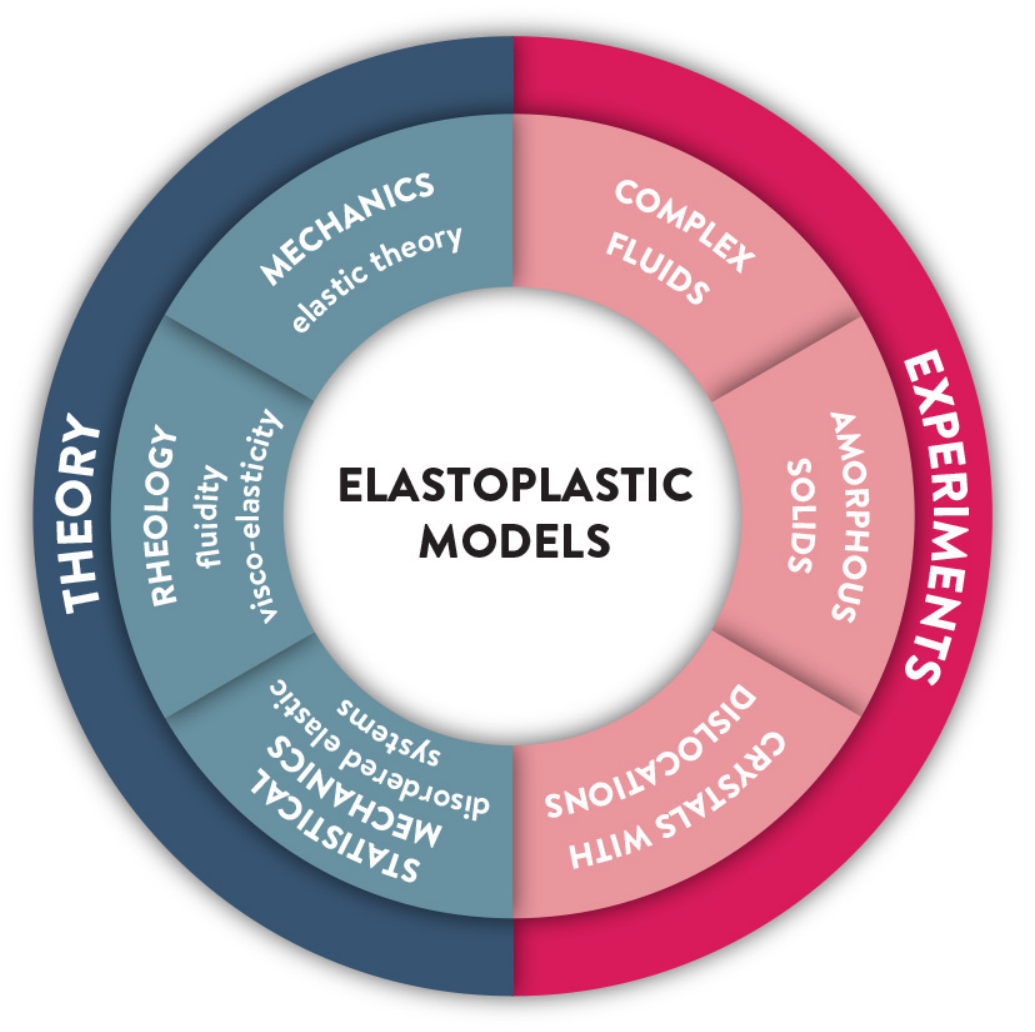} 
\end{centering}
\caption{\label{fig:sketchEPlocation} Scientific position of elastoplastic
modeling.}
\end{wrapfigure}%

Along this route, substantial assumptions are made to simplify the
problem. The prediction capability of models hinges on the
accuracy of these assumptions. Following their distinct interests
and objectives, different scientific communities have adopted different
modeling approaches. Material scientists tend to include
a large number of parameters, equations and rules, in order to reproduce
different aspects of the material behavior simultaneously . Statistical
physicists aspire for generality and favor minimal models, or even
toy models, in which the parameter space is narrowed down to a few
variables. At the interface between these approaches,
``elastoplastic'' models (EPM) consider an assembly of mesoscopic
material volumes that alternate between an elastic regime and plastic relaxation,
and interact among themselves. As simple models, they aim to describe a
general phenomenology for all amorphous materials, but they may also
include enough physical parameters to address material
particularities, in view of potential applications. They
rely on simple assumptions to connect the microsopic phenomenology to the macroscopic behavior
and therefore have a central position in the endeavor to bridge
scales in the field \citep{Rodney2011modeling}. To some extent,
EPM can be compared to classical lattice models of magnetic systems, which
permit the exploration of a number of fundamental and practical issues,
by retaining a few key features such as local exchange and long range
dipolar interactions, spin dynamics, local symmetries, etc., without
explicit incorporation of the more microscopic ingredients about the
electronic structure.

This review aims to articulate a coherent overview of the state of the art of these EPM,
starting in Sec.~\ref{sec:Gen_phenomenology} with the microscopic
observations that guided the coarse-graining efforts.
We will discuss several possible practical implementations of coarse-grained systems
of interacting elastoplastic elements, considering the possible attributes
of the building blocks (Sec.~\ref{sec:Building_blocks}) and the
more technical description of their mutual interactions (Sec.~\ref{sec:Elastic_couplings}).
Section~\ref{sec:Mechanical_noise} is then concerned with the widespread
approximations of the effect of the stress fluctuations resulting
from these interactions.
\blue{
In Sec.~\ref{sec:VI_Macroscopic_Shear_Deformation} we describe the 
current understanding of strain localization based on the study of EPM.
Section~\ref{sec:VIII_Avalanches} focuses on the statistical marks of criticality encountered when
the system is driven extremely slowly, especially in terms of the temporal and spatial organization of stress fluctuations
in `avalanches', while Sec. \ref{sec:VII_Bulk_rheology} describes the 
bulk rheology of amorphous solids in response to a shear deformation.
}
Section~\ref{sec:Relaxation} gives a short perspective on the much less studied
phenomena of creep and aging. 
The review ends on a discussion of the relation between EPM and several other descriptions
of mechanical response in disordered systems, in Sec.~\ref{sec:Related_topics}, and some final outlooks. 

\blue{
These sections are largely self-contained and can thus be read separately. 
Sections~\ref{sec:Gen_phenomenology} and \ref{sec:Building_blocks} are both particularly well suited as entry points for newcomers in the field,
while Sections ~\ref{sec:Elastic_couplings} and \ref{sec:Mechanical_noise} might be more technical and of greater relevance for the experts interested in
the implementation of EPM. Finally, Sections~\ref{sec:VI_Macroscopic_Shear_Deformation} to
\ref{sec:Relaxation} focus on applications of the models to specific physical phenomena and are largely independent from each other. 
}

\section{General phenomenology\label{sec:Gen_phenomenology}}

\subsection{What are amorphous solids? }

From a mechanical perspective, amorphous solids are neither perfect
solids nor simple liquids. 
Albeit solid, some of these materials are made of liquid to a large
extent and appear soft. Nevertheless, at rest they preserve a solid
structure,
% for example, challenging gravity or offering elastic resistance to deformations,
and will flow only if a sufficient load is
applied to them. Accordingly, in the rheology of complex fluids \citep{bonn2015yield},
they are often referred to as `yield stress materials'. Foams
and emulsions, that is, densely packed bubbles or droplets dispersed
in a continuous liquid phase, 
\blue{ are solid because
surface tension strives to restore the equilibrium shape of
their constituent bubbles or droplets upon deformation.}
Their elastic moduli,
\blue{ \emph{i.e.}, the coefficients of proportionality between the elastic strain and stress,}
are then approximately given by the surface tension divided
by the bubble or droplet \blue{radius} (between a few microns and several millimeters);
a few hundred Pascal would be a good order
of magnitude. Colloidal glasses, on the other hand, are dense suspensions of solid particles
 of less than a micron in diameter, which makes
them light enough for Brownian agitation to impede sedimentation.
They rely on entropic forces to maintain their reference structure and typically
have shear moduli of the order of $10-100\,\mathrm{Pa}$.

Poles apart from these soft solids, `hard'
amorphous solids comprise oxyde or metallic glasses, as well as glassy polymers.
 They are typically made of much smaller particles
than their soft counterparts. Indeed, very roughly speaking, the elastic
moduli are inversely proportional to the \blue{linear} size of the constituents.
(Granular media, in which the elastic moduli depend on the material
composing the grains and the applied pressure, are obviously an exception
to this vague rule of thumb.) For instance, the atoms that compose
the metallic or silica glasses live in the Angstr\"om scale, and these
materials have very large Young moduli, of order $100\,\mathrm{GPa}$
(somewhat below for silicate glasses, sometimes above for metallic
glasses). These atomic glasses are obtained from liquids
when temperature is lowered below the glass transition temperature 
while crystallization is avoided. To do
so,  high cooling rates of typically $10^{5}-10^{6}\mathrm{K\cdot s^{-1}}$
are required for metallic glasses \citep{greer1995metallic,greer2007bulk},
whereas values below $1\mathrm{K\cdot s^{-1}}$ may be used for oxide
glasses.
After a certain amount of deformation, brittle materials
 will break without incurring significant plastic (irretrievable)
deformation,
% (the typical example would be silica glasses, although this has been questioned \citep{lacroix2012plastic})
whereas ductile materials will deform plastically before breaking. 
\blue{We will discuss connections between these forms of deformation in Sec.~\ref{sec:VI_Macroscopic_Shear_Deformation}.}
%(For further experimental details on these materials, the reader is
%referred to \citep{bonn2015yield}, for instance.)

\subsection{What controls the dynamics of amorphous solids?\label{sub:what_turns_them_on}}

\begin{figure}
\includegraphics[clip,width=0.8\textwidth]{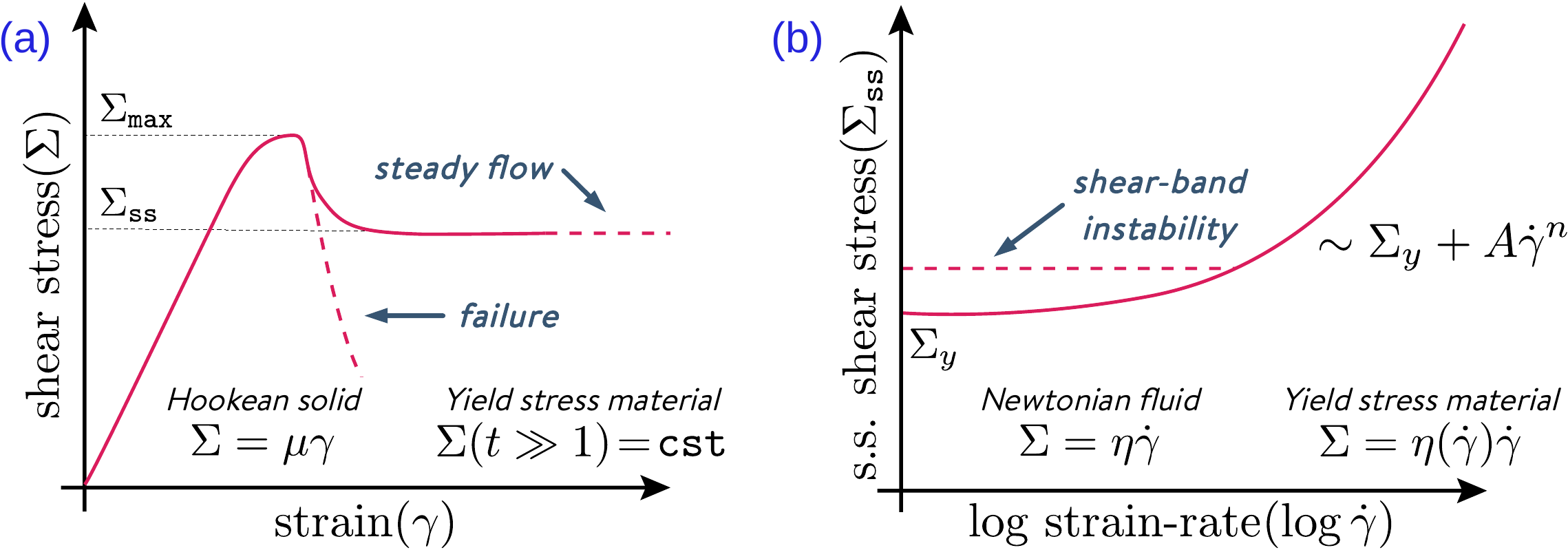}
\caption{\label{fig:schematiccurves_crop} \emph{Schematic macroscopic response of
amorphous solids to deformation.}
\textbf{(a)} Evolution of the shear
stress $\Sigma$ with the imposed shear strain $\gamma$, with a stress overshoot $\Sigma_\mathrm{max}$.
In the event of \blue{material failure, which is generally preceded by strain localization,} the stress
dramatically drops down.
\textbf{(b)} Steady-state
flow curve, i.e., dependence of the steady-state shear stress $\Sigma_{ss}$ on the shear rate $\dot{\gamma}$,
represented with semi-logarithmic axes.
If the \blue{flow is split into macroscopic shear bands,} a stress plateau is generally observed.}
\end{figure}

Another distinction regards the
nature of the excitations that can alter the structural configuration
of the system.

\subsubsection{Athermal systems}

When the elementary constituent sizes are large enough $(\gtrsim1{\rm \mu m}$)
to neglect Brownian effects (thermal fluctuations), the materials
are said to be athermal. Dry granular packings, dense granular suspensions,
foams, and emulsions (see Fig.~\ref{fig:amorphoussolids}) belong
in this category. An external force is required to activate their
dynamics and generate configurational changes. Typical protocols for
externally driving the system include: shearing it by rotating the
wall of a rheometer~\citep{barnes1989introduction}, deforming it
by applying pressure in a given direction, or simply making use of
gravity if the material lies on a tilted plane~\citep{coussot1995determination}.
Rheometers control either the applied torque $T$ or the angular velocity
$\Omega$ of the rotating part. In the former case, the applied macroscopic
shear stress is kept fixed, at a value $\Sigma=\frac{T}{2\pi hR^{2}}$ on a rotating cylinder
of radius $R$ and height $h$ \citep{fardin2014hydrogen}, while one monitors 
the resulting shear strain $\gamma$ or shear rate $\dot{\gamma}$
if the material flows steadily. Conversely, strain-controlled experiments
impose $\gamma(t)$ or $\dot{\gamma}$ and monitor
the stress response $\Sigma(t)$.

How do amorphous solids respond to such external forces? For
small applied stresses $\Sigma$, the deformation is elastic, \emph{i.e.},
mostly reversible.
Submitted to larger stresses, the material shows signs of plastic
(irreversible) deformation; but the latter ceases rapidly, unless
$\Sigma$ overcomes a critical threshold $\Sigma_{y}$ known as yield stress
(see Fig.\ref{fig:schematiccurves_crop}(a)).
For $\Sigma>\Sigma_{y}$, the material yields. This process
can culminate in macroscopic fracture; for brittle materials like
silica glass, it always does so.
Contrariwise, most soft amorphous solids will finally undergo stationary plastic flow.
The ensuing flow curve $\Sigma=f\left(\dot{\gamma}\right)$ in the steady state is often
 fitted by a Herschel-Bulkley law
\begin{equation}
\Sigma=\Sigma_{y}+A\dot{\gamma}^{n},\label{eq:II_Herschel-Bulkley}
\end{equation}
with $n>0$ (see Fig.~\ref{fig:schematiccurves_crop}(b)).

The transition between the solid-like elastic response and the irreversible
plastic deformation is known as
\blue{the} \textit{yielding} transition.
Statistical physicists \blue{are inclined to} regard it as a dynamical phase transition,
an out-of-equilibrium phenomenon with characteristics similar to equilibrium
phase transitions \citep{lin2014scaling,lin2015criticality,jaiswal2016mechanical}.

\subsubsection{Thermal systems \label{sub:II_thermal_systems}}

On the other hand, thermal fluctuations may play a role in materials
with small enough ($\lesssim 1\ \mathrm{\mu m}$) elementary constituents, such
as colloidal and polymeric glasses, colloidal gels, silicate and metallic
glasses.
%Experimentally, the latter types of glasses are obtained
%by quenching a high-temperature liquid (above the melting temperature
%$T_{m}$) to a very low temperature (below the glass transition temperature
%$T_{g}<T_{m}$). In general, the sample must be cooled fast enough
%to avoid crystallization. 
Still, these materials are out of thermodynamic equilibrium and
they do not sample the whole configuration space under the influence
of thermal fluctuations. It follows that different preparation routes (and in particular
different cooling rates)
tend to produce mechanically distinct systems. Even
the waiting time between the preparation and the experiment matters,
because the system's configuration evolves meanwhile, through activated
events. \blue{The evolution of the mechanical properties with the time since
preparation, usually making the system more solid, is called aging.}
In particular, the high cooling
rates used for quenching generate a highly heterogeneous internal
stress field in the material \citep{Ballauff2013}. In some regions,
particles manage to rearrange geometrically, minimizing in part the
interaction forces among them, but many other regions are frozen in
a highly strained configuration. Slow rearrangements will take place
at finite temperature and tend to relax locally strained configurations
(``particles break out of the cages made by their neighbors''),
along with the stress accumulated in them. 

That being said, the elastic moduli are usually only weakly affected by the
preparation route, i.e., the cooling rate \citep{Ashwin2013cooling} and
the waiting time \citep{Divoux2011a}, while other key features of
the transient response to the applied shear are often found to depend
on it. This sensitivity to preparation particularly affects the overshoot
in the stress \emph{vs.~}strain curve, depicted in Fig.~\ref{fig:schematiccurves_crop}
and used to define the static yield stress $\Sigma_{\mathrm{max}}$.
It is observed in experiments \citep{Divoux2011a} as well as numerical
simulations \citep{Rottler-PRL2005,patinet2016connecting}. In soft
materials amenable to stationary flow, this issue may be deemed
secondary; the flow creates a nonequilibrium stationary state, and
the memory of the initial preparation state is erased after a finite
deformation. On the other hand, in systems that break at finite deformation,
the amount of deformation before failure is of paramount importance,
and so is its possible sensitivity to the preparation scheme, due to different
abilities of the system to localize deformations (see Sec.~\ref{sec:VI_Macroscopic_Shear_Deformation}).

\subsubsection{Potential Energy Landscape}

The Potential Energy Landscape (PEL) picture offers an illuminating perspective
to understand the changes associated with aging in thermal systems 
\citep{goldstein1969viscous,doliwa2003does,Doliwa2003}. The whole configuration
of the system (particle coordinates and, possibly, velocities) is
considered as a `state point' $\Gamma$ that evolves on top of
a hypersurface $V(\Gamma)$ representing the total potential energy.
Despite the  high dimension of such a surface (proportional
to the number $N$ of particles), it can be viewed as a rugged landscape,
with hills and nested valleys; the number of local minima generally
grows exponentially with $N$ \citep{wales2006potential}. Contrary
to crystals, glassy (disordered) states do not minimize the free energy
of the system; aging thus consists in an evolution towards lower-energy
states (on average) through random, thermally activated jumps over
energy barriers, or more precisely saddle points of the PEL. As the
state point reaches deeper valleys, the jumps become rarer and rarer;
the structure stabilizes, even though some plasticity is still observed
locally \citep{Ruta2012}.

External driving 
\blue{restricts the regions of the PEL that can be visited by the state point to, say, those
with a (usually time-dependent) macroscopic strain $\gamma$.} Mathematically, this
constraint is enforced by means of a Lagrange multiplier, which effectively tilts
$V\left(\Gamma\right)$ into 
\begin{equation}
V_\sigma\left(\Gamma,\gamma\right)\equiv V\left(\Gamma\right)-\Omega_{0}\Sigma\gamma,\label{eq:tilted_eff_potential}
\end{equation}
where $\Omega_{0}$ is the volume of the system and $\Sigma$ the
macroscopic stress. 
\blue{The system's dynamics are then controlled by 
$\partial V_\sigma / \partial \Gamma$, instead of $\partial V / \partial \Gamma$, which
results in major changes, as we shall see next.
}
Typically, \blue{driven systems respond on much shorter times than (quiescent) aging ones.}
Accordingly, some thermal systems may
be treated as athermal, for all practical purposes. Nonetheless, interesting
physical behavior emerges when the aging and driving time scales compete, either
because temperature is high or because the driving is slow
\citep{Rottler-PRL2005,shi2005strain,Johnson2005,Vandembroucq2011,Chattoraj2010}.

\subsection{Jagged stress-strain curves and localized rearrangements\label{sub:Localised_rearrangements}}

The contrasting inelastic material responses to shear, ranging
 from failure to flow, may give the impression that
there is a chasm between `hard' and `soft' materials. They are indeed often seen as different fields, plasticity for hard solids versus rheology for  soft materials.
 Nevertheless,
the gap is not so wide as it looks. Indeed, some hard solids may flow
plastically to some extent without breaking, while soft solids retain
prominent solid-like features under flow at low enough shear rates,
unlike simple liquids.

To start with, consider the macroscopic response to a constant stress $\Sigma$
(or shear rate $\dot\gamma$) of a foam \citep{lauridsen2002shear} 
or a metallic glass \citep{wang2009self}: Instead of a smooth deformation, the evolution of strain $\gamma(t)$ with time
(or stress $\Sigma(t)$) is often found to be jagged.
The deeper the material lies in its solid phase, the more `serrated'
the curves \citep{dallaTorre2010stick,sun2012serrated}. \blue{This type of curves is not}
specific to the deformation of amorphous solids. It observed in all `stick-slip' phenomena, in which
the system
is \blue{repeatedly} loaded until a breaking point, where an abrupt discharge
(energy release) occurs. Interestingly, this forms the basis of the
elastic rebound theory proposed by \citet{Reid1910} after the 1906
Californian earthquake. Other elementary examples include pulling
a particle with a spring of finite stiffness in a periodic potential,
a picture often used in crystalline solids to describe the motion
of \blue{irregularities (defects) in the structure called dislocations} - the elementary mechanism of plasticity. 
In the plastic
flow of amorphous solids, potential energy $V$ is accumulated in the material
in the form of elastic strain, until some rupture threshold is passed.
At this point, a plastic event occurs, with a release of the stored
energy and a corresponding stress drop. 
%The precise nature of the
%plastic event that gives rise to the stress drop, however, will strongly
%depend on the scale of observation.

\blue{From the PEL perspective, the energy accumulation phase coincides with
the state point smoothly tracking the evolution of the local minimum in the effective potential $V_\sigma(\Gamma,\gamma)$ [Eq.~\eqref{eq:tilted_eff_potential}],}
as $\gamma$ increases. Meanwhile, some \blue{effective} barriers subside, until one
flattens so much that the system can slide into another valley
without energy cost. This \blue{topological change in $V_\sigma$ at a critical strain $\gamma=\gamma_{c}$
is a saddle-node bifurcation and}
marks the onset of a plastic event. For
smooth potentials, close to $\gamma_{c}$,
the effective barrier height scales as \citep{Gagnon2001,Maloney2006Energy}
\begin{equation}
V^\star \sim\left(\gamma_{c}-\gamma\right)^{\nicefrac{3}{2}}.
\end{equation}
Note that the instability
can \blue{be triggered prematurely}
if thermal fluctuations are present. In summary,
in the PEL, deformation is a succession of barrier-climbing phases
(elastic loading) and descents.
The first step in building a microscopic understanding of the flow
process is to identify the nature of these plastic events.

\begin{figure}
\begin{centering}
\includegraphics[width=\textwidth]{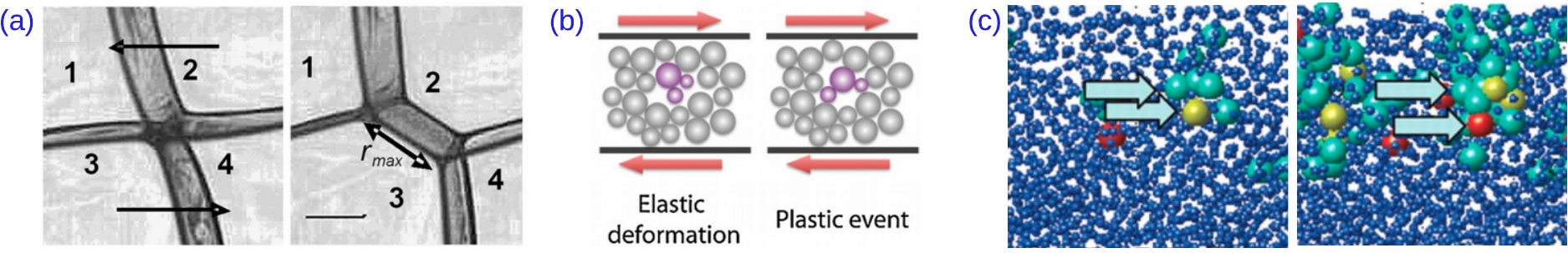}
\par\end{centering}

\caption{\label{fig:Localized_rearrangements}Localized rearrangements. 
\textbf{(a)}
T1 event in a strained bubble cluster. From \citep{Biance2009}.
\textbf{(b)} Sketch of a rearrangement. From \citep{Bocquet2009}.
\textbf{(c)} Instantaneous changes of neighbors in a slowly sheared colloidal
glass. Adapted from \citep{Schall2007}. Particles are magnified
and colored according to the number of nearest neighbors that they
lose.}
\end{figure}

But what can be said about the microscopic deformation of atomic or
molecular glasses when the motion of atoms and molecules remains virtually
invisible to direct experimental techniques? In the 1970s, inspiration
\blue{came from} the better known realm of crystals. As early as 1934,
with the works of Orowan, Polanyi and Taylor,
the motion of dislocations was known to be the main lever
of their (jerky) deformation. Could similar static structural defects
be identified in the absence of a regular structure? The question
has been vivid to the present day, so that it is at least fair to say that,
\blue{should they exist, such defects would be}
more elusive than in crystals 
\blue{(we will come back to this question later in this chapter).} In fact, the
main inspiration drawn from research on crystals was of more
pragmatic nature: \citet{Bragg1947} showed that bubble rafts,
\emph{i.e.}, monolayers of bubbles, could serve as upscaled models
for crystalline metals and provide insight into the structure of the
latter. The lesson was simple: If particles in crystals are too small
to be seen, let us make them larger. Some thirty years later, the
idea was \blue{transposed} to disordered systems by \citet{Argon1979}, who
used bidisperse bubble rafts as models for metallic glasses. Most
importantly, they observed prominent singularities in the
deformation: rapid rearrangements involving a few
bubbles. \citet{princen1986rheology} suggested that the elementary
rearrangement in \blue{these 2D} systems was a topological change 
involving four bubbles, termed T1 event  (see Fig.~\ref{fig:Localized_rearrangements}a).

\subsubsection*{Evidence}

Since 
%\citet{Argon1979}'s and \citet{Princen1983,princen1986rheology}'s 
\blue{these early} studies of foams and emulsions, evidence for swift localized rearrangements has been
amassed both experimentally and numerically, in very diverse systems, namely:

- simple numerical glass models like Lennard-Jones glasses \citep{Falk1998,Maloney2004,Maloney2006Amorphous,Tanguy2006}
and other systems \citep{gartner2015nonlinear},

- numerical models of metallic glasses \citep{Srolovitz1983,Rodney2009a},

- numerical models of silicon glasses (amorphous silicon) \citep{Fusco2014,albaret2016mapping}, 

- numerical models of polymer glasses \citep{Smessaert2013,papakonstantopoulos2008} 

- dense colloidal suspensions \citep{Schall2007,Chikkadi2012b,Jensen2014},

- concentrated emulsions \citep{desmond2015measurement},

- dry and wet foams \citep{Debregeas2001,kabla2003local,Biance2009,biance2011topological},

- granular matter \citep{Amon2012,Amon2012a,LeBouil2014,denisov2016universality}.

\blue{
Recently, \citet{poincloux2018knits} even proposed to extend the list to knits, where local slip events at inter-yarn contacts play the role of rearrangements.}

These are the essential events whereby the \blue{irreversible} macroscopic deformation is transcribed into the
material structure. As such, they are the building bricks of EPM and we will often refer to them as
`plastic events'\footnote{The reader should however be warned that the expression was also used
in the literature to refer to cascades of such localized rearrangements
\citep{Maloney2006Amorphous,Tanguy2006,Fusco2014}.}.
Since they must contribute to the externally imposed shear deformation, they
will retain part of its symmetry and
can thus be \blue{idealized} as localized shear deformations, or `shear transformations' (ST).
Admittedly, in reality, their details do somewhat vary \blue{across} systems (\emph{see below}),
but, compared to the crystalline case, their strong
spatial localization \blue{is a generic and remarkable feature.}

\blue{At this stage, no doubt should be left as to the solid nature of the materials considered here. Our review is concerned
with materials that are clearly solid at rest, and excludes systems at the fringe of rigidity, such as barely jammed packings of particles or emulsions. The latter are most probably governed by different physics, in which spatially extended rearrangements may occur 
under vanishing loading \citep{muller2015marginal}. }

\subsubsection*{Quantitative description}

Although rearrangements can sometimes be spotted visually, a
more objective and quantitative criterion for their detection is desirable.
Making use of the inelastic nature of these transformations, \citet{Falk1998}
pioneered the use of $D_{\mathrm{min}}^{2}$, a quantity that measures
how nonaffine the local displacements around a particle are. More
precisely, the relative displacements of neighboring particles between
successive configurations are computed, and compared to the ones that
would result from a locally affine deformation $\boldsymbol{F}$; $D_{\mathrm{min}}^{2}$
is the \blue{minimal square difference obtained by optimizing the tensor $\boldsymbol{F}$.} This quantity has been
used heavily since then \citep{Schall2007,Chikkadi2011,Chikkadi2012b,Jensen2014},
Generally speaking, very strong localization of events is observed \blue{at low enough shear rates,}
with spatial maps of $D_{\mathrm{min}}^{2}$ that consist of a few
active regions of limited spatial extension, separated by regions
of (locally) affine and elastic deformation. 

Other indicators of nonaffine transformations have also been used. For instance, different observables,
including the strain component along a neutral direction (say, $\epsilon_{yz}$
if the applied strain is along $\epsilon_{xy}$ in a 3D system) \citep{Schall2007},
the field of deviations \blue{of particle displacements with respect to a strictly uniform deformation,} the count of nearest-neighbor losses \citep{Chikkadi2012b}
or the identification of regions with large (marginal) particle velocities \citep{nicolas2018orientation},
are also good options to detect rearrangements. Up to differences
in their intensities, these methods were shown to provide similar
information about STs in slow flows of colloidal
suspensions \citep{Chikkadi2012b}. Alternative methods take advantage
of the irreversibility of plastic rearrangements, by reverting every
strain increment $\delta\gamma$ imposed on the system ($\gamma\rightarrow\gamma+\delta\gamma\rightarrow\gamma$)
in a quasistatic shear protocol and comparing the reverted configuration
with the original one \citep{albaret2016mapping}. Differences will
be seen in the rearrangement cores (which underwent plasticity) and
their surroundings (which were elastically deformed by the former).
To specifically target the anharmonic forces active in the core, shear
can be reverted partially, to harmonic order, by following the Hessian
upstream instead of performing a full nonlinear shear reversal \citep{lemaitre2015tensorial}.

Some reservations should now be made about the picture of
clearly separated localized transformations. First, the validity of
the binary vision distinguishing elastic and plastic regions has
been challenged for hard particles, such as grains \citep{bouzid2015non}.
\blue{\citet{de2017rheology} thus recently contended that, as particles become stiffer, this
binary picture fades into complex reorganizations of the network of contacts via cooperative motion of the particles.
It is very tempting to relate these changes to the surge of delocalized low-energy excitations in emulsions \citep{lin2016evidence} and
packings of frictionless spheres \citep{wyart2005effects,Andreotti2012} as pressure is reduced and they approach jamming from above: These
spatially extended and structurally complex soft modes are swept away upon compression and leave the floor to more
localized excitations at higher pressure and energies.
}

It is also clear that as the temperature or the shear rate are increased and the material
departs from solidity, thermal or mechanical noise may wash out the
picture of well isolated, localized events. Nevertheless, it has recently
been argued that localized rearrangements can still be identified
at relatively high temperatures. For instance, these rearrangements
leave an elastic imprint in supercooled liquids via the elastic field
that each of them induces; this imprint is revealed when one studies
suitable stress or strain correlation functions \citep{Chattoraj2013,lemaitre2014structural,illing2016strain}.

\subsubsection*{Variations}

The foregoing quantitative indicators of microscale plasticity have brought to light substantial
variations and differences between actual rearrangements and idealized STs.
Even though EPM will generally turn a blind eye to this variability,
let us shortly mention some of its salient features.

First, the sizes of rearranging regions vary
from a handful of particles in foams, emulsions and colloidal suspensions
(for instance, about 4 particles in a sheared colloidal glass, according
to \citet{Schall2007}) to a couple of dozens or hundreds in metallic
glasses (10 to 30 in the numerical simulations of \citet{fan2015crossover},
25 for the as-cast glass and 34 for its annealed counterpart in the
indentation experiment of \citet{choi2012indentation}, 200 to 700
in the shearing experiments of \citet{pan2008experimental}). Note
that, for metallic glasses, the indicated sizes are not backed out
by direct experimental evidence, but are based on activation energy
calculations \blue{and therefore strongly tied to $T_g$ \citep{Johnson2005,ZHANG2017294}}. 

\citet{albaret2016mapping} proposed a detailed numerical characterization
of plastic rearrangements in atomistic models for amorphous silicon by fitting
the actual particle displacements during plastic events 
\blue{
with the elastic displacement halos expected around spherical
STs, that is, by finding the size and spontaneous deformation $\boldsymbol{\epsilon}^{\star}$
of the inclusion which, upon embedding in an elastic medium, generates displacements that best match the observed ones 
(see Sec.~\ref{sub:gen_nonlocal_effects}).}
Although
rearrangements seem to have a typical linear size, around $3\overset{\circ}{A}$,
they found that the most robust quantity is actually the product of
$\boldsymbol{\epsilon}^{\star}$ with the inclusion volume $V_{in}$.
Furthermore, 
\blue{
the mean strain $\mathrm{Tr}(\boldsymbol{\epsilon}^{\star})/3$ 
is either positive or negative depending on the specificities of the implemented
potential (thus evidencing either a local dilation or a local compression) and represents
only  about 5\% of the shear component, which confirms that shear prevails in the
transformation.
 The orientations of the STs, \emph{i.e., } the directions of maximal shear,
were studied in greater detail in a more recent work, in 2D, where a fairly broad 
distribution of these orientations around the macroscopic shear direction was reported \citep{nicolas2018orientation}.}
Finally, \citet{albaret2016mapping} were able to reproduce the stress vs. strain
curve on the basis of the (strain-dependent) shear modulus and the
fitted local elastic strain releases $\epsilon^{\star}$. This proves
that localized plastic rearrangements \blue{surrounded by an elastic halo} are
the unique elementary carriers of the plastic response.

Secondly, the shape of the rearrangements is also subject to variations.
In quiescent systems rearrangements through string-like motion of
particles seem to be more accessible \citep{Keys2011}, even though
STs have also been claimed to be at the core of
structural relaxation in deeply supercooled liquids \citep{lemaitre2014structural}.
The application of a macroscopic shear clearly favors the latter type
of rearrangements. Albeit facilitated by the driving, in thermal systems these STs
may nonetheless be predominantly activated by thermal fluctuations
 \citep{Schall2007}. There is some (limited) indication
that the characteristics of the rearranging regions change as one
transits from mechanically triggered events to thermally activated
ones, for instance with a visible increase in the size of the region
in metallic glass models \citep{Cao2014}. 

Thirdly, owing to the granularity of the rearranging region (which
is not a continuum!), the displacements of the individual particles
in the region do not strictly coincide with an ST,
\emph{i.e.}, $\boldsymbol{r}\rightarrow\boldsymbol{r}+\boldsymbol{\epsilon}\cdot\left(\boldsymbol{r}-\boldsymbol{r}_{c}\right)$
(where $\boldsymbol{r}$ generically refers to a particle position);
incidentally, this is the major reason why the observable $D_{\mathrm{min}}^{2}$
detects plastic rearrangements.

\subsubsection*{Structural origins of rearrangements}

\begin{figure}
\begin{centering}
\includegraphics[width=0.9\textwidth]{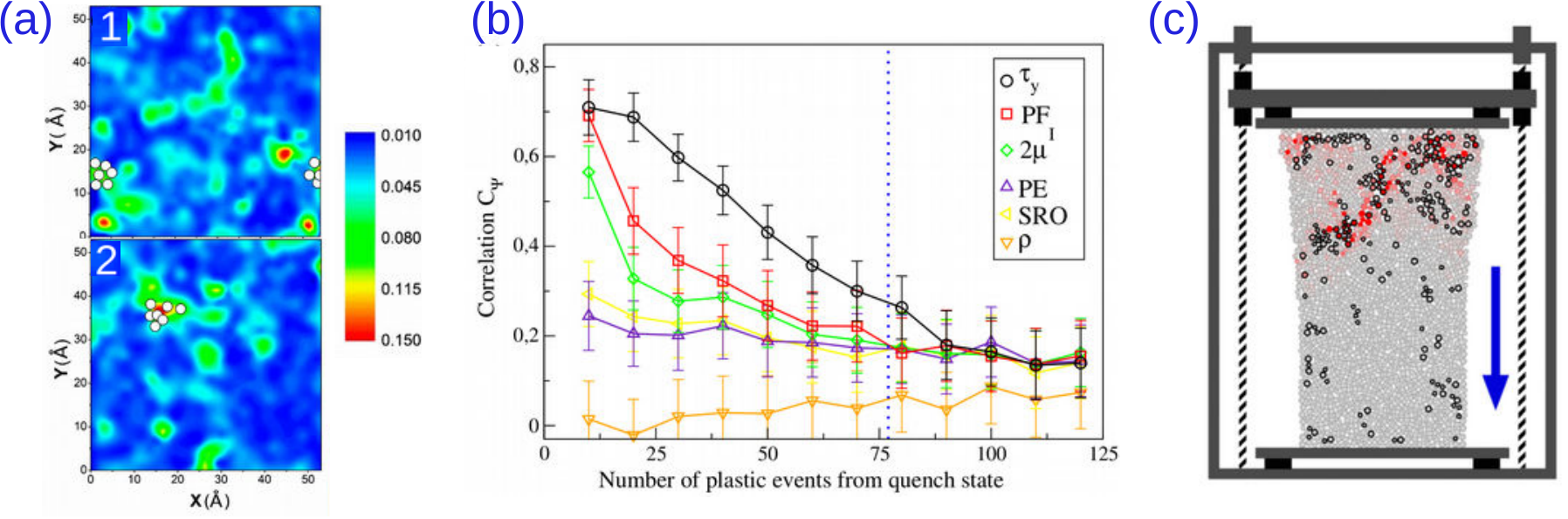}
\par\end{centering}

\caption{\label{fig:softness_vs_rearrangement}
\blue{ \emph{Comparison between different indicators based on the local structure
to predict future plastic rearrangements.}}
\textbf{(a)} Contour maps of the participation ratio in the 1\% softest vibrational modes for two 
numerical samples of a binary metallic glass \citep{ding2014soft}.
\textbf{(b)} Correlation between the locations of future rearrangements
and diverse local properties in an instantaneously quenched binary
glass model. The following properties have been considered: local
yield stress ($\tau_{y}$), participation in the soft vibrational
modes (PF), lowest shear modulus ($2\mu^{I}$), local potential energy
(PE), short-range order (SRO), local density ($\rho$). From~\citep{patinet2016connecting}.
\textbf{(c)} \blue{Identification of soft particles (thick black contours) by a machine-learning algorithm (SVM),} in a compressed granular pillar.
Particles are colored from gray to red according to their actual nonaffine motion ($D_{\mathrm{min}}^{2}$
value). From~\citep{Cubuk_PRL2015}.}
\end{figure}

What determines a region's propensity to \blue{rearrange}? \blue{Local} microstructural
properties underpinning the weakness of a region (i.e., how prone
to rearranging it is) have long been searched. In the first half of
the 20th century efforts were made to connect viscosity with the available
free volume $V_{f}$ per particle, notably by using (contested) experimental
evidence from polymeric materials \citep{batschinski1913untersuchungen,fox1950second,doolittle1951studies,williams1955temperature}.
The idea that local variations of $V_{f}$ control the
local weakness have then been applied widely to systems of hard particles
(metallic glasses, colloidal suspensions, granular materials) \citep{Spaepen1977}.
\citet{Falk1998}'s Shear Transformation Zone theory originally proposed
to distinguish weak zones prone to STs on the basis
of this criterion. \citet{hassani2016localized} have invalidated
criteria based on the strictly local free volume but showed that a
nonlocal definition distinctly correlates with the deformation field,
as do potential-energy based criteria \citep{shi2007evaluation}.
Paying closer attention to the microstructure, \citet{ding2014soft}
proved the existence of correlations between rearrangements and geometrically
unfavored local configurations (whose Voronoi cell strongly differs
from an icosahedra) in model binary alloys.
\blue{Beyond this particular example, the question of how to decipher the weakness
of a region from its local microstructure remains largely open.}

\blue{Looking beyond locally available information,} the linear response \blue{of the whole system has}
also been considered, with the hope that \blue{linearly soft regions}
will also be weak
in their nonlinear response. Regions with low elastic shear moduli
were indeed shown to concentrate most of the plastic activity \citep{Tsamados2009},
even though no yielding criterion based on the local stress or strain
is valid uniformly throughout the material \citep{Tsamados2008}.
\blue{One should mention that, albeit a local property, the \emph{local} shear modulus is best 
evaluated if the response of the \emph{global} system is computed. Indeed, approximations singling out 
a local region and enforcing an affine deformation of the outer medium are sensitive
to the size of the region and overestimate the shear modulus, because they hinder
nonaffine relaxation \citep{mizuno2013measuring}.}

Focusing on vibrational properties, \citet{brito2007heterogeneous}
and \citet{widmer-Cooper2008irreversible}
provided evidence that in hard-sphere glasses
as well as in supercooled liquids the particles that vibrate
most in the $M$ lowest energy modes (i.e., those with a high participation
fraction in the $M$ softest modes, where $M$ is arbitrarily fine-tuned),
are more likely to rearrange (note that for hard spheres the vibrational modes were computed using
an effective interaction potential). This holds true at zero temperature
\citep{manning2011vibrational} and also for metallic and polymer
glasses \citep{smessaert2015correlation} (see Fig.~\ref{fig:softness_vs_rearrangement}).
Note that this enhanced likelihood should be understood as a statistical
correlation, beating random guesses by a factor of 2 or 3 or up to
7 in some cases, rather than as a systematic criterion. However, in the cases where
the rearrangement spot is correctly predicted, the soft-mode-based prediction for the direction
of motion during the rearrangement is fairly reliable \citep{Rottler2014}.

If one is allowed to probe \emph{nonlinear} local properties, then \citet{patinet2016connecting}
showed that predictions based on the local yield stress, numerically
measured by deforming the outer medium affinely, outperform criteria
relying on the microstructure and the linear properties, as indicated
in Fig.~\ref{fig:softness_vs_rearrangement}c.

Leaving behind traditional approaches, a couple of recent
papers showed
that it is possible to train an algorithm to recognize the atomic-scale
patterns characteristic of a glassy state and spot its `soft'
regions ~\citep{Cubuk_PRL2015,Schoenholz2016,Cubuk_JPCB2016}.
In this Machine Learning approach, rather than focusing on typical
structure indicators, a large number of `features' quantified for
each particle is used, concretely M=166 `structure functions', indicating e.g.
the radial and angular correlations between an atom and its neighbors \citep{Behler_PRL2007}.
Adopting both an experimental frictional granular
packing and a bidisperse glass model, the authors focused on
the identification of local softness and its
relation with the dynamics of the glass transition. First, with
computationally costly shear simulations and measurements
of nonaffine displacements via $D_{\mathrm{min}}^{2}$,
the particles that `move'
(i.e., break out of the cages formed by their neighbors)
are identified as participating
in a plastic rearrangement and used  to
train a Support Vector Machine (SVM) algorithm.
Each particle's environment is handled as a point in the high-dimensional vector space parameterized
by the structure functions and the algorithm identifies the hyperplane that best
separates environments associated with `moving' particles and those associated with `stuck ones' in the
training set. Once trained, the algorithm is able to predict with
high accuracy if a particle  will `move' or not when the material is strained, depending
on its environment  in the \emph{quiescent} configuration, prior to shear.
%Further the `distance' to the
%hyperplane is defined as a quantity tagged "softness" and its average
%properties as a function of time after a quench are argued to characterize
%the glass transition~\citet{Cubuk_PRL2015,Schoenholz2016,Cubuk_JPCB2016}.
%This Machine Learning approach is an exciting path that opens plethora
%of opportunities in the field.

\subsection{Nonlocal effects\label{sub:gen_nonlocal_effects}}

Once a rearrangement is triggered, it will deform the medium over
long distances, in the same way as an earthquake is felt a large distance
away from its epicenter. This may trigger other rearrangements at
a distance, which rationalizes the presence of nonlocal effects in
the flow of disordered solids. Importantly, this mechanism 
relies on the solidity of the medium, which is key to the transmission
of elastic \blue{shear} waves.

These long-range interactions and the avalanches that they may generate
justify the somewhat hasty connection sketched above between the serrated
macroscopic stress curves and the abrupt localized events at the microscale.
The problem is that in the thermodynamic limit any one of these micro-events
should go unnoticed macroscopically. For sure, the thermodynamic limit
is not reached in some materials, notably those with large constituents,
such as foams and grains, but also in nanoscale experiments on metallic
glasses and numerical simulations. On the other hand,
\blue{if the sample is large compared to the ST size,}
the impact of microscopic events on the macroscopic
response could not be explained without collective effects and avalanches involving a large number
of plastic events. Since mesoscale plasticity models
intend to capture these collective effects, a description of the interactions
at play is required.

\subsubsection*{Idealized elastic propagator}

\begin{figure}
\begin{centering}
\includegraphics[width=0.9\textwidth]{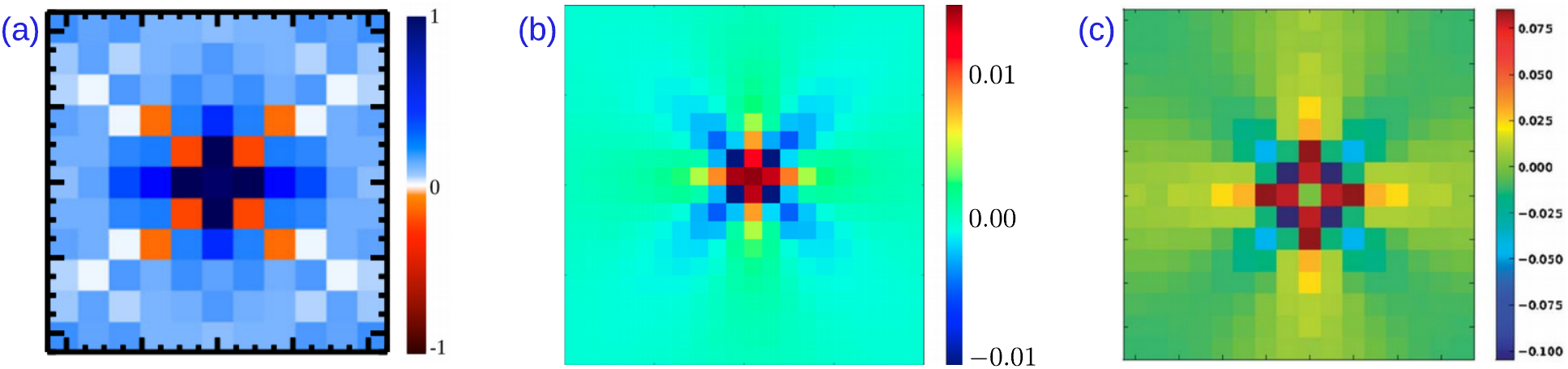}
\par\end{centering}

\caption{\label{fig:II_Average_stress_redistribution} \emph{Average stress redistribution
around a shear transformation}. 
\textbf{(a)} Experimental measurement in very dense
emulsions. Adapted from \citep{desmond2015measurement}. 
\textbf{(b)} Average response to an
imposed ST obtained in atomistic simulations with the binary Lennard-Jones glass used
by \citet{Puosi2014}. 
\textbf{(c)} Simplified theoretical form, given by Eq.~\eqref{eq:eps_xy_picard}. 
From \citep{Martens2012}. Note that the absolute
values are not directly comparable between the graphs and that in \textbf{(b-c)} the central blocks are artificially colored.}
\end{figure}

Let us start by focusing on the consequence of a single ST.
Its rotational part can be overlooked because its effect is negligible
in the far field, as compared to deformation, represented by the linear
strain tensor $\boldsymbol{\epsilon}=\frac{\nabla\boldsymbol{u}+\nabla\boldsymbol{u}^{\top}}{2}$,
where $\boldsymbol{u}$ stands for the displacement. Recall that a
shear deformation, say $\boldsymbol{\epsilon}\left(\boldsymbol{r}\approx\boldsymbol{0}\right)=\left(\begin{array}{cc}
0 & 1\\
1 & 0
\end{array}\right)$ in two dimensions (2D), consists of a stretch along the direction $\theta=\frac{\pi}{4}\,[\pi]$,
in polar coordinates, and a contraction along the perpendicular direction.
The induced displacement field $\boldsymbol{u}$ simply mirrors this
symmetry, with displacements that point outwards along $\theta=\frac{\pi}{4}\,[\pi]$
and inwards along $\theta=\frac{3\pi}{4}\,[\pi]$. This leads to a
dipolar azimuthal dependence for $\boldsymbol{u}$ and a four-fold (`quadrupolar')
one for its symmetrized gradient $\boldsymbol{\epsilon}$. More precisely,
by imposing mechanical equilibrium on the stress $\boldsymbol{\Sigma}$,
viz.,
\[
\nabla\cdot\boldsymbol{\Sigma}=0
\]
in an incompressible medium ($\nabla\cdot\boldsymbol{u}=0$) with
a linear elastic law, $\boldsymbol{\Sigma}^{\mathrm{dev}}\propto\boldsymbol{\epsilon}^{\mathrm{dev}}$
(where the superscript denotes the deviatoric part), \citet{Picard2004}
derived the induced strain field in 2D,
\begin{equation}
\epsilon_{xy}(r,\theta)\propto\frac{\cos\left(4\theta\right)}{r^{2}}.\label{eq:eps_xy_picard}
\end{equation}
Here, only one of the strain components is expressed, but the derivation
is straightforwardly extended to a tensorial form \citep{Nicolas2013b,budrikis2015universality}.
Experiments on colloidal suspensions \citep{Schall2007,Jensen2014}
and emulsions \citep{desmond2015measurement} as well as numerical
works \citep{kabla2003local,Maloney2006Amorphous,Tanguy2006} have
confirmed the relevance of Eq.~\eqref{eq:eps_xy_picard}, as illustrated
in Fig.~\ref{fig:II_Average_stress_redistribution}.

\subsubsection*{Exact induced field and variations }

The strain field of Eq.~\eqref{eq:eps_xy_picard} is valid in the far
field, or for a strictly pointwise ST. Yet, the
response can be calculated in the near field following \citet{Eshelby1957}'s
works, by modeling the ST as an elastic inclusion
bearing an \emph{eigenstrain} $\boldsymbol{\epsilon^{\star}}$, \emph{i.e.},
spontaneously evolving towards the deformed configuration $\boldsymbol{\epsilon^{\star}}$.
This handling adds near-field corrections to Eq.~\eqref{eq:eps_xy_picard}\blue{, whose
analytical expression is derived by \citet{jin2016explicit} and \citet{weinberger2005lecture} for an ellipsoidal inclusion in 3D on 
the basis of a method based on Green's function, which is probably more accessible than \citet{Eshelby1957}'s
original paper [see \citet{jin2017displacement} for the 2D version of the problem].}

Describing a \emph{plastic} rearrangement with an \emph{elastic} eigenstrain
is imperfect in principle, but the difference mostly affects the dynamics
of stress relaxation \citep{Nicolas2013b}. In fact, Eshelby's
expression perfectly reproduces the \emph{average} displacement field
induced by an ideal circular ST in a 2D binary Lennard-Jones
glass \citep{Puosi2014}, although significant fluctuations around
this mean response arise because of elastic heterogeneities. The numerical
study was then extended to the deformation of a spherical inclusion in
3D, and to the nonlinear regime, by \citet{priezjev2015plastic}. (Also
see \citep{puosi2016plastic} for the nonlinear consequences of artificially
triggered STs in a 2D glass).

Besides elastic heterogeneity, further deviations from the Eshelby
response result from the difference between an actual plastic rearrangement
and the idealized ST considered here. For instance,
\citet{Cao2014} report differences between the medium or far-field
response to rearrangements in the shear-driven regime as opposed to
the thermal regime; only the former would quantitatively obey Eshelby's
formula. It might be that the dilational component of the rearrangement,
discarded in the ideal ST, is important in the thermal
regime.

The salient points discussed above in the rheology of amorphous solids
seem to build a coherent scenario, consisting of periods of elastic
loading interspersed with swift localized rearrangements of particles.
These plastic events may interact via the long-range anisotropic elastic
deformations that they induce. These elements are the phenomenological
cornerstones of the EPM described in the following section.

\section{The building blocks of elastoplastic models (EPM)\label{sec:Building_blocks}}

\subsection{General philosophy of the models \label{sub:General_philosophy}}

The simplicity and genericity of the basic flow scenario described
above has led to the emergence of multiple, largely phenomenological,
coarse-grained models. These models are generally described as `elasto-plastic'
or `mesoscopic' models for
amorphous plasticity, or sometimes `discrete automata'. To mimick the basic flow scenario,
the material is split into
`mesoscopic' blocks, presumably of
the typical size of a rearrangement. 
\blue{These blocks are loaded elastically (R1) until a yield condition is met (R2),
at which point their stress relaxes plastically and
is redistributed to other blocks via the emission of an elastic stress field (R3);
finally, they become elastic again (R4).
Accordingly, EPM hinge on the following set of 
}
predefined rules
\citep{Rodney2011modeling}:
\begin{description}[before={\renewcommand\makelabel[1]{\bfseries ##1.}}]
\item [R1] a (default) elastic response of each mesoscopic block,
\item [R2] a local yield criterion that determines the onset of a plastic
event ($n:\,0\rightarrow1$),
\item [R3] a redistribution of the stress during plasticity that gives
rise to long-range interactions among blocks,
\item [R4] a recovery criterion that fixes the end of a plastic event
($n:\,1\rightarrow0$), 
\end{description}
where the activity $n$ is defined as $n=0$ if the block is purely elastic,
and $n=1$ otherwise.

\blue{To make it more concrete, consider the class of EPM
pioneered by \citet{Picard2005}. To fix rules R2 and R4, which define
when a region switches from elastic to plastic and conversely,
the model specifies the rates governing the transitions
\[
n:\,0\leftrightarrow1,
\]
whereas R1 and R3 are implemented via the following
equation of evolution for the stress $\sigma_{i}$ carried by block \emph{i }(where
$i$ is a \emph{d}-dimensional vector denoting the lattice coordinates
of the block),
}

\begin{equation}
\dot{\sigma}_{i}=\underset{\text{driving}}{\underbrace{\mu\dot{\gamma}}}
+\underset{\text{nonlocal contributions}}{\underbrace{\sum_{j\neq i}\mathcal{G}_{ij}n_{j}\frac{\sigma_{j}}{\tau}}}
-\underset{\text{local relaxation}}{\underbrace{|\mathcal{G}_{0}|n_{i}\frac{\sigma_{i}}{\tau}}}.\label{eq:gen_eq_of_motion}
\end{equation}
Here, the stress increment $\dot{\sigma}_{i}$ per unit time is the
sum of \blue{three contributions.} First, the externally applied simple shear contributes
a uniform amount $\dot{\sigma}=\mu\dot{\gamma}$ following Hooke's law,
with $\mu$ the shear modulus and $\dot{\gamma}$ the shear rate. 
Second, 
nonlocal plastic events (at positions \emph{j} with $n_j=1$)
release a stress of order $\sigma_{j}$ over a time scale $\tau$, which
is transmitted by the elastic propagator $\mathcal{G}_{ij}$.
Third, if the site is currently plastic (\emph{i.e.}, if $n_{i}=1$), 
\blue{its stress $\sigma_i$
relaxes. Spontaneously, it would do so at a rate $\tau^{-1}$, but, owing to the constraint of the
surrounding elastic medium, the efficiency of the process is not optimal and the rate drops to
$|\mathcal{G}_{0}| / \tau$, with $0<|\mathcal{G}_0|<1$. (During this relaxation, elastic
stress increments are still received, as are they in Maxwell's visco-elastic fluid model.)
}

\blue{
In many regards, the stress evolution described by Eq.~\eqref{eq:gen_eq_of_motion} is overly simplified:
It focuses on one stress component whereas stress is a tensor, it assumes instantaneous linear
transmission of stress releases in a uniform medium, the local stress relaxation rate is constant, etc.
In Sec.~\ref{sec:Elastic_couplings}, we will detail how these approximations
have been relaxed in part in some EPM and we will see how the phenomenological
set of rules R.1--4 can emerge in the framework of Continuum Mechanics (Sec.~\ref{sub:Continuous-approaches}).
But, for the time being, we find it important to delve into the philosophy of these models
and reflect on their goals.
}

In essence, EPM aspire to follow in the footsteps of the successes
of simplified lattice models in describing complex collective phenomena
in condensed matter and statistical physics.
\blue{Central is the} assumption that
most microscopic details are irrelevant 
for the main rheological properties and that
\blue{only} a few relevant parameters or processes \blue{really matter.}
Several reasons could be put forward to favor their use over more
realistic modeling approaches,\emph{ }e.g.,
\begin{enumerate}
\item to assess the validity of a theoretical scenario and 
identify the key physical processes in the rheology,
\item to provide an efficient simulation tool giving access to (otherwise
inaccessible) large statistics or long-time runs,
\item to \blue{facilitate} the derivation of macroscopic equations
and to bridge the gap between rheological models (constitutive laws)
and statistical physics models (sandpile models, depinning models,
Ising-like models).
\end{enumerate}
\blue{That distinct EPM highlight distinct physical ingredients
seems} to be a strong blow to the first objective.
But one should bear in mind that these materials are so diverse that
a given process (\emph{e.g.}, thermal activation) may be negligible
in some of them, and paramount in others. Less intuitive is perhaps
the role of the experimental conditions and the observables of interest
in determining the important physical ingredients,
\blue{but here come a couple of enlightening examples: 
There is no need to keep track of  previous configurations (e.g., past yield stresses)
to study steady shear, whereas this might be crucial
for oscillatory shear experiments in which the system cycles between a few configurations \citep{Fiocco2014encoding}.
}
Also, potentially universal
aspects of the yielding transition show little to no sensitivity
to the precise EPM rules, while the latter affect the details of the flow pattern.
Thus, as noted in \citep{bonn2015yield}, one should
not only select the relevant ingredients in a model only in light
of the \emph{intrinsic} importance of these effects (as quantified
for instance by dimensionless numbers), but also depending on their
bearing on the properties of interest. 

In the following, we list the physical processes that are put in the
limelight in diverse EPM and indicate for what type of materials
and in what conditions they are of primary importance.

\subsection{Thermal fluctuations\label{sub:Temperature}}

How relevant are thermal fluctuations and the associated Brownian motion of particles? This question \blue{brings us back} to the distinction
between thermal materials and athermal ones exposed in Section~\ref{sub:what_turns_them_on}. 

It is widely believed that thermal fluctuations largely contribute
to the activation of plastic events in metallic and molecular glasses,
as well as in suspensions of small enough colloids. 
\blue{For a suspension of $1.5\,\mathrm{\mu m}$-large colloids}, \citet{Schall2007} thus argue on the basis of an estimate of the activation
energy that transformations are mostly thermally activated,
with a stress-induced bias towards one direction. This will impact
the choice of the yield criterion (R2 above). EPM focusing on
thermal materials \citep{Bulatov1994,Ferrero2014} set a yield
rate based on a stress-biased Arrhenius law for thermal activation,
\emph{viz.},

\begin{equation}
\nu\left(\sigma\right)= \nu_{0}\,e^{\frac{-V^\star_\sigma}{k_{B}T}},\label{eq:activation_rate_simple}
\end{equation}
where $\nu_{0}$ is an attempt frequency, $V^\star_\sigma$ is the height of the
(smallest) potential barrier hindering the rearrangement. Recalling from Eq.~\eqref{eq:tilted_eff_potential}
that the potential is tilted by the stress $\sigma$, \emph{i.e.},
$V_\sigma=V-\Omega_{0}\sigma\gamma$, one immediately
 recovers the expression of $\nu(\sigma)$
used by \citet{Eyring1935} to calculate the viscosity of liquids
if $\sigma$ and $\gamma$ are treated as independent parameters.
\blue{To the contrary, if the local stress and elastic strain are
related by Hooke's law,} viz., $\sigma\propto\gamma$, one finds the $\gamma^{2}$-scaling
of the tilt used, e.g., in \citet{sollich1997rheology}'s Soft Glassy
Rheology model (Sec.~\ref{sub:V_SGR}). 

On the other hand, thermal activation plays virtually no role in foams
\citep{Ikeda2013} and granular materials. Consequently, EPM designed
for athermal materials \citep{Hebraud1998,Chen1991} favor a
\blue{a binary} yield criterion\emph{, viz.},
\[
\nu(\sigma)=\nu_{0}\Theta(-V^\star_\sigma)\text{ or equivalently }\nu(\sigma)=\nu_{0}\Theta(\sigma-\sigma_{y}),
\]
where $\sigma_{y}$ is the local stress threshold for yielding; a
deterministic yield criterion is recovered in the limit $\nu_{0}\rightarrow\infty$.
Incidentally, note that, in this 1D tilt picture, the
existence of favorable directions in the PEL is handled somewhat light-heartedly.
\blue{Indeed, in a high-dimensional PEL, the direction in which the loading pushes the system
may differ from that of the lowest saddle point; this fact may be particularly relevant for small systems,
whose granularity is more apparent.}

As far as  rheology is concerned,
\blue{thermal activation can be neglected if it does not trigger rearrangements much below the local
yield stress $\sigma_{y}$. By requiring that the activation rate $\nu(\sigma)$ in Eq.~\eqref{eq:activation_rate_simple} match the
 driving rate $\dot\gamma / \gamma_y$ (rescaled by the yield strain $\gamma_y$)
at a stress close to $\sigma_y$, we find that the athermal approximation is conditioned on}
\begin{equation}
\nu_0 \, e^{ \frac{-V^\star}{k_B T} } \ll \frac{\dot\gamma}{\gamma_y}.\label{eq:athermal_rheology_cond}
\end{equation}
\blue{
This criterion bears some resemblance with the limit of large P\'eclet number $\mathrm{Pe}\equiv\nicefrac{\dot{\gamma}a^{2}}{D}$,
where $a$ is the particle size and $D$ is the single-particle diffusivity
in the dilute limit \citep{Ikeda2013}, but duly takes into account the cage constraints which restrict diffusion in a dense system.
}

Now, some subtleties ought to be mentioned. An athermal system may
very well be sensitive to temperature variations, through changes
in their material properties (\emph{e.g}., dilation): For example,
the 
\blue{observation of creep} by \citet{Divoux2008} in a granular
heap submitted to cyclic temperature variations does not underscore
a possible importance of
thermal activation, but rather points to dilational effects.
Secondly, as already stressed, the relevance of thermal fluctuations
may depend on the considered level of detail: It has been argued that
they may \blue{expedite} the emergence of avalanches by breaking nano-contacts
between grains in very slowly sheared systems \citep{Zaitsev2014},
but it is very dubious that this may impact steady-shear granular
rheology.

\subsection{Driving\label{sub:III_Driving}}

Suppose that the material deforms under external
driving; how important are the specific driving conditions?

\subsubsection{Driving protocol}

\blue{
Numerical simulations have mostly considered strain-controlled (fixed $\dot{\gamma}$),
rather stress-controlled (fixed $\Sigma$), situations
(see Sec.~\ref{sub:what_turns_them_on}).
For strain-driven protocols,  the stress redistribution (R3)
operated by the elastic propagator $\mathcal{G}$
in EPM keeps the macroscopic strain fixed.
Meanwhile, the elastic response (R1) is generally
obtained by converting the macroscopic driving into local stress increments
$\mu\dot{\gamma}(t)dt$, where $dt$ is the time step and $\dot{\gamma}(t)$
is the current macroscopic strain rate.}
EPM often focus on steady shear situations, in which case $\dot{\gamma}(t)=\mathrm{cst}$.
But time-dependent
driving protocols $\dot{\gamma}=f(t)$ [or $\Sigma=f(t)$] are also
encountered, in particular step shear $\dot{\gamma}(t)=\gamma_{0}\delta(t)$ and
oscillatory shear $\dot{\gamma}(t)=\gamma_{0}\cos\left(\omega t\right)$,
which gives access to linear rheology for small $\gamma_{0}$. 

\blue{
Stress-controlled setups have
received less attention in the frame of EPM but examples can
be found in \citep{Picard2004,lin2014scaling,lin2015criticality,Homer2009,jagla2017elasto-plastic}.
In this case, the 0-Fourier
mode of the elastic propagator $\mathcal{G}$ is adjusted so that $\mathcal{G}$ keeps the macroscopic stress constant (see Sec.~\ref{sub:issues_with_approx}).
}
In creeping flows subjected to $\sigma(t)=\mathrm{cst}<\sigma_{y}$,
$\dot{\gamma}(t)$ eventually
decays to zero, often as a power law \citep{Leocmach2014}. Creep is further discussed in Sec.~\ref{sec:Relaxation}.

\subsubsection{Symmetry of the driving\label{sub:III_scalar_vs_tensorial}}

Plastic events are biased towards the direction of the external loading \citep{nicolas2018orientation}.
If the latter acts uniformly on the material, it is convenient to
focus on only one stress component, thus reducing the stress and strain
tensors to scalars. In particular, for simple shear conditions, with
a displacement gradient $\nabla\boldsymbol{u}=\left(\begin{array}{cc}
0 & \gamma(t)\\
0 & 0
\end{array}\right)$ (in the linear approximation in 2D), one may settle with
the $\epsilon_{xy}$ component \blue{of the linear strain tensor $\boldsymbol{\epsilon}= \frac{\nabla\boldsymbol{u}+\nabla\boldsymbol{u}^\top}{2}$}. 
It has the same principal strains (eigenvalues) $\pm\nicefrac{\gamma(t)}{2}$ as pure shear,
$\nabla\boldsymbol{u}=\left(\begin{array}{cc}
0 & \nicefrac{\gamma(t)}{2}\\
\nicefrac{\gamma(t)}{2} & 0
\end{array}\right)$,
but involves  a
rotational part $\boldsymbol{\omega}=\left(\begin{array}{cc}
0 & \nicefrac{\gamma(t)}{2}\\
-\nicefrac{\gamma(t)}{2} & 0
\end{array}\right)$, whereas the latter is rotationless.
These deformations are encountered \emph{locally} whenever volume
changes can be neglected; the cone-and-plate, plate-plate, and Taylor-Couette
rheometers \citep{larson1999structure} used to probe the flow of
yield-stress fluids fall in this category. For metallic glasses and
other hard materials, uniaxial compression tests (\emph{i.e}.,\emph{
}$\boldsymbol{\sigma}(t)=\sigma(t)\left(\begin{array}{cc}
1 & 0\\
0 & 0
\end{array}\right)$ in the bulk, with $\sigma(t)<0$) and tension ($\sigma(t)>0$) are
often performed \citep{priezjev2017structural}. 

Even though in several of these situations the macroscopic \blue{loading} is more
or less uniform and acts mostly on one component of the (suitably
defined) stress tensor, the other components reach finite values because
of stress redistribution. Full tensorial approaches may then be
justified \citep{Bulatov1994,Homer2009,sandfeld2014deformation}.
Recently, the influence of a tensorial, rather than scalar, description
on the flow and avalanche properties in these cases was evaluated;
it was found to be insignificant overall \citep{Nicolas2014u,budrikis2015universality},
and the effect of dimensionality to be weak \citep{liu2015driving}.
The reader is referred to Sec.~\ref{sec:VIII_Avalanches} for
more details. However, there exist a wide range of experimental setups
in which the loading is intrinsically heterogeneous, in particular
the bending, torsion, and indentation tests on hard glasses (see \citep{budrikis2015universality}
for an implementation of these tests in a finite-element-based EPM)
or the microchannel flows of dense emulsions \citep{Nicolas2013b}.
\blue{For these heterogeneous driving conditions, EPM has emerged
as a promising alternative to atomistic simulations. Nevertheless,
further upscaling is needed whenever the length scale associated with the
heterogeneous driving is very large compared to the particle size.
}

\blue{
\subsection{Driving rate and material time scales\label{sub:III.D.Rearrangement-duration-and}}
}

To resolve the flow temporally, the simplest approach is an Eulerian
method, which computes the strain increments on all blocks at each
time step from Eq.~\eqref{eq:gen_eq_of_motion}. Kinetic Monte-Carlo
methods have also been employed. They are particularly efficient in
stress-controlled slow flows, insofar as the long elastic loading phases
without plastic events are bypassed: 
\blue{The activation rate $\nu_{i}$ is calculated
for every block $i$} using a refined version of Eq.~\eqref{eq:activation_rate_simple}
and the time lapse before the next plastic event is deduced from the
cumulative rate $\nu=\sum_{i}\nu_{i}$ \citep{Homer2009}. 

In various models the finite duration of plastic
events plays a major role in the $\dot{\gamma}$-dependence of the
rheology \citep{Picard2005,Martens2012,nicolas2014rheology,liu2015driving}
or in the intrinsic relaxation of the system \citep{Ferrero2014}.
Suppose that, under slow driving, a rearrangement takes a typical
time
\footnote{A subtlety may arise if the destabilization process is dawdling, due
to the flattening of a \emph{smooth} local potential: In this particular case, $\tau_{\mathrm{pl}}$ may diverge as $\dot\gamma \to 0$.}
 $\tau_{\mathrm{pl}}$. For overdamped dynamics, one expects this
time scale to be the ratio between an effective microscopic viscosity
$\eta_{\mathrm{eff}}$ and the elastic shear modulus $\mu$ \citep{Nicolas2013b},
\emph{viz.}, 
\[
\tau_{\mathrm{pl}}\sim\nicefrac{\eta_{\mathrm{eff}}}{\mu},
\]
while for underdamped systems $\tau_{\mathrm{pl}}$ is associated
with the persistence time of localized vibrations. If
$\tau_{\mathrm{pl}}$ competes with the driving time scale $\tau_{\dot{\gamma}}\equiv \gamma_{y} / \dot{\gamma}$,
where $\gamma_{y}$ is the local yield strain, then plastic events
will be disrupted by the driving. The rate dependence of the macroscopic
stress may then stem from this disruption \citep{nicolas2014rheology}.

\blue{At even lower driving rates,} one reaches a regime where
$\tau_{\mathrm{pl}}\ll\tau_{\dot{\gamma}}$ and individual
rearrangements \blue{become insensitive to the driving.} But the latter
may still affect \emph{avalanches} of rearrangements (\emph{i.e.},
the series of plastic events that would still be triggered by an initial
event, were the driving turned off). 
\blue{Indeed, since the size of avalanches is expected to diverge in an athermal system
in the limits of vanishing shear rate and infinite system size $L$, their duration $\tau_\mathrm{av}(L)$, bounded below by the signal propagation time between rearranging regions, may become arbitarily large as $\dot{\gamma}\to 0$.}
\blue{While most EPM turn a blind eye to the delays due to shear wave propagation,}
some works have bestowed them a central role in the finite shear-rate
rheology \citep{Lemaitre2009,lin2014scaling} and there have been endeavors to represent
this propagation in a more realistic way in EPM \citep{nicolas2015elastic,karimi2016continuum} (see Sec.~\ref{sub:IV_Finite_Elements}).
Sections~\ref{sub:VIII_finite_strain_rates} and \ref{sub:VII_flow_curve_mean}
will provide more details on the influence of vanishingly small shear rates on the flow curve.

The quasistatic limit is reached when 
\begin{equation}
\frac{\tau_{\mathrm{pl}}}{\tau_{\dot{\gamma}}}\to 0\text{ and }\frac{\tau_{\mathrm{av}}(L)}{\tau_{\dot{\gamma}}}\to 0
\label{eq:III_quasistatic}
\end{equation}
and the athermal criterion of Eq.~\eqref{eq:athermal_rheology_cond}
is satisfied. In that case, the material remains in mechanical equilibrium at all times
and its trajectory in the PEL is rate-independent. 
\blue{Atomistic simulations can then be simplified by} 
applying a small strain increment at each
step and letting the system relax athermally to the local energy minimum
\citep{Maloney2004}. 
\blue{The EPM counterparts of these quasistatic simulations are called
\emph{extremal or quasistatic models} and have been studied intensively}
\citep{baret2002extremal,talamali2012strain,lin2014scaling,jagla2017elasto-plastic}.
In these models, the algorithm identifies the least stable
site at each step and increases the applied stress  enough to
destabilize it. From this single destabilization an avalanche of plastic
events may ensue.
\blue{The material time scales are then naturally brushed aside, while 
connection to real time is lost.}

\text{}

\subsection{Spatial disorder in the mechanical properties}

\begin{figure}
\begin{centering}
\includegraphics[width=1\textwidth]{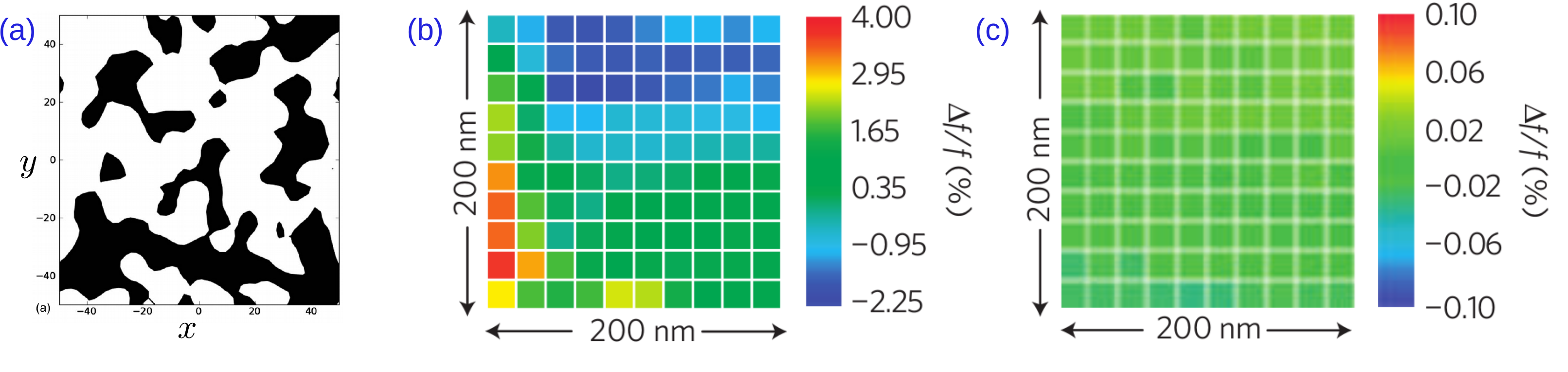}
\par\end{centering}

\caption{\label{fig:III_Spatial-variations-of} \emph{Spatial variations of the mechanical
and configurational properties of glasses.} 
\textbf{(a)} Maps of the weaker
local shear modulus in a 2D Lennard-Jones glass. Black (white) represents
values larger (smaller) than the mean value. Distances are in particle
size units. From \citep{Tsamados2009}. 
\textbf{(b-c)}
Maps of the local contact-resonance frequency, which is related to
the indentation modulus, measured by atomic force acoustic microscopy
in \textbf{(b)} a bulk metallic glass (PdCuSi) and 
\textbf{(c)} a crystal ($\mathrm{SrTiO}_{3}$). The latter clearly appears to be mechanically more homogeneous. 
The radius of contact is of order $10\,\mathrm{nm}$.
From \citep{wagner2011}.}

\end{figure}

Glasses, and more generally amorphous solids, are 
\blue{mechanically heterogeneous.} Indeed, there have been both
experimental and numerical reports on the heterogeneity of the local
elastic moduli (see Fig.~\ref{fig:III_Spatial-variations-of}) and
the energy barriers on the mesoscale \citep{Tsamados2008,Zargar2013}.
Yet, the extent to which this disorder impacts the rheology remains
unclear. This uncertainty is reflected in EPM. Some models feature
no such heterogeneity \citep{Hebraud1998,Picard2005}, while it 
\blue{plays a central role in others} \citep{sollich1997rheology,Langer2008}. In the latter case, heterogeneity
is generally implemented in the form of a disorder on the yield stresses
or energy barriers. Let us mention a couple of examples. \citet{sollich1997rheology}'s
Soft Glassy Rheology model 
\blue{introduces exponentially distributed energy barriers;
this translates into an even broader distribution of activation rates
via Eq.~\eqref{eq:activation_rate_simple} and leads to 
a transition from
Newtonian to non-Newtonian rheology as temperature is reduced
(see Sec.~\ref{sub:V_SGR} for more details on the model).
}
In their EPM centered on metallic glasses, \citet{Li2013} modify the
free energy required for the activation of an event depending on the
free volume created during previous rearrangements. Finally, amorphous composite
materials, \emph{i.e.}, materials featuring meso/macro-inclusions
of another material, can be \blue{modeled} as a patchwork of regions of
high and low yield stresses \citep{tyukodi2015finite} or high and
low elastic moduli \citep{chen2015elasticity}. In the latter case,
macroscopic effective shear and bulk moduli can be derived.

More generally, for single phase materials, the survey of the above
results gives the impression that disorder has bearing on the rheology
when thermal activation plays an important role.
On the other hand, the impact of a yield stress disorder may be less
important in athermal systems. 
In fact,
\citet{agoritsas2015relevance} showed that disorder
is irrelevant in the mean-field description of athermal plasticity
originally proposed by \citet{Hebraud1998}, in the low shear rate
limit: It only affects the coefficients of the
rheological law, \blue{and not the functional shape.}

\subsection{Spatial resolution of the model}

On a related note, how important is it to spatially resolve
an EPM? In what cases can one settle with a mean-field
approach blind to spatial information? Clearly, there are situations
in which mean field makes a bad candidate, in particular when the
driving or flow is macroscopically heterogeneous, when the focus is
on spatial correlations \citep{Nicolas2014s} or \blue{even on some} critical properties
\citep{lin2014scaling,liu2015driving}. But a mean field analysis
could suffice in many other situations. For example, \citet{Martens2012}
showed that the flow curve obtained with their spatially resolved
EPM \blue{at finite shear rates} can be predicted on the basis of mean-field reasoning,
\blue{whereas spatial correlations and avalanches are thought to
impact the macroscopic stress at vanishing shear rates \citep{liu2015driving,roy2015rheology}.} 
Similarly, \citet{Ferrero2014}'s EPM-based simulations
confirmed mean-field predictions by \citet{bouchaud2001anomalous}
regarding thermal relaxation of amorphous solids in some regimes;
but not without finding discrepancies in others. In the latter regimes,
spatial correlations thus seemed to play a significant role.

The discussion about whether spatial resolution is required to describe
global quantities is not settled yet. It has been argued that, owing
to the long range of the elastic propagator (which decays radially
$r^{-d}$ in \emph{d} dimensions), mean-field arguments should generically
hold in amorphous solids \citep{dahmen1998gutenberg,Dahmen2009}.
However, it has been realized that the non-convex nature of the propagator
(alternatively positively and negatively along the azimuthal direction)
undermines this argument \citep{Budrikis2013} and results in much
larger fluctuations than the ones produced by a uniform stress redistribution
\citep{Talamali2011,Nicolas2014u,Lin2014Density}. Mean-field predictions have
been tested against the results of lattice-based
models simulations of a sheared amorphous solid close to (or in) the
limit of vanishing driving, with a focus on the statistics of stress-drops
or avalanches, and \emph{non-mean-field} exponents were found for the power-law
distribution of avalanche sizes \citep{Talamali2011,Budrikis2013,lin2014scaling,liu2015driving}.
This question is addressed in greater depth in Sec.~\ref{sec:VIII_Avalanches}.

In this review, we will put the spotlight on spatially resolved models,
which are not exactly solvable in general and require a numerical
treatment. When relevant, we will discuss how a mean-field treatment
can be performed to obtain analytical results.

\subsection{Bird's eye view of the various models}

To conclude this section, some of the main EPM are classified in Table~\ref{tab:classification_EPM}.

\begin{table}[h!]

\caption[Classification of some of the main EPM in the literature.]{Classification of some of the main EPM in the literature.\\
\mbox{ }\label{tab:classification_EPM}
}

  \centering
  \begin{tabular}{>{\bfseries\centering}m{24mm} | m{48mm} | l | >{\raggedright\arraybackslash\hspace{0pt}\footnotesize}m{48mm} | >{\raggedright\arraybackslash\hspace{0pt}\footnotesize\scshape}m{36mm} } 
  % \begin{tabular}{>{\centering}m{20mm}>{\centering}m{37mm}>{\centering}m{20mm}>{\centering}m{50mm}>{\centering}m{32mm}}
 \hline
 \hline
 \textbf{Yielding}   &  \textbf{Reference}   &    \textbf{Features} &   \normalfont{\textbf{Remarks}} & \normalfont{\textbf{Proposed for}} \tabularnewline 
 \hline
        \multirow{4}{*}{Activated} & 
        \citet{Bulatov1994} \emph{et seq.} & \includegraphics[width=4mm]{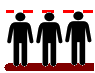}  \includegraphics[width=4mm]{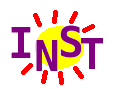} \includegraphics[width=4mm]{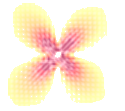} &
         Propagator computed on hexagonal lattice & amorphous solids, in particular glasses and glass-forming liquids \\ 
        %\cline{2-5}
         
         & \citet{Homer2009} \emph{et seq.} &
          \includegraphics[width=4mm]{single_barrier_icon}  \includegraphics[width=4mm]{inst_icon} \includegraphics[width=4mm]{quadrupole_icon}    &
        Stress redistribution computed with Finite Elements &  metallic glasses \\
        %\cline{2-5}

         & \citet{Ferrero2014} &
          \includegraphics[width=4mm]{single_barrier_icon} \includegraphics[width=4mm]{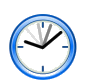} \includegraphics[width=4mm]{quadrupole_icon}&
          Pl. events of finite duration &  amorphous solids \\
         %\cline{2-5}
          
         & \citet{sollich1997rheology} [SGR model]& \includegraphics[width=4mm]{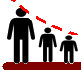}  \includegraphics[width=4mm]{inst_icon} \includegraphics[width=4mm]{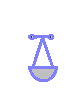}    &
         Effective activation temperature accounts for mechanical noise & soft materials (foams, emulsions, etc.) \\ 
         
         & \citet{merabia2016creep}& \includegraphics[width=4mm]{exp_barrier_icon}  \includegraphics[width=4mm]{inst_icon} \includegraphics[width=4mm]{quadrupole_icon}    &
         Variant of SGR, with a \emph{bona fide} (instead of mechanical) temperature  & Polymeric and metallic glasses under creep \\

  \hline      
        \multirow{7}{*}{Threshold} &     
          \citet{Chen1991} & \includegraphics[width=4mm]{single_barrier_icon}  \includegraphics[width=4mm]{inst_icon} \includegraphics[width=4mm]{quadrupole_icon}    &
          Propagator computed on square spring network & earthquakes \\
          %\cline{2-5}
       
          & \citet{baret2002extremal}$^\dag$~\emph{et seq.}, \citet{Talamali2011}$^\S$, \citet{Budrikis2013}$^\ddag$~\emph{et seq.} & \includegraphics[width=4mm]{exp_barrier_icon}  \includegraphics[width=4mm]{inst_icon} \includegraphics[width=4mm]{quadrupole_icon}  &  Uniform distribution of barriers; coupled to a moving spring ($\S$,$\ddag$); or stress controlled with extremal dynamics ($\dag$) or adiabatic driving
          ($\ddag$) & amorphous solids, notably metallic glasses \\                 
         % stress controlled ($\dag$: extremal dynamics, $\ddag$: adiabatic driving); or strain controlled by spring coupling ($\S$ and $\ddag$) 
          %\cline{2-5}

          & \citet{Dahmen2011} & \includegraphics[width=4mm]{single_barrier_icon}  \includegraphics[width=4mm]{inst_icon}  \includegraphics[width=4mm]{other_interaction_icon}  &
           `Narrow' distribution of thresholds; static and dynamic thresholds differ; mean-field approach &  granular matter and akin \\
           %\cline{2-5}
            
          & \citet{Hebraud1998} & \includegraphics[width=4mm]{single_barrier_icon}  \includegraphics[width=4mm]{inst_icon}  \includegraphics[width=4mm]{other_interaction_icon}   &
           Finite yield rate above threshold; stress redistributed as white noise & soft materials (dense suspensions) \\
           %\cline{2-5}
         
         & \citet{Picard2005}, \citet{Martens2012} & \includegraphics[width=4mm]{single_barrier_icon} \includegraphics[width=4mm]{finite_icon} \includegraphics[width=4mm]{quadrupole_icon}    &
          Finite yield rate above threshold; pl. events of finite duration & amorphous solids \\
          %\cline{2-5}
           
         & \citet{nicolas2014rheology} \emph{et seq} & \includegraphics[width=4mm]{exp_barrier_icon} \includegraphics[width=4mm]{finite_icon} \includegraphics[width=4mm]{quadrupole_icon}  &
           Pl. events end after finite \emph{strain}& soft athermal amorphous solids \\
           %\cline{2-5}
           
         & \citet{lin2014scaling}& \includegraphics[width=4mm]{single_barrier_icon}  \includegraphics[width=4mm]{inst_icon} \includegraphics[width=4mm]{quadrupole_icon}    &
			  Stress- and strain- control protocols & soft amorphous solids \\
  \hline      	  		
  
  		\multirow{3}{24mm}{`Continuous' approaches} &     
 	  		\citet{onuki2003plastic} & \includegraphics[width=4mm]{single_barrier_icon} \includegraphics[width=4mm]{finite_icon}  \includegraphics[width=4mm]{quadrupole_icon}    &
 	  		  Dynamical evolution on a periodic potential; dipolar propagator due to opposite dislocations
 	  		  & 2D crystalline and glassy solids \\
 	  		  %\cline{2-5}
 	  		  
 	  		&     
 	  		\citet{Jagla2007} & \includegraphics[width=4mm]{exp_barrier_icon} \includegraphics[width=4mm]{finite_icon} \includegraphics[width=4mm]{quadrupole_icon}    &
 	  		  Dynamical evolution on random potential; propagator computed via compatibility condition & amorphous solids \\
 	  		  %\cline{2-5}
 	  		  
 	  		& \citet{Marmottant2013} & \includegraphics[width=4mm]{single_barrier_icon} \includegraphics[width=4mm]{finite_icon} \includegraphics[width=4mm]{other_interaction_icon} & 
 	  		Overdamped evolution in a periodic potential; pl. events of finite duration; no stress redistribution & foams \\
  \hline      
  \multicolumn{5}{l}{\textsc{Symbols -- } \emph{Barrier distribution}: \includegraphics[width=4mm]{single_barrier_icon}  single value /
  \includegraphics[width=4mm]{exp_barrier_icon} distributed (exponentially, unless otherwise specified).} \\
  \multicolumn{5}{l}{\hspace{17mm} \emph{Plastic events}: \includegraphics[width=4mm]{inst_icon} instantaneous / \includegraphics[width=4mm]{finite_icon} finite duration.}\\
   \multicolumn{5}{l}{\hspace{17mm} \emph{Interactions}: \includegraphics[width=4mm]{quadrupole_icon} Quadrupolar elastic propagator /
   \includegraphics[width=4mm]{other_interaction_icon} other (uniform redistribution, noise temperature, etc.)} \\
  \hline      
  \hline      
    \end{tabular}
\end{table}

\section{Elastic couplings and the interaction kernel\label{sec:Elastic_couplings}}

A key feature of EPM is to allow plastic events to interact via an
elastic deformation field, which can generate avalanches. In this
respect, the choice of the elastic interaction kernel may significantly
impact the results of the simulations \citep{Martens2012,Budrikis2013}.
%In particular, the physically motivated differences between the elastic
%kernels used in related problems, such as the depinning of an elastic
%line on a disordered substrate or the propagation of a crack in a
%solid presented in Sec~\ref{sec:Related_topics},
%may be at the origin of the quantitative discrepancies in the scalings
%of their avalanches (see Section~\ref{sec:VIII_Avalanches}). 
This fairly technical section presents the various idealizations of the interaction kernel
that have been used in the literature on amorphous solids, by increasing
order of sophistication. We relate this level of sophistication
with the nature of the developments that were sought.

\subsection{Sandpile models and first-neighbor stress redistribution\label{sub:IV.Sandpile-models}}

\begin{figure}
\begin{centering}
\includegraphics[width=1\textwidth]{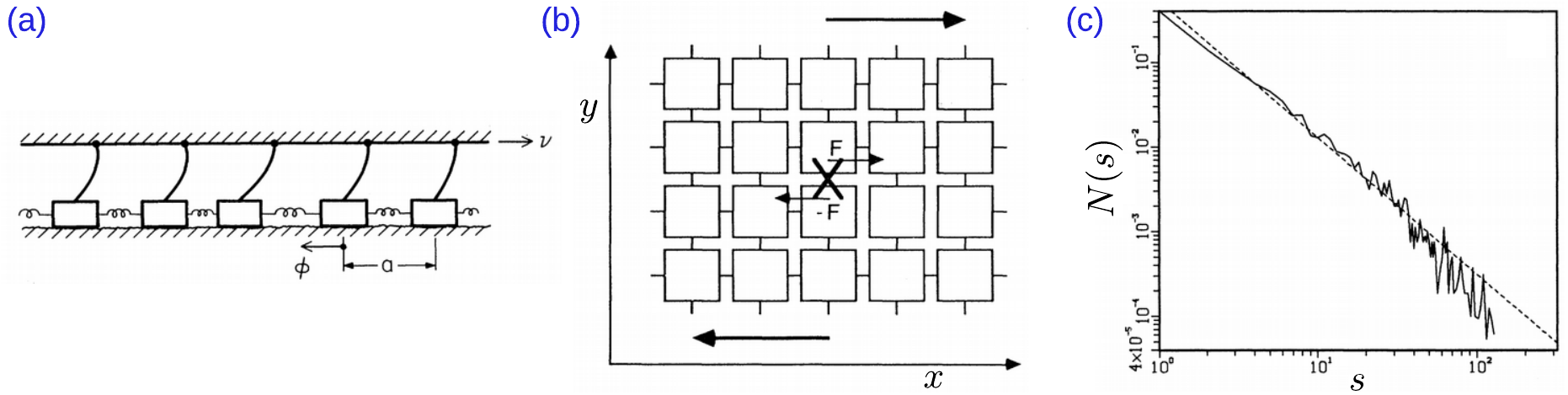}
\par\end{centering}

\caption{\label{fig:Burridge_Knopoff} \emph{Blocks-and-springs models.} 
\textbf{(a)} Sketch of the discrete 1D Burridge
and Knopoff model. From \citep{carlson1994dynamics}. 
\textbf{(b)} Sketch of the effect
of a bond rupture in \citet{Chen1991}'s spring network model. 
\textbf{(c)} Distribution
of avalanche sizes, in terms of number of broken bonds, in the model sketched in (b). Adapted from
\citep{Chen1991}. }
\end{figure}

\blue{
EPM owe much to the quake-ridden
scientific grounds on which they burgeoned at the beginning of the 1990s, marked by the advent of seminal models for
earthquakes and avalanches.
}

As a paradigmatic example for earthquakes, consider the celebrated model
by \citet{burridge1967model}, whose main features are concisely reviewed
in \citep{carlson1994dynamics}. It focuses on the fault separating
two slowly moving tectonic plates. This region is structurally weak
because of the gouge (crushed rock powder) it is made of; thus, failure
tends to localize \blue{along its length.} In the model, the contact points
across the fault are represented by massive blocks and the compressive
and shear forces acting along it are modeled as springs, as sketched
in Fig.~\ref{fig:Burridge_Knopoff}a. Due to these forces, the initially
pinned (stuck) blocks may slide during avalanches. More precisely,
in the continuous, nondimensional 1D form, the displacement $U(x,t)$
at time $t$ of the material at position $x$ reads
\begin{equation}
\ddot{U}=\xi^{2}\frac{\partial^{2}U}{\partial x^{2}}+vt-U-\phi(\dot{U}).\label{eq:burridge_EoM}
\end{equation}
Here, the left-hand side (lhs) is related to inertia, the second derivative
on the right-hand side (rhs) is of compressive origin,
and the loading term $vt$ due to the motion of the plate as well
as the displacement $-U$ contribute to a shear term. Finally, $\phi(\dot{U})$
is a velocity-dependent frictional term. Had Coulomb's
law of friction been used, it would have been constant for $|\dot{U}|\neq0$,
but the original model assumed velocity weakening, \emph{i.e.}, a decrease
of $|\phi(\dot{U})|$ with $|\dot{U}|$. At $\dot{U}=0$, the function
$\phi$ is degenerate, which allows static friction to exactly cancel
the sum of forces on the rhs of Eq.~(\ref{eq:burridge_EoM}), so
the blocks remain pinned at a fixed position $U$ until the destabilizing
forces $\xi^{2}\frac{\partial^{2}U}{\partial x^{2}}+vt$ exceed a
certain threshold. Phenomenologically, simulations of the model show
frequent small events (with a power-law distribution of cumulative
slip) and rare events of large magnitude, in which the destabilization
of a number of sites close to instability results in a perturbation
of large amplitude \citep{otsuka1972simulation,carlson1994dynamics}.

Important in the above model is the effect of the pinning force $\phi$
at $\dot{U}=0$. It entails that the destabilizing action caused by
the depinning of a site (via the diffusive term in Eq.~\eqref{eq:burridge_EoM})
is fully screened by its neighbors, unless they yield too. Such first-neighbor
redistribution of strain is readily simulated using cellular automata,
which can be interpreted as sandpile models: Whenever a column of
sand, labelled $(i,j)$, gets too high with respect to its neighbors
(say, for convenience, whenever $\sigma_{i,j}\geqslant4$), some grains
at its top are transferred to the neighboring columns, with the following
discharge rules in 2D: 
\begin{eqnarray}
\sigma_{i,j}\geqslant4:\,\sigma_{i,j} & \rightarrow & \sigma_{i,j}-4\nonumber \\
\sigma_{i\pm1,j} & \rightarrow & \sigma_{i\pm1,j}+1\label{eq:IV_sandpile_redistribution}\\
\sigma_{i,j\pm1} & \rightarrow & \sigma_{i,j\pm1}+1\nonumber 
\end{eqnarray}
where $\sigma$ is the height difference between columns. The sandpile
is loaded by randomly strewing grains over it in a quasistatic manner.
The study of these systems soared in the late 1980s and early 1990s,
whence the concept of self-organized criticality emerged
\citep{Bak1987}. According to the latter, avalanches naturally
drive the sandpiles toward marginally stable states, with no characteristic
lengthscale for the regions on the verge of instability, hence the
observation of scale-free frequency distributions of avalanche sizes.
As an aside, let us mention that this approach has not been used only
for earthquakes \citep{carlson1989properties,sornette1989self,bak1989earthquakes,ito1990earthquakes}
and avalanches in sandpiles, it has also been transposed to the study
of integrate-and-fire cells \citep{corral1995self} and forest fires
\citep{chen1990deterministic}, \emph{inter alia}.

In seismology, these models have been fairly successful in reproducing
the \citet{gutenberg1944frequency} statistics of earthquake. This
empirical law states that the frequency of earthquakes of (energy)
magnitude 
\begin{equation}
M_{e}=\frac{2}{3}\log(E)-2.9,\label{eq:IV_magnitude_scale}
\end{equation}
where $E$ is the energy release, in a given region obeys the power
law relation, $\log\,P(m\geqslant m_{0})\simeq-bm_{0}+\mathrm{cst}$,
where $b\simeq0.88$, or equivalently 
\[
p(E)\sim E^{-\tau}\text{, with }\tau=1+\frac{2}{3}b\approx1.5.
\]
For the sake of accuracy, we ought to say that there exist several
earthquake magnitude scales besides that of Eq.~\eqref{eq:IV_magnitude_scale}.
They roughly coincide at not too large values; in fact, $M_{e}$ is
not the initial Richter scale. More importantly, the value of the
exponent $b\in[0.8,1.5]$ depends on the considered earthquake catalog,
and notably on the considered region. For sandpile-like models, various
exponents have been reported: $\tau\approx1$ in 2D and $\tau\approx1.35$
in 3D, with no effect of disorder of the yield stresses \citep{bak1989earthquakes},
whereas the exponent for the mean-field democratic fiber bundle close
to global failure is $\tau=\nicefrac{3}{2}$ (see Sec.~\ref{sub:X_Fibre_bundles}).
More extensive numerical simulations led to the values $\tau\simeq1.30$
\citep{lubeck1997numerical}, or $\tau\simeq1.27$ \citep{chessa1999universality},
for the 2D \citet{Bak1987} sandpile model. 

\citet{olami1992self} modified the model to make the redistribution
rule of Eq.~\eqref{eq:IV_sandpile_redistribution} non-conservative.
In this sandpile picture, this would correspond to a net loss of grains,
which seems unphysical; but in \citet{burridge1967model}'s block-and-spring
model the non-conservative parameter simply refers to the fraction
of strain which is absorbed by the driving plate during an event,
instead of being transferred to the neighbors. Interestingly, as non-conservativeness
increases, criticality is maintained, insofar as the avalanche distribution
$p(E)$ remains scale-free, even though the critical exponent $\tau$
gradually gets larger. Only when less than 20\% of the strain is transferred
to the neighbors does a transition to an exponential distribution
occur. The dynamics then become more and more local with increasing
dissipation, until the blocks completely stop interacting, when the
\blue{transfer} is purely dissipative.

However, unlike the redistribution of grains in the sandpile model,
elastic interactions are actually long-ranged, as we wrote in Section~\ref{sub:gen_nonlocal_effects}.
In particular, in the deformation of amorphous solids, no pinning
of the region surrounding an event can be invoked to justify the restriction
of the interaction to the first neighbors.

\subsection{Networks of springs}

Accordingly, a more realistic account of the long-ranged elastic propagation
is desirable. Unfortunately, the complexity of the \emph{bona fide}
Eshelby propagator obtained from Continuum
Mechanics hampers its numerical implementation and use, so most
studies have relied on simplified \blue{variants thereof.}

First, in the spirit of the classical description of a solid as an
assembly of particles confined to their positions by interactions
with their neighbors, the material was modeled as a system of blocks
connected by ``springs'' of stiffness $\kappa$ and potential energy
\[
\frac{1}{2}\kappa\left(\boldsymbol{u}_{i}-\boldsymbol{u}_{j}\right)^{2},
\]
where $\boldsymbol{u}_{i}$ is the displacement of block \emph{i.}
Note that this expression for the potential energy entails noncentral
forces, so that the ``springs'' can bear shear forces; some details
about the difference with respect to networks of conventional springs
are presented in Section~\ref{sub:X_Fibre_bundles}. The pioneering
steps towards EPM followed from the application of such spring network
models to the study of rupture. For this purpose, each bond is endowed
with a random threshold, above which it yields and redistributes the
force that it used to bear. In their study of a 2D triangular lattice
with central forces, \citet{hansen1989rupture} measured the evolution
of the applied force $F$ with the displacement $u$; this evolution
starts with a phase of linear increase, followed by a peak and a smooth
decline until global failure. The $F(u)$ curves for different linear
lattice sizes $L$ roughly collapsed onto a master curve if $F$ and
$u$ were rescaled by $L^{-\nicefrac{3}{4}}$. In addition, just before
failure, the distribution of forces in the system was ``multifractal'',
with no characteristic value. 

\citet{Chen1991} considered a square lattice of blocks and ``springs'', 
sketched in Fig.~\ref{fig:Burridge_Knopoff}b.
%, in which the force
%propagator, for a point force directed along $\boldsymbol{e_{x}}$
%and located at the origin, reads
%\[
%\mathcal{G}_{x}\left(\boldsymbol{r}\right)=\int_{0}^{2\pi}dk_{x}\int_{0}^{2\pi}dk_{y}\frac{1-e^{ik_{x}}}{4-2\left[\cos\left(k_{x}%\right)+\cos\left(k_{y}\right)\right]}e^{i\left(\boldsymbol{r}\cdot\boldsymbol{k}\right)}.
%\]
The rupture of a spring triggers the release of a \emph{dipole} of opposite
point forces (generating vorticity) on neighboring blocks. \blue{In passing, note
the subtle difference with respect to}
the double force-dipole \blue{(often called quadrupole)} describing \emph{irrotational}
local shear (see Sec.~\ref{sub:Pointwise-idealisation-of}), \blue{which has
a distinct anisotropy.} Contrary to \citet{hansen1989rupture},
they allowed broken springs to instantly regenerate to an unloaded
state, after redistribution of their load. Physically, this discrepancy
parallels a change of focus, from brittle materials to earthquakes,
for which the external loading due to tectonic movements is assumed
to be by far slower than the healing of bonds. For a quasistatic increase
of the load, the model displays intermittent dynamics and scale-free
avalanches, and a power-law exponent $\tau=1.4$ was reported in 2D, in semi-quantitative
agreement with the Gutenberg-Richter earthquake statistics.

\subsection{Elastic propagators\label{sub:Pointwise-idealisation-of}}

\blue{
After the discrete vision promoted by block-and-spring models, let us momentarily turn to a continuous
description of the amorphous solid. The free energy of the material can be expressed in terms
of the displacement field $\boldsymbol{u}(\boldsymbol{r})$ as 
\begin{equation}
F[\boldsymbol{u}] = \int \psi(\boldsymbol{\epsilon}(\boldsymbol{r})) \,d^d\boldsymbol{r}, \label{eq:IV_free_energy}
\end{equation}
where the free energy density  $\psi$ depends on the strain tensor $\boldsymbol{\epsilon}(\boldsymbol{r})=\frac{\nabla \boldsymbol{u}(\boldsymbol{r}) +\nabla \boldsymbol{u}(\boldsymbol{r})^\top}{2}$ (in the linear approximation), 
because rigid transformations cost no energy. In the quiescent system, $\psi$ reaches its minimum in the reference
strain state $\boldsymbol{\epsilon}(\boldsymbol{r})= \boldsymbol{0}$, viz.,
\begin{equation}
\frac{\delta F}{\delta \boldsymbol{u}(\boldsymbol{r})} = \nabla \cdot \sigma (\boldsymbol{r})= \boldsymbol{0}, \label{eq:IV_mech_eq}
\end{equation}
where $\boldsymbol{\sigma} = d\psi / d \boldsymbol{\epsilon}$ is the stress. 
But a plastic rearrangement taking place at, say, 
$\boldsymbol{r}=\boldsymbol{0}$ will shift the reference state to $\boldsymbol{\epsilon}= \boldsymbol{\epsilon^{\mathrm{pl}}}$
in a small region around $\boldsymbol{0}$. Because of its embedding, this region cannot deform freely. Therefore, the plastic strain $\boldsymbol{\epsilon^{\mathrm{pl}}}$ (more generally
known as eigenstrain) will induce a nonlocal elastic response in the surrounding medium, which was worked out
by \citet{Eshelby1957} for ellipsoidal inclusions in a linear elastic solid.
} 

\subsubsection{Pointwise transformation in a uniform medium}

\blue{
\citet{Picard2004} simplified the calculation of the elastic response
by supposing that the shear transformation (ST) has vanishing linear size $a \to 0$. Prior to the
ST, the material is linear elastic, viz.,
\begin{equation}
\boldsymbol{\sigma}=-p\boldsymbol{I}+2\mu\boldsymbol{\epsilon},\label{eq:IV_elastic_stress}
\end{equation}
where $p$ is the pressure, and incompressible, $\nabla\cdot\boldsymbol{u}=0$.
After the ST, the reference state is shifted and Eq.~\eqref{eq:IV_mech_eq}
turns into
\begin{equation}
\nabla\cdot\boldsymbol{\sigma}\left(\boldsymbol{r}\right)+\boldsymbol{f}^{\prime}\left(\boldsymbol{r}\right)=0.\label{eq:IV_mechanical_equilibrium}
\end{equation}
where the source term generated by the plastic strain at the origin reads
$\boldsymbol{f}^{\prime}\left(\boldsymbol{r}\right)=-2\mu\nabla\cdot\left[\boldsymbol{\epsilon^{\mathrm{pl}}}a^{d}\delta\left(\boldsymbol{r}\right)\right]$.
}
 
The solution of Eq.~\eqref{eq:IV_mechanical_equilibrium}
is well known in hydrodynamics and involves the Oseen-Burgers
tensor $\boldsymbol{\mathcal{O}}(\boldsymbol{r})=\frac{1}{8\pi\mu r}\left(\mathbb{I}+\frac{\boldsymbol{r}\otimes\boldsymbol{r}}{r^{2}}\right)$ in 2D,
with $\mathbb{I}$ the identity matrix, viz.,
\begin{equation}
\boldsymbol{u}\left(\boldsymbol{r}\right)=\int\boldsymbol{\mathcal{O}}(\boldsymbol{r}-\boldsymbol{r^{\prime}})\boldsymbol{f}^{\prime}\left(\boldsymbol{r}^{\prime}\right).
\label{eq:IV_Oseen_displ_tensor}
\end{equation}
In the unbounded 2D plane,
setting coordinates such that $\boldsymbol{\epsilon^{\mathrm{pl}}}=\left(\begin{array}{cc}
0 & \epsilon_{0}\\
\epsilon_{0} & 0
\end{array}\right)$, the response to $\boldsymbol{f}^{\prime}\left(\boldsymbol{r}\right)$
in terms of \emph{xy}-component of the stress reads
\begin{equation}
\sigma_{xy}\left(\boldsymbol{r}\right)=2\mu\epsilon_{0}a^{2}\mathcal{G}^{\infty}\left(\boldsymbol{r}\right)\text{ with the propagator }\mathcal{G}^{\infty}\left(\boldsymbol{r}\right)\equiv\frac{\cos\left(4\theta\right)}{\pi r^{2}},\label{eq:propagator_real_space}
\end{equation}
where $(r,\theta)$ are polar coordinates. This field is shown in
Fig.~\ref{fig:II_Average_stress_redistribution}c. Reassuringly,
in the far field it coincides with the response to a cylindrical
Eshelby inclusion. 

As a short aside, let us mention a variant to these calculations,
which \blue{underscores} the connection with deformation processes
in a crystal. This variant is reminiscent of Eshelby's a cut-and-glue
method, \blue{whereby} an ellipsoid is cut out of the material, deformed,
and then reinserted. Following earlier endeavors by \citet{benzion1993earthquake},
\citet{tuzes2016disorder} carved out a square around the rearrangement,
instead of an ellipsoid, displaced its edges to mimick shear, and
then glued it back. This is tantamount to inserting four edge dislocations
in the region and also yields an Eshelby-like quadrupolar field.

Rather than focusing on unbounded media, it is convenient to work
in a bounded system with periodic boundary conditions and with a general
plastic strain field $\boldsymbol{\epsilon^{\mathrm{pl}}}(\boldsymbol{r})$.
Switching to Fourier space ($\boldsymbol{r}\leftrightarrow\boldsymbol{q}\equiv\left(q_{x},q_{y}\right)$),
the counterpart of Eq.~\eqref{eq:propagator_real_space} is then
\begin{equation}
\sigma_{xy}\left(\boldsymbol{q}\right)=2\mu\mathcal{G}\left(\boldsymbol{q}\right)\boldsymbol{\epsilon^{\mathrm{pl}}}(\boldsymbol{q})\text{ where }\mathcal{G}\left(\boldsymbol{q}\right)=-\frac{4q_{x}^{2}q_{y}^{2}}{q^{4}}.\label{eq:elastic_propagator_Fourier}
\end{equation}
Note that the frame is sometimes defined such that $\boldsymbol{\epsilon^{\mathrm{pl}}}=\left(\begin{array}{cc}
\epsilon_{0} & 0\\
0 & -\epsilon_{0}
\end{array}\right)$; in this case, $\mathcal{G}\left(\boldsymbol{q}\right)=\frac{-\left(q_{x}^{2}-q_{y}^{2}\right)^{2}}{q^{4}}$.
In practice, the system will generally be discretized into a (square)
lattice, which allows one to use a Fast Fourier Transform routine and
restrict the considered wavenumbers to $q_{x},\,q_{y}=\frac{2\pi n}{L},\,n\in\left\{ \left\lceil \frac{-L}{2}\right\rceil ,\ldots,\left\lfloor \frac{L}{2}\right\rfloor \right\} $.

Besides, because of dissipative forces, quantified by an effective
viscosity $\eta_\mathrm{eff}$, the strain rate $\boldsymbol{\dot{\epsilon}}$
in the ST cannot be infinite and a rearrangement
will last for a finite time $\tau_\mathrm{pl}\sim\nicefrac{\eta_\mathrm{eff}}{\mu}$
(see Sec.~\ref{sub:III.D.Rearrangement-duration-and}). Therefore,
in each numerical time step, the plastic strain $\boldsymbol{\epsilon^{\mathrm{pl}}}$
implemented in Eq.~\eqref{eq:elastic_propagator_Fourier} will be the
strain increment $\boldsymbol{\delta\epsilon^{\mathrm{pl}}}$
during that step. This amounts to saying that, locally, 
\blue{dissipative forces make the rearrangement
gradual, while stress is redistributed instantaneously to the
rest of the medium} (because of the assumption of mechanical equilibrium),
so that there is no time dependence in the elastic propagator
in Eq.~\eqref{eq:elastic_propagator_Fourier}.

\subsubsection{Technical issues with pointwise transformations and possible remediations \label{sub:issues_with_approx} }

The idealized elastic propagator in Eq.~\eqref{eq:elastic_propagator_Fourier}
brings on some technical issues. Firstly, its slow ($\propto r^{-d}$)
radial decay raises convergence problems in periodic space.
Indeed, the fields created by the periodic images of each plastic
event have to be summed, but the sum converges only conditionally
in real space, \emph{i.e.}, depends on the order of summation. This is reflected
by the singularity of $\mathcal{G}\left(\boldsymbol{q}\right)$ near
$\boldsymbol{q}=\boldsymbol{0}$. In polar crystals, such a difficulty
also arises, when computing the Madelung energy, but may be solved
with the \citet{ewald1921berechnung} method. Here, we make
use of the conserved quantities to state that $\mathcal{G}\left(\boldsymbol{q}=\boldsymbol{0}\right)=0$
in a stress-controlled system and $\mathcal{G}\left(\boldsymbol{q}=\boldsymbol{0}\right)=-1$
in a strain-controlled system. Another possibility is to sum the images
in an arbitrary order that is compatible with convergence. These distinct
implementations match in the far field, but differ in the near field,
which leads to different organizations for the flow \citep{Budrikis2013}.

Secondly, on a periodic lattice, one should in principle compute the
periodic sum
\[
\mathcal{G}^{sum}\left(\boldsymbol{q}\right)\equiv\sum_{\boldsymbol{n}\in\mathbb{Z}^{d}}\mathcal{G}\left(\boldsymbol{q}+2\pi\boldsymbol{n}\right)
\]
if, at the lattice nodes, one wishes the backward discrete Fourier
transform of $\mathcal{G}^{sum}\left(\boldsymbol{q}\right)$ to coincide
with the solution $\mathcal{G^{\infty}}(\boldsymbol{r})$ for an unbounded
medium. However, the high-frequency components in $\mathcal{G}\left(\boldsymbol{q}\right)$,
due to the spurious singularity of $\mathcal{G^{\infty}}(\boldsymbol{r})$
at $\boldsymbol{r}=\boldsymbol{0}$ [Eq.~\eqref{eq:elastic_propagator_Fourier}],
make the periodic sum diverge. In practice, wavenumbers outside the
first Brillouin zone $]-\pi,\pi]^{d}$ are plainly discarded, which
comes down to solving Eqs.~\eqref{eq:IV_elastic_stress}-\eqref{eq:IV_mechanical_equilibrium}
on the periodic lattice, rather than in the continuum. Nevertheless,
spurious fluctuations in the response field are sometimes observed;
the problem is mitigated by using a finer grid and smoothing the obtained
field \citep{Nicolas2014u}.

\subsubsection{Variations: Soft modes and lattice symmetries; tensoriality; convection \label{sub:IV_soft_modes}}

All in all, many technical details of the implementation of the elastic
propagator appear to affect the spatial organization of the flow \citep{Talamali2011},
but leave the qualitative picture and (apparently) the scaling laws
unaltered. However, an aspect that seems to be crucial is the need
to preserve the eigenmodes of the propagator $\mathcal{G}(\boldsymbol{q})$
associated with zero energy. These so called soft modes \blue{(or null modes)}
$\boldsymbol{\epsilon^{pl}}$, satisfying 
\[
\forall\boldsymbol{q},\,\mathcal{G}\left(\boldsymbol{q}\right)\boldsymbol{\epsilon^{pl}}\left(\boldsymbol{q}\right)=0,
\]
cost no elastic energy; their deployment is thus favored
by the dynamics \citep{tyukodi2015depinning}. Their importance is
further explained in Sec.~\ref{sub:V_Validity-of-MF}. It turns
out that the eigenmodes of $\mathcal{G}(\boldsymbol{q})$ in Eq.~\eqref{eq:elastic_propagator_Fourier}
are the Fourier modes (plane waves); among these, the soft
modes are those with wavevectors $\boldsymbol{q}$ making an angle $\pm\nicefrac{\pi}{4}$
with respect to the principal direction of the plastic strain tensor
$\boldsymbol{\epsilon^{pl}}$. 

In particular, under simple shear with velocity
direction $x$ and velocity gradient along $y$, the emergence of
a uniform shear band along $x$ should produce no elastic stress in
the medium, at least if such a band emerges uniformly. However, misaligned
lattice axes (not directed along $x$ or $y$) are incompatible
with such a shear band (which would then have sawtooth-like edges)
and artificially suppress the soft modes \citep{tyokodi2016thesis}.
More generally, the use of a regular lattice in EPM may be questioned, insofar
as the localization of plastic events is sensitive to variations of stress redistribution
in the near field \citep{Budrikis2013}. The scalings of avalanche sizes, however, seem to be mostly insensitive
to these details. Indeed, these details do not affect the long-range \blue{interactions between blocks.}

On another note, the foregoing calculations focused on the \emph{xy}-shear
stress component, because of the macroscopic stress symmetry, thus
promoting a scalar description. It is straightforward to generalize
the reasoning to a fully tensorial form; but it turns out that, for
setups with uniform loading, the tensorial extension has virtually
no effect \citep{Nicolas2014u}. The 
statistics of avalanches of plastic events found in tensorial models and in scalar models
\blue{are similar, up to tenuous differences:} 
The values of the critical exponents at the yielding transition reported for scalar
models \citep{SandfeldJSTAT2015} are close to those obtained in the
corresponding tensorial models \citep{budrikis2015universality}. Similarly,
moving from 2D to 3D does not introduce qualitative changes and scaling
relations are preserved \citep{budrikis2015universality,liu2015driving}.

\blue{
Another refinement consists in taking into account the possible anisotropy of the solid.
\citet{cao2018soft} argued that anisotropic shear moduli ($\mu_2 \neq \mu_3)$
may result from the shear softening of well annealed solids just before their macroscopic
failure. The displacement field induced by an ST,
\begin{equation}
u_r(r,\theta) = \frac{ \cos(2\theta) }{1+\delta \cos(4\theta)},\ u_\theta(r,\theta)=0, 
\end{equation}
where $\theta=0$ denotes the principal direction of positive stretch, then tends to
concentrate along `easy' axes as the anisotropy parameter
$\delta= (\mu_3 - \mu_2) / (\mu_3 + \mu_2)$ increases.
}

\blue{Translational invariance may also be broken, if the system is 
confined between walls, as in a microchannel, instead of being periodic.
If there is no slip at the wall, a method of images allows 
the derivation of the elastic propagator for the bounded medium \citep{Picard2004,Nicolas2013b}.
Plastic events are found to} relax stress faster, for a given eigenstrain, if they occur
close to the walls. Such changes in boundary conditions
affect the spatial organization of the flow, 
but not the critical properties at the yielding transition \citep{SandfeldJSTAT2015}.

Finally, despite the convenience of using a fixed
lattice grid with static elasto-plastic blocks, physically these blocks
should be advected by the flow. In a bounded medium, a coarse version
of advection can be implemented by incrementally shifting the blocks
along the streamlines without altering the global shape of the lattice
\citep{Nicolas2013b}. On the other hand, with periodic boundary conditions,
the deformation of the frame results in the shift of the periodic
images with respect to the simulation cell; advection thus requires
to compute the elastic propagator afresh, in the deformed frame \citep{Nicolas2014u}.

\subsection{Approaches resorting to Finite-Element methods\label{sub:IV_Finite_Elements}}

\begin{figure}
\begin{centering}
\includegraphics[width=0.9\textwidth]{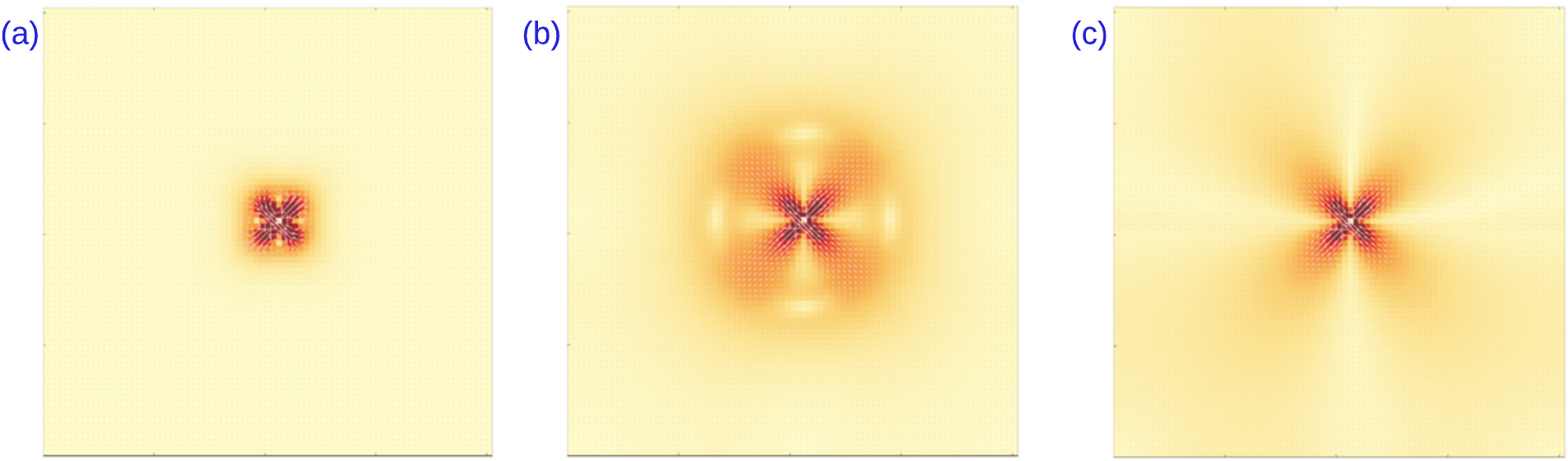}
\par\end{centering}

\caption{\label{fig:IV_FE_propagation}Average displacement field induced by
an ST in an underdamped elastic medium, computed
with a basic Finite Element routine. The plotted snapshots correspond
to different delays after the transformation was (artificially) triggered
at the origin: 
\textbf{(a)} $\Delta t=2$, 
\textbf{(b)} $\Delta t=10$,
\textbf{(c)} $\Delta t=1000$. Red hues indicate larger displacements. 
Adapted from \citep{nicolas2015elastic}}

\end{figure}

Albeit computationally more costly, Finite-Element (FE)-based computations
of stress redistribution overcome some limitations of the foregoing
approaches and offer more flexibility. The FE method solves the continuum
mechanics equation 
\blue{associated with the free energy of Eq.~\eqref{eq:IV_free_energy}}
by interpolating the strain $\boldsymbol{\epsilon}$ and stress $\boldsymbol{\sigma}$ within each element of a meshgrid  from the values of the displacements and
point forces at the nodes of the element. 
\blue{As far as EPM are concerned,
the default elastic response of each block is generally assumed linear,
so that the free energy density in Eq.~\eqref{eq:IV_free_energy} reads
$\psi(\boldsymbol{\epsilon})= \frac{1}{2} \, \boldsymbol{\epsilon} \cdot \boldsymbol{C} \cdot\boldsymbol{\epsilon}$ and
$\boldsymbol{\sigma}= \boldsymbol{C} \cdot \boldsymbol{\epsilon}$.
}

If mechanical equilibrium is maintained at all times, as in Eq.~\eqref{eq:IV_mech_eq}, the 
response to an ST is obtained by equilibrating the elastic
stress $\boldsymbol{C} \boldsymbol{\epsilon^\mathrm{pl}}$ that it releases.
Using a triangular mesh refined around the
ST-bearing element, \citet{SandfeldJSTAT2015} demonstrated
that the computed stress field coincides with the elastic propagator
of Eq.~\eqref{eq:propagator_real_space} in the limit of a pointwise ST. 
But these researchers also found that a coarser mesh made of
uniform square elements gives results that are almost as good, except
in a near-field region of a handful of elements' radius. The flexibility
of the method was then exploited to study the quasistatic deformation
of the system beyond the periodic boundary conditions, e.g., in a
bounded medium and with free surfaces, and with inhomogeneous loading
conditions (indentation, bending, etc.). Universal, but non-mean-field,
statistics of avalanches of plastic events were reported in these
diverse conditions \citep{budrikis2015universality} (also see Sec.~\ref{sec:VIII_Avalanches}).

In an earlier endeavor \citep{Homer2009,Homer2010}, each ST
zone consisted of \emph{several} elements of a triangular mesh which
all bore an eigenstrain. As the size of this zone increases, the redistributed
stress field accurately converged to the theoretical Eshelby field.
Zones made of 13 elements were deemed quite satisfactory in this respect. \citet{Homer2010three} 
later extended the approach to 3D.
Dynamics were brought into play via the implementation of an event-driven
(Kinetic Monte Carlo) scheme determining the thermal activation of
STs, in the wake of the pioneering works of \citet{Bulatov1994}.
The cooling of the system, its thermal relaxation and its rheology
under applied stress were then studied. Macroscopically homogeneous
flows were observed at low stresses and/or high temperatures, whereas
the strain localized at low temperature for initially unequilibrated
(zero residual stress) systems, which was not necessarily supported
by experimental data. More systematic strain localization at low temperature
was found by \citet{Li2013}, who incorporated the processes of free
volume creation during plastic rearrangements and subsequent free
volume annihilation (see Sec.~\ref{sub:VI_Structural-softening}).

The capabilities of FE methods were further exploited by \citet{nicolas2015elastic}
to go beyond the assumption of elastic homogeneity
\blue{and capture the fluctuations caused by elastic disorder,
notably those evidenced in the response to an ideal ST
in MD simulations of a binary glass \citep{Puosi2014}.
For this purpose, stiffness matrices $\boldsymbol{C}$
were measured \emph{locally} in that model glass, in mesoscale} regions of 5
particles in diameter. The authors found a broad distribution $p(\mu)$ of
local shear moduli, with a relative
dispersion of around 30\%, and marked anisotropy (i.e., one direction
of shear being much weaker than the other one).
\blue{The dispersion of $p(\mu)$ explains the fluctuations observed in MD. Indeed,
an FE-based model in which the shear modulus of each element was
randomly drawn from $p(\mu)$ displayed comparable fluctuations in the response
to an ideal ST, triggered by suitably moving the nodes of
a chosen element. On the other hand, accounting for anisotropy was less critical.
Furthermore, in the FE description inertial and
viscous terms were restored in Eq.~\eqref{eq:IV_mechanical_equilibrium}.
This gives access to the transient elastic reponse,
involving the propagation of shear waves.} Exploiting this opportunity, \citet{karimi2016inertia}
analyzed the effect of inertia on the avalanche statistics
and compared it with results from atomistic simulations. (Note that the effect
of a delay in signal propagation had already been contemplated in
an \emph{effective} way by \citet{Lin2014Density}, while, for the
same purpose, \citet{papanikolaou2016shearing} introduced a pinning
delay in his EPM based on the depinning framework.) It was then possible
to investigate the influence of the damping strength on the rheology
of the elastoplastic system, which was indeed done by \citet{karimi2016continuum}.
Using a Maxwellian fluid description for blocks in the plastic regime
and an unstructured mesh, these researchers found trends qualitatively
very similar to what is observed in MD when the friction
coefficient is varied. 
%The authors noted the possible relevance of
%their study for earthquake dynamics. 

\section{Mean-field treatments of mechanical noise\label{sec:Mechanical_noise}}

The previous section has shed light on the modeling of the elastic
propagator, \emph{i.e.}, the effect of a single rearrangement on the
surrounding elastic medium. In practice, however, several rearrangements
may occur simultaneously. The rate $\zeta_{i}(t)$ of stress increment
experienced by a given block (say, site $i$) at time $t$ is then a
sum of contributions from many sites, i.e., using Eq.~\eqref{eq:gen_eq_of_motion},
$\zeta_{i}(t)=\sum_{j\neq i}n_{j}\mathcal{G}_{ij}\frac{\sigma_{j}}{\tau}$,
where $n_{j}$ denotes the plastic activity of site $j$. Due to its
fluctuating nature, this quantity is often referred to as \emph{mechanical
noise}. By rewriting Eq.~\eqref{eq:gen_eq_of_motion} as 
\begin{equation}
\frac{\partial}{\partial t}\sigma_{i}(t)=\mu\dot{\gamma}- n_{i}\frac{ |\mathcal{G}_0|  \sigma_{i}(t)}{\tau}+\zeta_{i}(t),\label{eq:gen_eq_of_motion2}
\end{equation}
one can readily see that, in combination with the external loading and the dynamical rules
governing $n_{i}$, the mechanical noise signal $\left\{ \zeta_{i}(t)\right\} $
fully determines the local stress evolution. All one-point properties (such
as the flow curve, the density of plastic sites, the distribution
of local stresses, etc.) can be obtained by averaging the local properties
at $i$ over time \blue{(and over $i$ if ergodicity is broken).}
%, instead of averaging over configurations and over space. 
This shows the central role of $\left\{ \zeta_{i}(t)\right\} $
in determining these properties. Unfortunately, this signal is complex,
as it stems from interacting plastic events throughout the system;
nevertheless, mean-field approaches suggest to substitute it with
a simpler `mean' field.

\subsection{Uniform redistribution of stress\label{sub:uniform_stress_redistrib}}

The mechanical noise can be split into :
\begin{itemize}
\item a constant background $\left\langle \zeta_{i}\right\rangle $, which
contributes to a drift term $\mu\dot{\gamma}_{i}^{\mathrm{eff}}\equiv\mu\dot{\gamma}+\left\langle \zeta_{i}\right\rangle $
in Eq.~\eqref{eq:gen_eq_of_motion2}, and
\item zero-average fluctuations $\delta\zeta_{i}(t)$. 
\end{itemize}
Owing to the infinite range and slow decay of the elastic propagator
($\propto r^{-d}$ in d-dimensional space, see Sec.~\ref{sub:gen_nonlocal_effects}),
site $j$ is significantly coupled to a large number of other sites.
This large connectivity has led some researchers to \blue{brush aside} fluctuations
in favor of the average drift term. Along these lines, in the framework of
Picard's EPM, which features a constant rate $\tau^{-1}$ of yield above a uniform
threshold and a constant rate $\tau_\mathrm{res}^{-1}$ of elastic recovery, viz., 
\begin{equation}
n:\,0\underset{\tau_{\mathrm{res}}^{-1}}{\overset{\tau^{-1}\Theta(\sigma-\sigma_{y})}{\rightleftharpoons}}1,\label{eq:V_Picard_rates}
\end{equation}
\citet{Martens2012} averaged Eq.~\eqref{eq:gen_eq_of_motion2}
over time and found an analytical expression for the flow
curve, which reproduces the simulation results to a large extent, \blue{at least at reasonably large shear rates.}
It also correctly predicts the destabilization of the homogeneous
flow leading to shear-banding for a range of model parameters, in
particular at large $\tau_{\mathrm{res}}$.

In fact, the neglect of fluctuations would be rigorously justified
if the system were infinite and the propagator $\mathcal{G}$ were positive. The latter criterion is for instance
fulfilled in a simple quasistatic model in which sites yield past
a threshold $\sigma_{y}$ and redistribute the released stress ($\delta\sigma_{i}$)
\emph{uniformly} to the other $N-1\approx N$ sites \citep{dahmen1998gutenberg},
viz.,

\begin{eqnarray*}
\sigma_{i}>\sigma_{y} & : & \sigma_{i}\rightarrow\sigma_{i}-\delta\sigma_{i}\\
 &  & \sigma_{j}\rightarrow\sigma_{j}+\frac{\delta\sigma_{i}}{N},\,\forall j\neq i.
\end{eqnarray*}
The simplicity of the model allows analytical progress. A first approach
consists in treating the distances $x_{i}=\sigma_{y}-\sigma_{i}$
to the threshold $\sigma_{y}$ as independent variables in the system
and sorting them in ascending order ($i\rightarrow1,2,\ldots$). An
avalanche will persist as long as the stress increment $\frac{\delta\sigma_{1}}{N}$
due to the yielding of the most unstable site suffices to make the
second most unstable fail, viz., $\frac{\delta\sigma_{1}}{N}>x_{2}$.
Using an argument along these lines in a model featuring disorder
in the yield thresholds ($\sigma_{y}\rightarrow\sigma_{y,i}$) and
post-failure weakening (i.e., when site \emph{i} yields, the threshold
is restored to a lower value $\sigma_{y,i}(t+1)<\sigma_{y,i}(t)$),
\citet{dahmen1998gutenberg} were able to rationalize the existence
of a regime of power-law distributed avalanches and a regime of runaway,
system-spanning avalanches.

Alternatively, owing to the similarity of the simplified problem with
force-driven depinning, one can make use of the machinery developed
in the latter field. Transversal scaling arguments and renormalization
group expansions \citep{fisher1997statistics} then allow one to derive
scalings for different properties of the system in the quasistatic
limit, such as the size of avalanches. Note that this method was initially
applied to the depinning problem and to earthquakes. Only later
on was it claimed to be much more general and to have bearing on very
diverse systems exhibiting intermittent dynamics or ``crackling noise''
\citep{Sethna2001}, in particular the yielding transition of amorphous
solids. Recently, these
mean-field scaling predictions about avalanche sizes, shapes, and
dynamics have been used to fit experimental data, in metallic
glasses subjected to extremely slow uniaxial compression \citep{Antonaglia2014,Dahmen2009}
as well as in compacted granular matter \citep{denisov2016universality}. We will come back
to this point in Sec.~\ref{sec:VIII_Avalanches}.

\subsection{Random stress redistribution\label{sub:random_stress_redistribution}}

\subsubsection{Deviations from uniform mean field}

The underpinning of the foregoing mean-field
approach has been called into question. Theoretically, the argument
based on the long range of the interactions is undermined by the fact
that these interactions are sometimes positive and sometimes negative \citep{Budrikis2013}.
The ratio of fluctations over mean value of the
stress increments, estimated by \citet{Nicolas2014u} in an EPM, diverges at low shear
rates $\dot{\gamma}$, which points to the failure of the mean-field
theory, according to Ginzburg and Landau's criterion.
Numerically, some lattice-based simulations do indeed reveal departures
from mean-field predictions for the critical exponents \citep{Budrikis2013,Lin2014Density,liu2015driving}.
For instance, in these simulations, near $\dot{\gamma}\rightarrow0$,
the distribution of avalanche sizes $S$ follows a power law $P(S)\sim S^{-\tau}$
with an exponent $\tau$ that deviates from the $\tau=\nicefrac{3}{2}$
value predicted by mean field (see Sec.~\ref{sec:VIII_Avalanches}
for details).

\subsubsection{The H\'ebraud-Lequeux model\label{sub:V_Hebraud_Lequeux}}

To improve on the hypothesis of a constant mean field $\dot{\gamma}^{\mathrm{eff}}$,
fluctuations of the mechanical noise need to be accounted for. In
the crudest approximation, they can be substituted by random white
noise $\zeta^{\left(w.n.\right)}(t)$, with $\left\langle \zeta^{\left(w.n.\right)}\right\rangle =0$.
This turns Eq.~\eqref{eq:gen_eq_of_motion2} into a biased Brownian
walk for the local stresses, in the elastic regime $n_{i}=0$. \citet{Hebraud1998}'s
 model was developed along these lines.
The ensuing stochastic equation (Eq.~\eqref{eq:gen_eq_of_motion2}
with $\zeta_{i}(t)\rightarrow\dot{\gamma}^{\mathrm{eff}}+\zeta^{\left(w.n.\right)}(t)$ and $\tau\rightarrow 0$)
can be recast into a probabilistic Fokker-Planck-like equation operating
on the distribution $P(\sigma,t)$ of local stresses $\sigma$, viz.,
\begin{equation}
\frac{\partial P(\sigma,t)}{\partial t}=-\mu\dot{\gamma}\frac{\partial P(\sigma,t)}{\partial\sigma}+D(t)\frac{\partial^{2}P(\sigma,t)}{\partial\sigma^{2}}-\frac{\Theta\left(|\sigma|-\sigma_{y}\right)}{\tau_c}P(\sigma,t)+\Gamma(t)\delta(\sigma),\label{eq:HL_equation}
\end{equation}
where the diffusive term $D\frac{\partial^{2}P}{\partial\sigma^{2}}$
on the rhs arises from the fluctuations acting on $\sigma_{i}$, with
a coefficient $D(t)$ assumed to be proportional to the number of
plastic sites $\Gamma(t)\equiv\tau_c^{-1}\int_{|\sigma^{\prime}|>\sigma_{y}}P(\sigma^{\prime},t)d\sigma^{\prime}$,
viz., $D(t)=\alpha\Gamma(t)$. The first term on the rhs of Eq.~\eqref{eq:HL_equation} is a drift
term, which amalgamates $\dot{\gamma}^{\mathrm{eff}}$ with $\dot{\gamma}$;
the last two terms correspond to the failure of overloaded sites (above
$\sigma_{y}$) on a timescale $\tau_c$ and their rebirth
at $\sigma=0$ due to stress relaxation. The resulting mean-field
equations can be solved in the limit of vanishing shear rates $\dot{\gamma}$
\citep{olivier2011glass,agoritsas2015relevance}. For a coupling constant
$\alpha<\nicefrac{1}{2}$, diffusion vanishes at low shear rates,
with $D\propto\dot{\gamma}$, a yield stress $\Sigma_{y}>0$
is obtained and the average stress obeys $\Sigma\simeq\Sigma_{y}+k\dot{\gamma}^{\nicefrac{1}{2}}$,
with $k>0$, in the low shear rate limit. 
For $\alpha>\nicefrac{1}{2}$,
the system behaves like a Newtonian liquid.

\subsubsection{Fraction of sites close to yielding \label{sub:V_distances_to_yield}}

The diffusive term introduced in Eq.~\eqref{eq:HL_equation} impacts the distribution
%  with respect to the constant mean-field approximation
%(Sec.~\ref{sub:uniform_stress_redistrib})
of sites close to yield, i.e., at distances $x\equiv|\sigma|-\sigma_{y}\ll1$ from the yield
threshold $\sigma_{y}$. On these
short distances, or, equivalently, in the limit of short time scales
$\Delta t$, the back-and-forth diffusive motion over typical distances
$\propto\sqrt{\Delta t}$ prevails over the drift in the random walk.
Therefore, for $\dot{\gamma}\rightarrow0$, determining the distribution
$P(x)$ is tantamount to finding the concentration of Brownian particles
near an absorbing boundary at $x=0$ (yielding): The well-known solution
is a linear vanishing of the concentration near $x=0$, viz., $P(x)\sim x$
for $x\approx0$ \citep{Lin2014Density,lin2016mean}. This result
ought to be compared with $P(x)\sim x^{0}$ for drift-dominated problems,
such as depinning. \citet{Lin2014Density} further claim that this
discrepancy is at the origin of the differences in scaling behavior
between the depinning transition [$v\propto\left(F-F_{c})^{\beta}\right)$
with $\beta<1$] and the flow of disordered solids [$\dot{\gamma}\propto\left(\Sigma-\Sigma_{Y})^{\beta}\right)$
with $\beta>1$ generally].

\subsection{Validity of the above `mean-field' approximations\label{sub:V_Validity-of-MF}}

The foregoing paragraphs have presented distinct levels of `mean-field'
approximations. Now we enquire into their range of validity and record the results in Table~\ref{tab:fluct_handling}.

\subsubsection{Uniform mean field \label{sub:V_uniform_mean_field}}

\blue{Neglecting} fluctuations in the constant mean-field approach
\blue{makes sense} in the drift-dominated regime, \emph{i.e.}, when
\blue{
$|\dot{\gamma}^{\mathrm{eff}}| \Delta t \gg|\int_{0}^{\text{\ensuremath{\Delta}t}}\delta\zeta(t^{\prime})dt^{\prime}|$
}
on the considered time window $\Delta t$, with the notations of Sec.~\ref{sub:uniform_stress_redistrib}.
With interactions that change signs, this excludes vanishing shear rates
or too small $\Delta t$. But at higher shear rates, this
approach appears to correctly predict the avalanche scaling exponents
in the EPM studied by \citet{liu2015driving}.

\subsubsection{White-noise fluctuations}

Complemented with Gaussian fluctuations, the approximation is valid
beyond the drift-dominated regime. In fact,
\blue{if the distribution of global mechanical noise $\delta \zeta$ (i) is Gaussian-distributed}
% 1) is bounded by 
% 2) is a Gaussian or any narrow-tailed power-law distribution 
%in particular $w_1(\delta\zeta)=o(|\delta\zeta|^{-3})$,
and (ii) has no significant time correlations, the noise fluctuations $\delta \zeta$
can be replaced by Gaussian
white noise in Eq.~\eqref{eq:gen_eq_of_motion2} \citep{lin2016mean}.
\blue{Note that, if a single plastic event releases stress fluctuations with a distribution $w_1$ such as $w_1(\delta\zeta)\propto |\delta\zeta|^{-1-k}$ with $k>2$,
condition (i) will be fulfilled as soon as condition (ii) is (i.e., events are uncorrelated). 
}

\blue{Provided that} the mechanical
noise fulfils the above criteria (i) and (ii), all models based on similar rules for plasticity thus
fall in the universality class of the H\'ebraud-Lequeux model in the limit of large systems.
In particular, for
coupling parameters $\alpha$ such that the diffusivity $D(t)$ goes
to zero at $\dot{\gamma}\rightarrow0$, their flow curves will follow
a Herschel-Bulkley behavior $\Sigma=\Sigma_{y}+k\dot{\gamma}^{n}$
with $n=\nicefrac{1}{2}$ in the low shear rate limit. This holds
true in the presence of disorder on the local yield thresholds $\sigma_{y}$ 
\citep{agoritsas2015relevance}
and for plastic events that do not relax the local stress strictly
to zero, but to a low random value \citep{agoritsas2016nontrivial}.
On the other hand, should the shear modulus of elastic blocks or the
relaxation time \blue{depend on $\dot\gamma$,} the exponent $n$
will deviate from $\nicefrac{1}{2}$ \citep{agoritsas2016nontrivial}.

\subsubsection{Heavy-tailed fluctuations\label{sub:V_Heavy_tailed_fluctuations}}

\blue{
However, the decay of the elastic propagator as $\mathcal{G}\sim\frac{\cos4\theta}{r^{d}}$
}
casts doubt on the Gaussian nature of the random stress increments
$\delta\zeta$ and would rather suggest a broad density function  for the mechanical noise,
\begin{equation}
w_1(\delta\zeta)\sim |\delta\zeta|^{-1-k}\text{ with }k=1 \label{eq:mech_noise_dist_1}
\end{equation}
in the limit of sparse plastic events, with an upper cut-off $\delta\zeta_{M}$
proportional to the volume of a rearranging region. For such a heavy-tailed
distribution, the biased random Brownian walk of $\sigma_{j}$ is
replaced by a L\'evy flight of index $k=1$ for $\sigma_{j}$. On
the basis of a simple extremal model, \citet{Lemaitre2007II} demonstrated
that this change altered the avalanche statistics as well as the distribution
of distances to yielding $P(x)$. To be explicit, their model was based
on plastic yielding above a uniform yield strain $\gamma_{y}$, which
resets the local stress to zero and increments the stresses at other
sites by random values drawn from $w_1(\delta\zeta$). \blue{Somewhat surprinsingly, the use of a Gaussian
distribution $w_1$ gave better agreement with quasistatic atomistic simulations
than a heavy-tailed distribution, in terms of
the scaling of avalanches with system size.}

\blue{
Further insight is gained by understanding that the distribution $w_1$ in Eq.~\eqref{eq:mech_noise_dist_1}
describes the instantaneous stress released by a single event, whereas a material region will yield under the cumulated effect of a \emph{sum} of such
contributions. Assuming that this sum is random,   
\citet{lin2016mean} arrived at the following probabilistic equation
\[
\frac{\partial P}{\partial \gamma} = v \frac{\partial P}{\partial x} + \int_{-\infty}^{\infty} [P(y)-P(x)] w(y-x) dy + \delta(x-1),
\] 
where $P(x)=0$ for $x \notin [0,2]$, $v>0$ drives sites towards their yielding point under the action of stress, and the distribution $w\,\delta\gamma$ accounts for
the sum of stress releases (drawn from $w_1(\delta\zeta)$) taking place over $\delta\gamma$: $w \sim w_1$
if $k<2$; 
otherwise, it is a Gaussian. This model is a variant of the H\'ebraud-Lequeux model which differs from it
in that (i) broad distributions $w_1(\delta\zeta)$ are allowed,
(ii) time is measured in terms of plastic strain $\gamma$, and (iii)
sites that cross the yield threshold $\sigma_y=1$ abruptly relax to 0 as $\gamma$ is
incremented. In this framework,
the authors derived an asymptotic analytical expression for the distribution
$P(x)$ (see Sec.~\ref{sub:V_distances_to_yield}),
which scales as $x^{\theta}$ for $x\ll1$. Interestingly, if the noise distribution is heavy-tailed, with $k<2$,
its breadth $k$ affects the value of $\theta>0$ in this mean-field model; $\theta$ turns into
a non-universal exponent that depends on the loading and the amplitude
of the noise, in line with atomistic simulations, and it supplements the other two exponents characterizing
the depinning transition. While these predictions accurately match the value measured in lattice-based EPM 
relying on the genuine elastic
propagator in dimension $d=4$, some discrepancy was noticed in 2D and to a lesser extent 3D.
This points to the progressive failure of the criterion of no-correlation as dimensionality
is lowered.
These results can be confronted
}
with \citet{Chen1991}'s early speculation of an upper
critical dimension of 3 for the applicability of \emph{constant} mean
field in their model. Leaving aside mean-field concerns for a minute,
we find quite noteworthy that the $\theta$ exponents measured in
the lattice-based EPM are quite compatible with their (indirectly)
measured value in atomistic simulations in the quasistatic regime,
in 2D and 3D \citep{Lin2014Density}.

\blue{More recently, \citet{aguirre2018critical} underlined the need to improve on the foregoing approximation
of mechanical noise as a random sum of single events. In reality, the noise results from avalanches 
displaying a linear spatial structure. 
A schematic argument
taking into account the statistics of avalanche sizes $S$ [namely, $P(S) \sim S^{-\tau}$, see Sec.~\ref{sec:VIII_Avalanches}]
then suggests to use the cumulative noise distribution
$W(\delta \sigma) \sim |\delta \sigma|^{-1-k}$ with $k=3-\tau$ (due to multiple avalanches),  instead of the instantaneous single-event distribution  $w_1(\delta \zeta)$.  If the density of stability $P(x)\sim x^\theta$ coincides with the probability of presence of random walker subjected to the noise $\delta \sigma$ near the aborbing boundary $x=0$, they further speculate that 
$\theta = k - 1$, which is consistent with their data.
}

\subsubsection{Structure of the elastic propagator and soft modes}

Coming back to \blue{the validity of mean-field estimates,
we note that the latter become accurate} (even for $d<4$)
if the elastic propagator $\mathcal{G}_{ij}$ is shuffled, that is, replaced by $\mathcal{G}_{\tau(i)\tau(j)}$,
where $\tau$ is a random permutation of indices which changes at each time step \citep{lin2016mean}. This shows that the temporal correlations
in the mechanical noise signal are due to the spatial structure
of $\mathcal{G}$. Of particular importance in this structure, claim
\citet{talamali2012strain} and \citet{tyukodi2015depinning}, are
the soft deformation modes of the propagator (recall that these are the uniform shear bands
described in Sec.~\ref{sub:Pointwise-idealisation-of} which
create no elastic stress in the material). To clarify this importance,
the authors focused on the evolution of the cumulative plastic strain
$\epsilon^{\mathrm{pl}}$ in extremal dynamics and recast the EPM
equation of motion [Eq.~\eqref{eq:gen_eq_of_motion}] into a depinning-like
equation (also see Sec.~\ref{sub:X_Depinning-transition}), viz.,
\[
\eta\partial_{t}\epsilon^{\mathrm{pl}}=\mathcal{P}\left(\sigma^{ext}+2\mu\mathcal{G}*\epsilon^{\mathrm{pl}}-\sigma_{y}\right),
\]
where $\eta=\mu\tau$
is a viscosity, $\sigma_{y}$ is the local stress threshold, and $\mathcal{P}(x)=x$
if $x>0$ and 0 otherwise. The deformation of a disordered solid in
$d$ dimensions is then regarded as the advance of an
elastic hypersurface in a $(d+1)$-dimensional space, where the
additional dimension is $\epsilon^{\mathrm{pl}}$. Under the influence
of the driving, the hypersurface moves forward along $\epsilon^{\mathrm{pl}}$.
In so doing, it gets deformed \blue{as a result of} the disorder in the
thresholds $\sigma_{y}$ seen by different \blue{sites.}

\blue{Although EPM and depinning models share a formally comparable framework,} \citet{tyukodi2015depinning} showed
that \blue{their results will differ}
because of the existence of soft modes in the EPM kernel $\mathcal{G}$,
while nontrivial soft modes are prohibited by the definite positiveness
of the \emph{depinning} propagator. As time goes on, the width $W^{\epsilon}\equiv\left\langle \overline{\left(\epsilon^{\mathrm{pl}}-\overline{\epsilon^{\mathrm{pl}}}\right)^{2}}\right\rangle $
of the elastic hypersurface (where the overbar denotes a spatial average
and the brackets indicate an ensemble average over the disorder) saturates
in the depinning problem. This saturation is due to the higher elastic
stresses released by regions of higher $\epsilon^{\mathrm{pl}}$,
which destabilize regions of lower $\epsilon^{\mathrm{pl}}$ and therefore
act as restoring forces to homogenize $\epsilon^{\mathrm{pl}}$ over
the hypersurface. On the contrary, in EPM, $W^{\epsilon}$ (the variance
of $\epsilon^{\mathrm{pl}}$) grows endlessly by populating the soft
modes of plastic deformation, which generate no elastic restoring force,
and diverges in a diffusive fashion at long times.

\subsection{A mechanical noise activation temperature?}

\subsubsection{The Soft Glassy Rheology model (SGR)\label{sub:V_SGR}}

The Soft Glassy Rheology model of \citet{sollich1997rheology}
proposed an alternative way to handle mechanical noise fluctuations
$\left\{ \delta\zeta(t)\right\} $. In the SGR spirit, these random
stress `kicks' operate as an effective temperature $x$ that can
activate plastic events, in the same way as thermal fluctuations do.
Accordingly, the diffusive term in Eq.~\eqref{eq:HL_equation}
is replaced by an Arrhenius rate to describe activated effects. More
precisely, in SGR, the material is divided into mesoscopic regions
that evolve in a landscape of traps
whose depths are randomly drawn from an exponential distribution  \citep{Bouchaud1992}
\[
\rho(E)\propto\mathrm{exp}(-E/E_{g}).
\]
Here, $E_{g}$ is a material parameter that will be set to unity.
The external driving facilitates hops from trap to trap (over the local energy barrier
$E$) by elastically deforming
each region at a rate $\dot{l}=\dot{\gamma}$, where $l$ is the local
strain. This lowers the barrier:
$E\rightarrow E-\frac{1}{2}kl^{2}$. (The stiffness parameter $k$
is such that $kl$ is the local stress.) Finally, SGR assumes that
the random `kicks' due to mechanical noise activate hops at a
rate $\omega(E,l)$ given by an Arrhenius law, viz.,
\begin{equation}
\omega(E,l)=\omega_{0}\,\mathrm{exp}\left(\frac{-E+\frac{1}{2}kl^{2}}{x}\right),\label{eq:SGR_t_esc}
\end{equation}
where $\omega_{0}$ is the attempt frequency and $x$ quantifies the
intensity of the mechanical noise. After a hop, $l$ is set back to
zero and a new trap depth $E$ is randomly picked from $\rho$.

The low-$\dot{\gamma}$ rheology that emerges from this simple model
is quite interesting. As the effective temperature $x$ decreases,
the system transits from a Newtonian regime $\Sigma\propto\dot{\gamma}$,
for $x>2$, to a power-law regime $\Sigma\propto\dot{\gamma}^{x-1}$
for $1<x<2$. A yield stress emerges for $x<1$ and the stress
follows the Herschel--Bulkley law $\Sigma-\Sigma_{y}\propto\dot{\gamma}^{1-x}$.
Indeed, for $x<1$, the ensemble average of the time spent in a trap,
viz., 
\[
\left\langle \tau\right\rangle =\int\rho\left(E\right)\omega^{-1}(E,l)dE
\]
diverges at $\dot{\gamma}=0$. The system ages and falls into deeper
and deeper traps on average; it follows that there is no typical material
time for the relaxation of the cumulated stress. Moreover, the wealth
of timescales afforded by an Arrhenius law also leads to
interesting linear viscoelastic properties, in accordance with experimental
data on colloidal pastes and emulsions.

\subsubsection{Mechanical noise v. thermal fluctuations}

\begin{figure}

\begin{centering}
\includegraphics[width=1\textwidth]{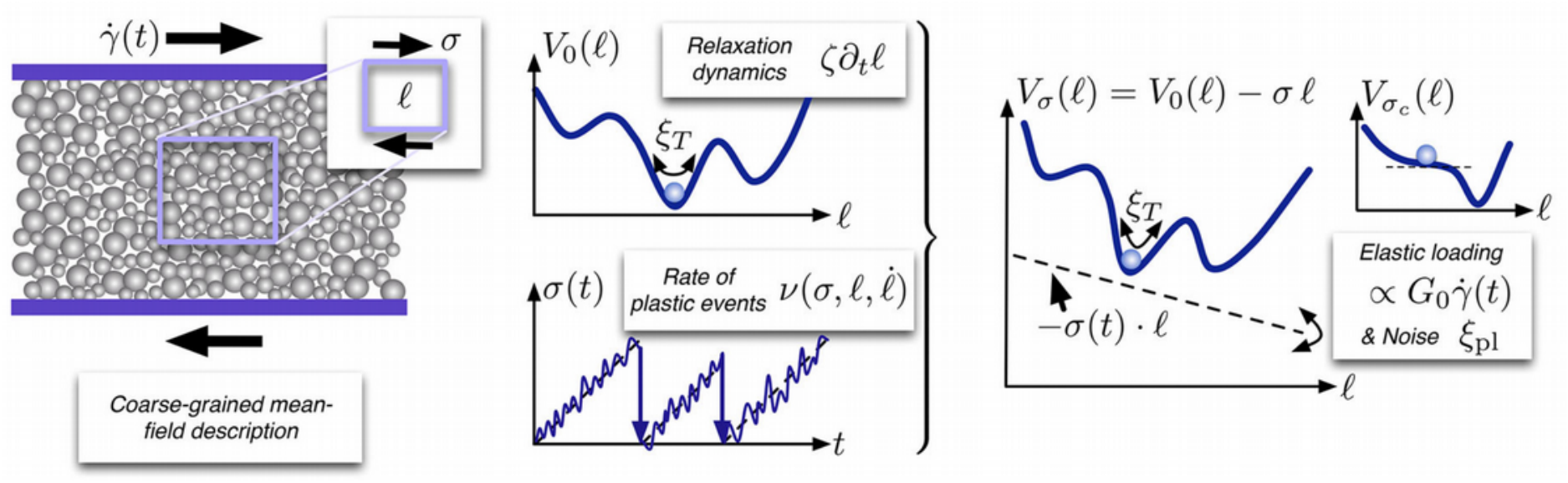}
\par\end{centering}

\caption{\label{fig:V_thermal_vs_mech_Agoritsas}Sketches illustrating the
difference between thermal fluctuations $\xi_{T}$ and mechanical
noise $\xi_{\mathrm{pl}}$; 
\blue{the variable $l$ represents the local strain configuration. $\xi_{T}$ are thermal kicks within a 
fixed potential energy landscape (PEL), whereas $\sigma(t)$ and its fluctuations $\xi_{\mathrm{pl}}$ tilt the local PEL up and down 
}
From \citep{agoritsas2015relevance}.}

\end{figure}

However, 
\blue{the grounds for using an effective activation temperature to describe the effect of mechanical noise have been contested
in recent years
} \citep{nicolas2014rheology,agoritsas2015relevance}.
The bone of contention is that, contrary to thermal fluctuations,
mechanical noise fluctuations persistently modify the energy landscape
of the region, insofar as the plastic events that trigger them are
mostly irreversible.

\blue{
More precisely, the argument runs as follows. The motion of particles in a 
mesoscopic region, of unit volume, is controlled by their interaction energy $V_0$, subject to
some constraint enforced at the boundary of the region. This constraint tilts the potential $V_0$ into $V_\sigma(t) \equiv V_0- l(t) \sigma(t)$, 
where $\sigma(t)$ is the time-dependent stress imposed by the outer medium and $l(t)$ is the strain associated with the 
internal configuration (see Fig.~\ref{fig:V_thermal_vs_mech_Agoritsas}). While the region responds elastically, $l(t)$
is a sensible proxy for its configuration. Assuming overdamped dynamics (with viscosity $\eta$), one can then write
}
\begin{eqnarray*}
0 & = & -\eta \dot{l}(t)-\frac{\partial V_\sigma}{\partial l}\left(t\right)+\xi_{\mathrm{T}}(t)\\
 & = & -\eta \dot{l}(t)-\frac{d}{dl}V_0[l(t)]+ \sigma(t)+\xi_{\mathrm{T}}(t).
\end{eqnarray*}
\blue{
Thermal fluctuations, denoted by $\xi_{\mathrm{T}}$ here, differ from mechanical noise in that they have a short-lived effect: $\langle\xi_{\mathrm{T}}(t)\xi_{\mathrm{T}}(t^{\prime})\rangle\propto\delta(t-t^{\prime})$ in the case of white noise. Exceptional sequences of fluctuations $\xi_{\mathrm{T}}$ are then required to climb up and overcome energy barriers. In contrast, changes to $\sigma(t)$, due to either the external
driving or distant plastic events, are persistent (hence cumulative). Even if one subtracts a constant drift term from
$\sigma(t)$, as we did in Sec.~\ref{sub:uniform_stress_redistrib}, the effect of mechanical noise \emph{fluctuations}  
$\xi_{\mathrm{pl}}(t) \equiv \int_{0}^{t}\delta\zeta\left(t^{\prime}\right)dt^{\prime}$ is cumulative,  \emph{viz.},
\[
\langle\xi_{\mathrm{pl}}(t)\xi_{\mathrm{pl}}(t^{\prime})\rangle\ =\int_{0}^{t}d\tau\int_{0}^{t^{\prime}}d\tau^{\prime}C(\tau-\tau^{\prime}) \sim\mathrm{min}(t,t^{\prime}),
\]
where the autocorrelation function $C(\Delta t)\equiv\langle\delta\zeta(t)\delta\zeta(t+\Delta t)\rangle$ was assumed to
decay quickly to zero. It follows that, under the sole
influence of $\xi_{\mathrm{pl}}$, the energy barrier $V_\sigma$ flattens out 
after a diffusive time $T\sim(\mathrm{max}\ dV/d\gamma)^{2}\equiv\sigma_{y}^{2}$, hence much faster than the Arrhenius law [Eq.~\eqref{eq:activation_rate_simple}] encountered
in activated processes. 
}

The diffusive growth of local stress fluctuations with
time has been confirmed by molecular dynamics simulations of model glasses
at least at very low shear rates, where \citet{puosi2015probing} have reported that
\[
\left\langle \left(\xi_{\mathrm{pl}}(t+\Delta t)-\xi_{\mathrm{pl}}(t)\right)^{2}\right\rangle \propto\Delta t.
\]
One should however mention that in this context the observation of stochastic resonance induced by mechanical noise in lattice-Boltmann simulations of 
emulsions is puzzling \citep{benzi2015internal}.

\blue{
The question of the mechanical noise has also sparked intense debate on the experimental side. 
In granular matter, it is now clear that locally shearing a region of the sample can affect
distant, unsheared regions: The applied shear facilitates the penetration
of an intruder \citep{Nichol2010} or the motion of a rodlike probe
}
\citep{Reddy2011}, presumably by agitating the grains in the distant
region, as if they were thermally agitated. An Eyring-like
activation picture
\blue{
using the magnitude of force fluctuations as temperature
may indeed account for the 
observed fluidization.}
But \citet{bouzid2015microrheology}
have argued that other
nonlocal models can replicate \blue{this observation} as well. Studying a related effect, \citet{pons2015mechanical}
have shown that applying small oscillatory stress modulations to a
granular packing subjected to a small loading can dramatically fluidize
it: Steady flow is then observed even though the loading is below the yield stress. This effect 
presumably stems from the \blue{cumulative impact of the stress modulations; the secular enhancement of
the fluidity in the proposed rationalization is at odds with the expectations for any activated process.}

\begin{table}[b]
\caption{\label{tab:fluct_handling}Synthetic view of the distinct types of fluctuations at play and the
methods with which they can be handled.
}
\centering
  \begin{tabular}{>{\centering\slshape}p{2.8cm} | >{\centering\arraybackslash\hspace{0pt}\footnotesize}p{4.05cm} >{\centering\arraybackslash\hspace{0pt}\footnotesize}p{3.6cm} >{\centering\arraybackslash\hspace{0pt}\footnotesize}p{3.0cm} >{\centering\arraybackslash\hspace{0pt}\footnotesize}p{3.6cm}}
 \hline
 \hline
\rule{0pt}{4ex} & \multicolumn{3}{c|}{\normalfont{\centering\bfseries Fluctuation-dominated regime}} & \normalfont{\centering\bfseries Drift-dominated regime}  \tabularnewline
 \hline

\rule{0pt}{4ex}     Mechanical noise fluctuations $\delta\zeta$ & Strong correlations and broad distribution & No time correlations but broad dist.
$ w(\delta\zeta)\!\propto\!\delta\zeta^{-1-k}$  & \multicolumn{2}{c}{\footnotesize{Gaussian white noise $w(\delta\zeta)\in o(\delta\zeta^{-3})$}} \tabularnewline

\rule{0pt}{4ex}     Dynamics of $\sigma$ in elastic regime & Correlated evolution & L\'evy flight & \multicolumn{2}{c}{\footnotesize{(biased) Brownian motion}}\tabularnewline

\rule{0pt}{4ex}     Applicable `mean field' treatment & None known so far & Reasoning on $P(\sigma)$ taking into account $w(\delta\zeta)$ &
 Eq. on $P(\sigma)$ with dif- -fusive term due to $\delta\zeta$ & Uniform mean-field approx. may be valid\tabularnewline
% Reasoning on $P(\sigma)$ where $\delta\zeta$ becomes a diffusive term & Uniform mean-field approx. may be valid\tabularnewline

\rule{0pt}{4ex}     $P(x)\text{ for }x\approx0$ & $\sim x^{\theta}$ with unrationalized exponent & $\sim x^{\theta}$ with dimension dependent $\theta$ & $\sim x$ & $\sim x^{0}$ \tabularnewline

\rule{0pt}{4ex}     References and examples & \citet{lin2016mean} for $d\leqslant 3$; \citet{liu2015driving} at low $\dot{\gamma}$;
\citet{budrikis2015universality} & \citet{lin2016mean}'s shuffled model and EPM in $d\geqslant 4$ & \citet{Hebraud1997} 
& \citet{liu2015driving} at high $\dot{\gamma}$\tabularnewline
 \hline
 \hline
\end{tabular}
\end{table}

\subsection{Connection with the diffusion of tracers\label{sub:V_tracers}}

\begin{figure}[t!]
\begin{center}
\includegraphics[width=\textwidth, clip]{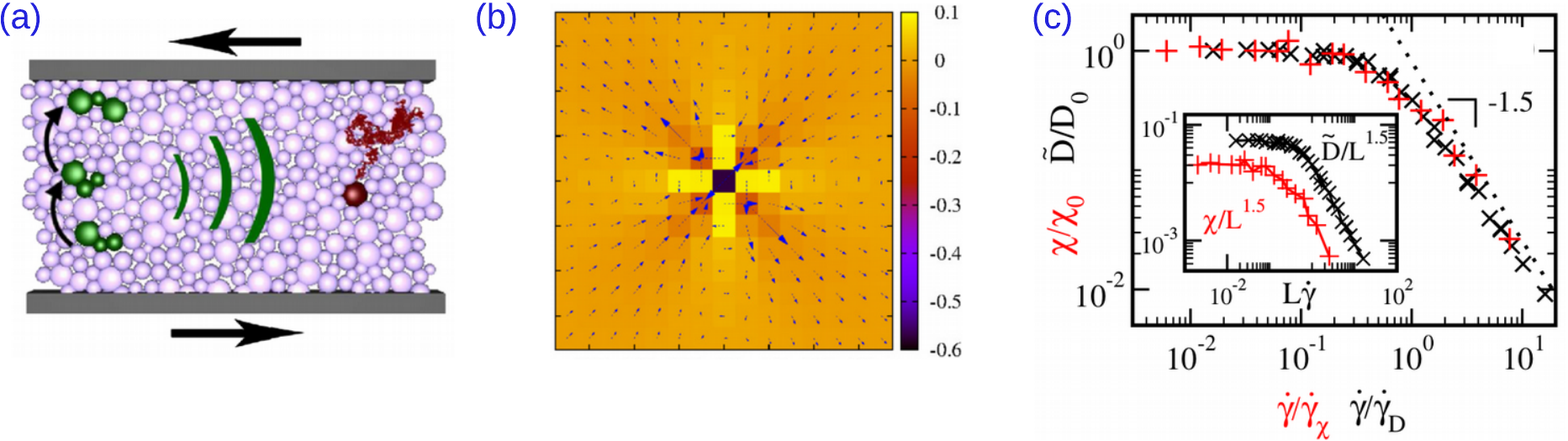}
\end{center}
\caption{\emph{Origins of cooperative effects.} 
\textbf{(a)} Schematic illustration of the long-range effects of plastic avalanches (in green) on the diffusion of a tracer (in red). From \citep{Martens2011} 
\textbf{(b)} Color map of the stress redistributed by a plastic event at the origin and associated displacement field (arrows). 
\textbf{(c)} Comparison of the scaling of the rescaled dynamical susceptibility $\chi/\chi_0$ for different system sizes and  shear rates with the scaling of the rescaled long-time  diffusion coefficient $\tilde{D}/D_0$; the inset shows the individual scalings. From \citep{Martens2011}.}
\label{fig:tracers}
\end{figure}

Rather than the local stresses, many experimental works  have access to observables related to particle displacements, in particular dynamic light scattering or particle tracking techniques. It is thus interesting to be able to connect the local stress dynamics to the diffusion of tracer particles. Single events as well as plastic avalanches are expected to contribute to the tracers' motion even far away from the plastic zone due to their long-range effects, as sketched in Fig.~\ref{fig:tracers}(a) \citep{Lemaitre2009, Nichol2010}.

The displacement field induced by a single plastic event can be calculated using Eq.~\eqref{eq:IV_Oseen_displ_tensor} and is displayed in Fig.~\ref{fig:tracers}(b). To mimic diffusion, \citet{Martens2011} introduced imaginary tracers that follow the displacement field generated by the ongoing plastic events and were able to rationalize the relation between the nonaffine part of the self-diffusion coefficient $D$ and dynamical heterogeneities (characterized by the four-point stress susceptibility $\chi_4^\star$),  as shown in Fig.~\ref{fig:tracers}(c). 
\blue{In particular, $D$ was found to decrease as $\dot\gamma^{-\nicefrac{1}{2}}$ at high $\dot\gamma$, while
at low shear rates saturation at a 
value depending on the linear system size $L$
as $L^{\nicefrac{3}{2}}$ was reported, contrasting with the $D \sim L$ scaling 
measured in atomistic
simulations \citep{maloney2008evolution}. \citet{Nicolas2014u} later argued that including convection in the EPM 
favored linear structures in the flow along the velocity direction and altered the 
dependence of $\chi_4^\star$ on the system size,
but no proper scaling of the data could be achieved. The dispute was very recently settled by \citet{tyukodi2018diffusion}:
It turns out that the roughly linear scaling of $D$ with $L$ observed in atomistic simulations is also present in EPM, but 
when one looks at \emph{short} strain intervals $\Delta \gamma$. More precisely, \citet{tyukodi2018diffusion}'s extremal model 
yields a $D\sim L^{1.05}$ scaling, which is robust to variations of implementation, and can be explained by considering that in each window $\Delta \gamma$ a roughly \emph{linear}
cascade of plastic events is either present or not. At longer times, the diffusivity enters the
 regime previously
identified ($D\sim L^{1.6}$), which is very sensitive to the presence of soft modes in the elastic propagator (see Sec.~\ref{sub:IV_soft_modes}).
}

Quantities comparable to the self-intermediate scattering function in purely relaxing systems are also accessible, as discussed in Sec.~\ref{sub:IX_relaxation}.

\subsection{Continuous approaches based on plastic disorder potentials\label{sub:Continuous-approaches}}

Notwithstanding their variable sophistication, all above methods rest
on a clearcut distinction between plastic rearrangements and elastic
deformations. This binary distinction is relaxed in continuous approaches,
\blue{ which intend to stay closer to the schematic dynamics in the Potential
Energy Landscape (PEL) outlined in Sec.~\ref{sub:Localised_rearrangements}.
The PEL is reduced to a free energy functional akin to that of Eq.~\eqref{eq:IV_free_energy},
except that the strain tensor $\boldsymbol{\epsilon}$ is conveniently traded
off for new strain variables:
one volumetric strain $e_1=\frac{\epsilon_{xx} + \epsilon_{yy}}{2}$ and two `shear' strains $e_2=\frac{\epsilon_{xx} - \epsilon_{yy}}{2}$ and $e_3=\epsilon_{xy}$, in 2D (5 in 3D,  $e_2, \dots, e_6$). The multiple valleys in the PEL between
which the system jumps during plastic rearrangements are reflected by multiple equilibrium values for the
shear strains in the free energy functional.  
}

For instance, \citet{Marmottant2013} 
\blue{focused on the shear strain $e_3$, split it into a cumulated plastic strain $\epsilon_p$ and a complementary
elastic strain, and proposed a minimalistic mean-field model
based on an effective energy $U_{\mathrm{eff}}$ that depends \emph{periodically} on $\epsilon_p$, viz.}
\[
U_{\mathrm{eff}}(e_3,\epsilon_{p})=\frac{E}{2}\left(e_3-\epsilon_{p}\right)^{2}+E\epsilon_{y}\frac{\epsilon_{0}}{2\pi}\cos\left(\frac{2\pi\epsilon_{p}}{\epsilon_{0}}\right),
\]
where $E$ is an elastic modulus, $\epsilon_{y}$ is a yield strain
and $\epsilon_{0}$ is the period of the pinning potential. If this
prescription is coupled with a dynamical equation of the form 
\[
\tau\dot{\epsilon}_{p}=\frac{1}{E}\left(-\frac{\partial U_{\mathrm{eff}}}{\partial\epsilon_{p}}\right),
\]
with $\tau$ the characteristic relaxation timescale (leading to the
Prandtl--Tomlinson model for stick--slip), a serrated
stress \emph{vs}. strain curve is obtained under constant driving.
The finite time needed by the plastic deformation $\epsilon_{p}$
to jump between energy valleys implies that, at high driving rates,
$\epsilon_{p}$ will not be able to instantaneously jump between,
say, $\epsilon_{p}^{(-)}$ and $\epsilon_{p}^{(+)}$. Therefore, the
elastic strain $(e_3-\epsilon_p)$ will keep increasing in the valley
around $\epsilon_{p}^{(-)}$ for some time, although a new equilibrium value, $\epsilon_{p}^{(+)}$, has appeared.
This is similar
to having a finite latency time before relaxation once the threshold
is exceeded in \citet{Picard2005}'s model. Similar equations of motion
in a random potential have been proposed for solid friction; the occurrence
of stick-slip dynamics owes to the ``pinning'' of the system in
one potential valley, up to some threshold, while there exists another
stable position \citep{tyukodi2015depinning}.

To go beyond the mean-field level, this type of continuous approach
can be resolved spatially. In an inspiratonial 
study, \citet{onuki2003nonlinear} introduced a free energy
of the form
\begin{equation}
F[\boldsymbol{u}]=\int d\boldsymbol{r}\,B\,e_{1}^{2}\left(\boldsymbol{r}\right)+\mathcal{F}\left(e_{2}(\boldsymbol{r}),e_{3}(\boldsymbol{r})\right),\label{eq:free_energy_Onuki}
\end{equation}
where $B$ is the bulk modulus and \blue{$e_{1}$, $e_{2}$, and $e_{3}$ are explicit functions
of the displacement field $\boldsymbol{u}(\boldsymbol{r})$.} Here,
$\mathcal{F}$ is an arbitrarily chosen function that is invariant
under rotations of the reference frame $\theta\rightarrow\theta+\nicefrac{\pi}{3}$
(because a 2D triangular lattice is assumed) and periodic in its arguments.
Introducing $F$ in the equation of motion
\begin{equation}
\rho\boldsymbol{\ddot{u}} (\boldsymbol{r})=-\frac{\delta F}{\delta\boldsymbol{u}(\boldsymbol{r})} + \eta_{0}\nabla^{2}\dot{\boldsymbol{u}}(\boldsymbol{r}) + \nabla\cdot\boldsymbol{\sigma}^{R}(\boldsymbol{r}),\label{eq:eq_of_motion_Onuki}
\end{equation}
where $\rho$ is the density, $\eta_{0}$ is the viscosity and $\boldsymbol{\sigma}^{R}$
is a random stress tensor due to thermal fluctuations, suffices to
obtain qualitatively realistic stress vs. strain curves. The framework
was then extended to study the effect of an interplay between the
volumetric strain $e_{1}$ and the density $\rho$, and to capture
the elastic effects of edge dislocations, if the material is crystalline
\citep{onuki2003plastic}. 

\blue{To avoid keeping track of the displacement field $\boldsymbol{u}$,
one may handle} the strain components $e_{1}$, $e_{2}$ and $e_{3}$
as independent primary variables, writing for instance  
\begin{equation}
F[e_1,e_2,e_3]=\int d\boldsymbol{r}\,B\,e_{1}(\boldsymbol{r})^{2} + \mu \, e_{2}(\boldsymbol{r})^2 + V(e_3(\boldsymbol{r})),
\label{eq:IV_free_energy_disorder}
\end{equation}
if the only deviation from linear elasticity is borne by $e_3$ and encoded in a `plastic disorder potential'
$V$. 
\blue{
Close to the reference state, $V$ will not deviate much from linear elasticity, viz., $V(e_3) \approx \mu e_3^2$, but
more globally it should have a corrugated shape that allows the system to reach new equilibria after each ST. }
The compatibility of $(e_1,e_2,e_3)$
as differentials of a displacement field should then be ensured by the Saint-Venant
condition 
\[
S\left[e_{1},e_{2},e_{3}\right]=0\text{ where }S\left[e_{1},e_{2},e_{3}\right]\equiv\left(\frac{\partial^{2}}{\partial x^{2}}+\frac{\partial^{2}}{\partial y^{2}}\right)e_{1}-\left(\frac{\partial^{2}}{\partial x^{2}}-\frac{\partial^{2}}{\partial y^{2}}\right)e_{2}-2\frac{\partial^{2}}{\partial x\partial y}e_{3}.
\]
This constraint is implemented by means of a Lagrange multiplier in
the total free energy $F$, viz., $F \to F+\lambda S$ \citep{Jagla2007}.
\blue{ It couples the different
strain components. In particular, in an incompressible linear elastic solid ($B \to \infty$ and $V(e_3)= \mu e_3^2$), a plastic strain arising at $\boldsymbol{r}=\boldsymbol{0}$ (for example, if the potential $V$ is chosen different at this position) will eventually unfold into the `quadrupolar' elastic field given by Eq.~\eqref{eq:elastic_propagator_Fourier} \citep{kartha1995disorder,cao2018soft}.
But, contrary to binary EPM, this unfolding is not instantaneous. Instead, it is generally governed by overdamped dynamics, 
$\dot{e}_i(\boldsymbol{r}) \propto -\delta F / \delta e_i(\boldsymbol{r})$ ($i=1,\,2,\,3)$.
An additional difference with respect to binary EPM is that the potential $V$
affects the tangential shear modulus $\mu_3=\nicefrac{1}{2} V''(e_3)$ as $e_{3}$ varies and may therefore
alter the destabilization dynamics. In this case, the system becomes elastically heterogeneous,
which precludes the use of Green's functions to calculate stress redistribution. 
}

\citet{jagla2017non} examined
the influence of different functional choices for $V$
on the flow curve and critical exponents, and reported differences between smooth and cuspy potentials.

\blue{
In this chapter, different levels of detail in the description of the elastic interactions have been considered. 
We will see that the specific form of these interactions may impact the low-shear-rate rheology (see  Sec.~\ref{sec:VII_Bulk_rheology})
and the local stress fluctuations (discussed in the next chapter), while
many flow properties at high shear rates do not require as exquisite a description.
}

\section{Strain localization: From transient heterogeneities to permanent
shear bands\label{sec:VI_Macroscopic_Shear_Deformation}}

\blue{
In Sec.~\ref{sec:Gen_phenomenology}, we brought to light latent similarities in the deformation of amorphous solids.
These, however,} should not mask the widely
different macroscopic consequences of applying shear to these materials.
The elastoplastic viewpoint helps to understand these differences
in a common framework.

\subsection{Two opposite standpoints}

\begin{figure}
\begin{centering}
\includegraphics[width=0.6\textwidth]{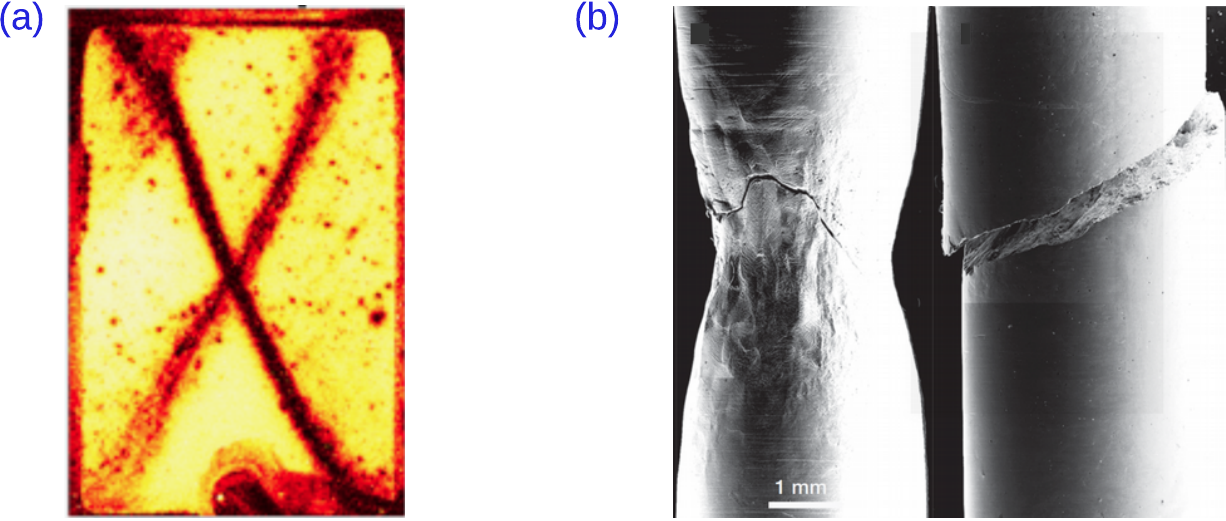}
\par\end{centering}
\caption{\label{fig:VI_experimental_shear_bands} \emph{Experimental observations of
shear bands and material failure}. 
\textbf{(a)} In a granular packing of $90\,\mathrm{\mu m}$ glass
beads under biaxial compression. 
\textbf{(b, right)} In a bulk metallic glass
under uniaxial tension. The composite glass shown in 
\textbf{(b,
left)}, reinforced with dendrites, displays a more ductile response to tension.
From \citep{LeBouil2014} and \citep{hofmann2008designing},
respectively.}
\end{figure}

In the common sense, there is a chasm between (i) foams and other
soft solids, that flow, and (ii) metallic or silicate glasses that
break/fracture after a certain amount of deformation (see Fig.~\ref{fig:VI_experimental_shear_bands}b(right)).

To start with the
\blue{far end of the latter category, perfectly brittle materials will 
deform elastically and then break, without going
through a stage of plastic deformation.
}
In daily life, this situation is exemplified by the soda-lime glasses
routinely used to make windowpanes, bottles, etc., and more generally
silicate glasses. Nevertheless, at small scales plastic deformations, 
\blue{resulting in a denser material,
}
were revealed in indentation experiments with a diamond tip \citep{yoshida2007indentation}
as well as experiments of uni-axial compression of micropillars of
amorphous silica \citep{lacroix2012plastic} (which overall behaves
comparably to soda-lime glass \citep{perriot2011plastic}) and simulations
of extended shear \citep{rountree2009plasticity}. However, in many situations, plasticity plays virtually
no role, in particular when failure is initiated by a crack: No evidence
of plasticity-related cavities was seen by \citet{guin2004fracture}
(also see references therein) and, with the help of simulations, \citet{fett2008finite}
claimed that the surface displacements experimentally observed at
crack tips are compatible with theoretical predictions discarding
plasticity. (It should however be mentioned that a minority of works
support the existence of plasticity near the crack tip).

In metallic glasses, global failure is preceded by substantial plastic
deformation. The latter is generally localized in thin shear bands,
that appear as clear bands in \emph{post-mortem} scanning electron
micrographs. These bands are typically 10 to 50nm or even $100\,\mathrm{nm}$-thin
\citep{Bokeloh2011,Schuh2007}, \emph{i.e.}, much thinner than the adiabatic
shear bands encountered in crystalline metals and alloys, which are
about $10-100\,\mathrm{\mu m}$-thick. Despite these plastic deformations,
brittleness remains a major industrial issue for metallic glasses.
Added to their cost and the difficulty of obtaining large samples,
this drawback may outshine their advantageous mechanical properties,
such as their high elastic limit \citep{wang2012elastic}. As a consequence,
much effort has been devoted to improving their ductility.

By contrast, foams, emulsions and various other soft solids can undergo
permanent shear flow without enduring irretrievable damage. This conspicuous
discrepancy with hard molecular glasses can however be lessened by
noticing that, even among soft solids, the flow sometimes localizes
in shear bands \citep{Becu2006,Lauridsen2004}. Still, the distinction
between hard solids that deform and break and soft solids that deform
and flow is overly caricatural. The case of gels, which consist of
long entangled (and often cross-linked) chains, demonstrates that
soft solids, too, may break upon deformation. But, then,\emph{ what
distinguishes a material that flows from one that fails? What determines
whether the deformation will be macroscopically localized in shear
bands or homogeneous (on the macroscopic scale)?}

\subsubsection{The shear-banding instability from the standpoint of rheology}

To start with, let us consider the rheological perspective. Shear-banding
in complex fluids is interpreted as the consequence of the presence
of an instability in the constitutive curve, \emph{i.e.}, the flow
curve $\Sigma_{0}=f(\dot{\gamma})$ that would be obtained if the
flow were macroscopically \emph{homogeneous}.
Indeed, it is easy to show that homogeneous flow in \emph{decreasing} portions
of the constitutive curve is unstable to perturbations and \blue{gives in to}
co-existing bands.
% In some cases, the shear rates $\dot{\gamma}_{h}$ and $\dot{\gamma}_{l}$
%in the high-shear-rate and low-shear-rate bands, respectively, may
%be predicted by a so called Maxwell construction (see Fig.~\ref{fig:VI_Maxwell_criterion}).
The actual flow curve displays a stress plateau $\Sigma\left(\dot{\gamma}\right)=\mathrm{cst}$
for $\dot{\gamma}$ between two values $\dot{\gamma_{l}}$ and $\dot{\gamma}_{h}$.
Shear localization corresponds to the particular case $\dot{\gamma}_{l}\simeq 0$, \emph{i.e.}, that of a non-flowing band.
%In the Maxwell picture, shear localization 
In other words, it
will occur if the constitutive curve already starts decreasing at $\dot{\gamma}=0$. 

\blue{Accordingly, the shear-banding criterion based on the slope of the constitutive curve $\Sigma_{0}(\dot{\gamma})$
can be studied at the mean-field level (see, for instance, \citet{Coussot2010}'s analysis of a simple model).
Incidentally, note that this}
is somehow counterintuitive, given the manifest spatial heterogeneity
associated with the phenomenon. Nevertheless, mean-field calculations obviously
leave aside the spatial organization of the flow (its banded structure),
which hinges on the shape of the elastic propagator in simulations:
In EPM, with similar dynamical rules, the banded flow structure obtained
with the long-ranged elastic propagator of Eq.~\eqref{eq:elastic_propagator_Fourier}
is not preserved if the propagator is substituted by a stress redistribution
to the first-neighbors, even if the latter is anisotropic \citep{Martens2012}. 

The simple criterion based on the steady-state constitutive curve
needs to be somewhat adjusted for amorphous solids, which often exhibit
aging effects. Then, the yield stress of the quiescent material may
vary with the waiting time since preparation \citep{Varnik2003}.
Consequently, even if the flow curve obtained by ramping down $\dot\gamma$
from a high value is strictly monotonic, shear-banding
may arise in non presheared samples. This will happen if an initially
undeformed band gradually solidifies and thus further resists deformation,
while the rest of the material is sheared. The solid band is `trapped'
in its solid state because of the aging at play \citep{Moorcroft2011,martin2012transient}.

\subsubsection{The mechanics of bands in a solid}

Turning to the viewpoint of solid-state mechanics, as emphasized in
Sec.~\ref{sub:Pointwise-idealisation-of}, uniform strain bands
inclined by $\pm\nicefrac{\pi}{4}$ with respect to the principal
directions of the strain tensor are soft modes of the elastic propagator
[Eq.~\eqref{eq:elastic_propagator_Fourier}], which means that they
do not generate elastic stresses in the system. Should there be a
weak stripe in the material (in the sense of low elastic moduli or
low yield thresholds), it will then be energetically beneficial to
accommodate part of the macroscopic strain in it in the form of a
slip line. Such an energy-based argument is especially relevant in
a quasistatic protocol in which the system always reaches the local
energy minimum between strain increments. If the stripe in which the
strain localizes displays ideal plasticity, the macroscopic stress-strain
curve $\Sigma=f(\gamma)$ stops increasing due to the banding instability.

But this continuum-based approach ignores the granularity of the
material at the scale of plastic rearrangements by postulating the
spontaneous and synchronous creation of a strain band all at once.
Contrasting with this postulate, some experimental evidence in colloidal
glasses \citep{Chikkadi2011} and granular matter \citep{Amon2012,LeBouil2014}
indicates that shear bands actually consist of disconnected, non-simultaneous
localized plastic rearrangements, as implemented in EPM. Therefore,
only \emph{on average} is a strain band uniform; its granularity
(as a patchwork of localized plastic rearrangements) as well as the
time fluctuations in its plastic activity have no reason to be overlooked.
The sequential emergence of the band may explain its sensitivity to
details in the implementation of the elastic propagator \citep{talamali2012strain}.

Taking the granularity of the band into account, \citet{dasgupta2012microscopic,dasgupta2013shear}
proposed to explain the existence and the direction of shear bands
by an argument based on the minimization of the elastic energy of
a collection of Eshelby inclusions in a uniform elastic medium over
their possible configurations in space. The neglect of the elastic
heterogeneity of glasses in the reasoning was justified by the authors
by the specific consideration of carefully quenched (hence, more homogeneous)
glasses. An additional concern could be raised as regards the use
of a global one-step minimization, whereas plastic events occur sequentially
and the elastic deformation field in the material evolves during the
process. \blue{Nevertheless, in a similar endeavor, \citet{karimi2018correlation}
rationalized the observed deviation between the direction of the 
macroscopic shear band in a deformed granular medium and that of the microscopic
correlations between rearrangements. The authors contended that the former direction is the direction
of maximal instability with respect to the Mohr-Coulomb failure criterion, rather than that
of maximal increase of the shear stress.
} 

More generally, the strain bands described in the context of solids
probably differ from the long-lived or permanent shear bands observed
experimentally in steadily sheared materials. The former might be
more accurately referred to as transient `slip lines' and some
reports of `shear bands' in atomistic simulations should probably
rather be interpreted as slip lines, as already noted by \citet{Maloney2006Amorphous}.
However, it has been suggested that the transient banding instability
can act as a \emph{precursor} to the formation of a shear band \citep{fielding2014shear}. 

In fact, transient banding is a matter of interest \emph{per se, }as
it can be long-lived \citep{divoux2010transient}. \citet{moorcroft2013criteria}
proposed a way to rationalize its occurrence on the basis of a generic
banding criterion involving the transient constitutive curves $\Sigma_{0}=f(\dot{\gamma},\gamma)$,
where $\gamma$ is the cumulative strain since shear startup, in a
fictitious system constrained to deform homogeneously. The rheological
criterion $\frac{d\Sigma_{0}}{d\dot{\gamma}}<0$ is recovered at infinite
times $\gamma\rightarrow\infty$, while a purely elastic banding instability
is predicted if
%\[
$A\frac{\partial\Sigma_{0}}{\partial\gamma}+\dot{\gamma}\frac{\partial^{2}\Sigma_{0}}{\partial\gamma^{2}}<0$,
%\]
with a model-dependent prefactor $A>0$, provided that the material
is sheared much faster than it can relax ($\dot{\gamma}\rightarrow\infty$).
In the light of this, the authors claim that there is a generic tendency
to transient banding in materials that exhibit a stress overshoot
in shear startup. 

\blue{Coincidence between a stress overshoot and an emerging shear band
has effectively been noticed in EPM and atomistic simulations \citep{lin2015criticality,ozawa2018random,popovic2018elasto}. 
Moreover, EPM unveiled the 
major impact of the system's preparation on these features. Within a depinning-like model,
\citet{ozawa2018random} proved that a broad initial distribution $P_0(x)$ of site stabilities
(\emph{i.e.}, distances to yield, as defined in see Sec.~\ref{sub:V_distances_to_yield}), reflecting poor annealing, suppresses the overshoot. Contrariwise, the latter and the associated stress drop grow as $P_0(x)$ becomes more sharply peaked, 
as expected for a well aged glass.
At some point, the stress-strain curve even becomes discontinuous, as the system undergoes a spinodal instability. A critical
point separates the continuous and discontinuous regimes.
The authors observed very good agreement between these mean-field results and atomistic simulations of ultrastable glasses
\citep{ozawa2018random}. In a parallel paper, \citet{popovic2018elasto} showed that these features survive 
in a spatially-resolved
EPM. In addition, they related the presence of run-away avalanches
to the curvature of the distribution of site stabilities $P(x)$ at $x=0$. The stress drops caused by these 
avalanches grow as failure is approached and can thus signal the imminence of failure. But there may exist
an alternative mechanism for failure (without diverging avalanches), namely the nucleation of a shear band in a 
fortuitously weak region. We will discuss an earlier work by \citet{Vandembroucq2011} in Sec.~\ref{sub:VI_Ingredients_shear_loc}, where we review the mechanisms promoting shear banding.
}

\blue{For the time being, let us enquire about the fate of the material after 
the overshoot. \citet{ozawa2018random} and \citet{popovic2018elasto} argue that 
the transition from a continuous stress-strain curve to a discontinuous one, as the initial preparation $P_0(x)$
is varied, marks a transition from a ductile response to a brittle one. Accordingly, the discontinuity
in the stress curve is interpreted as irreversible material failure, in the very spirit
of fiber bundle models, where fibers break irreversibly (see Sec.~\ref{sub:X_fiber_bundles}).
In this case, the finite stress signal displayed by their systems (EPM and atomistic simulations)
\emph{after} failure must be considered spurious: The model is no longer valid after failure.}

\blue{
On the other hand, if the material does preserve some cohesion after the stress drop accompanying the overshoot,
one may wonder whether the transient band will 
}
convert into a steady-state
band, under homogeneous loading.
What is required for this purpose is a mechanism that explains how the transient
`slip lines', instead of being dispersed, concentrate in the same region
of the shear-banded material as time goes on.
The distinction between the situation at finite strains and in the steady state
should perhaps be emphasized.
The first-order yield transition in the statistics of low-energy barriers observed
by~\citet{Karmakar2010} at a \emph{finite strain} $\gamma_{c}$ is not necessarily
associated with a first-order (banding) transition in the \emph{steady-state}
flow curve $\Sigma\left(\dot{\gamma}\right)$.
Similarly, \citet{jaiswal2016mechanical} numerically observed that, in a batch
of finite-size samples subjected to a strain $\gamma$, about half of the samples
will have irreversibly yielded when $\gamma=\gamma_{c}$, while the other half
come back to their initial configuration upon unloading;
but it is not straightforward to conclude from this interesting observation that,
if one stitched a `yielding' sample together with a `recovering' one, a shear
band would localize in the `yielding' part at longer times.

\subsection{Spatial correlations in driven amorphous solids}

EPM help bridge the time and length scale
gap between transient slip lines and permanent shear-banding. 
\blue{At short to intermediate time scales and under slow enough driving,
the organization of the flow is complex and exhibits}
strong intermittency and marked spatial correlations between rearrangements
even in driven amorphous solids that are not susceptible to macroscopic shear localization.

\subsubsection{Spatial correlations}

The spatial extent of correlations in the flow \blue{can be quantified by 
cooperativity or correlation lengths $\xi$ in bulk flows, brought within reach by
the computational efficiency of EPM.}
 The Kinetic Elastoplastic
(KEP) Theory of \citet{Bocquet2009}, an extension of the H\'ebraud-Lequeux
model (see Sec.~\ref{sub:V_Hebraud_Lequeux}) that includes heterogeneities,
predicts a decrease of $\xi$ with the shear rate as 
\[
\xi\sim\left(\Sigma-\Sigma_{y}\right)^{\nicefrac{-1}{2}}\sim\dot{\gamma}^{\nicefrac{-1}{4}},
\]
in contrast with \citet{Lemaitre2009}'s theoretical prediction $\xi\sim\dot{\gamma}^{\nicefrac{-1}{2}}$
in 2D, beyond which independent avalanches
are supposedly triggered.

Simulations of homogeneous shear flow \blue{in} spatially resolved EPM have generally shown
results departing from the $\xi\sim\dot{\gamma}^{\nicefrac{-1}{4}}$
scaling. \citet{Picard2005} reported a correlation length
that scales with $\dot{\gamma}^{\nicefrac{-1}{2}}$ in 2D (see Fig.\ref{fig:spatialcorrelations}a),
\blue{on the basis of a study of the variations of the average
stress drop $\langle \delta\sigma\rangle $ with $\dot\gamma$ for different linear system sizes $L$; indeed,
the data can be collapsed onto a master curve by rescaling $\dot\gamma$ into $L^2\dot\gamma$.}
\citet{Nicolas2014u} related this scaling to the average spacing
between simultaneous plastic events, which scales as $\dot{\gamma}^{\nicefrac{-1}{d}}$
in $d$ dimensions, and several definitions of correlation lengths
were shown to follow this dependence in EPM. The variable sign of
the elastic propagator enters the reasoning, insofar as plastic events
are able to screen each other, because the sign of their contributions
may differ. 

Nevertheless, the $\dot{\gamma}^{\nicefrac{-1}{d}}$ scaling
is not generic. In particular, the correlation length derived from
the four-point stress correlator $\mathcal{G}_{4}\left(\boldsymbol{r}\right)$,
exploited by \citet{Martens2011} (see Fig.\ref{fig:spatialcorrelations}b),
is more sensitive to the avalanche shape and was shown to depend on
the chosen EPM dynamical rules. 
\blue{Along similar lines, \citet{roy2015thesis}'s atomistic simulations of soft disks in 2D point
to a sensitivity of the correlation length $\xi_D$ derived from finite-time particle diffusion to the damping scheme; 
more precisely, they measured $\xi_D \sim \dot\gamma^{\nicefrac{-1}{3}}$ for mean-field drag and $\xi_D \sim \dot\gamma^{\nicefrac{-1}{2}}$
if the drag force depends on the relative particle velocities.}

Below the yield stress, \citet{lin2015criticality}
claim that the system is critical, with system-spanning avalanches
in the transient, which is supported by a study of the cutoffs in
the avalanche size distributions in EPM simulations. This implies
a diverging correlation length $\xi=\infty$ in the whole $\Sigma<\Sigma_{y}$
phase - not unlike what is seen in 2D dislocation systems, at all
applied stresses \citep{IspanovityPRL2014}. 
However, the divergence $\xi \to \infty$ 
observed in athermal EPM, e.g., in the quasistatic limit $\dot{\gamma} \to 0$, will be strongly cut
off in systems at a finite temperature, where thermal noise
stifles the correlations \citep{Hentschel2010}. More generally, one should say that
EPM tend to overestimate the absolute magnitude and
and extent of the correlations between plastic events, e.g., compared
to particle-based simulations \citep{Nicolas2014s}. \blue{We surmise that
the overestimation is due not only to the neglect of elastic heterogeneities, but also to the
regular lattice generally used in EPM, which standardizes the interactions between blocks.}

\subsubsection{Cooperative effects under inhomogeneous driving}

\begin{figure}
\begin{centering}
\includegraphics[width=1\textwidth]{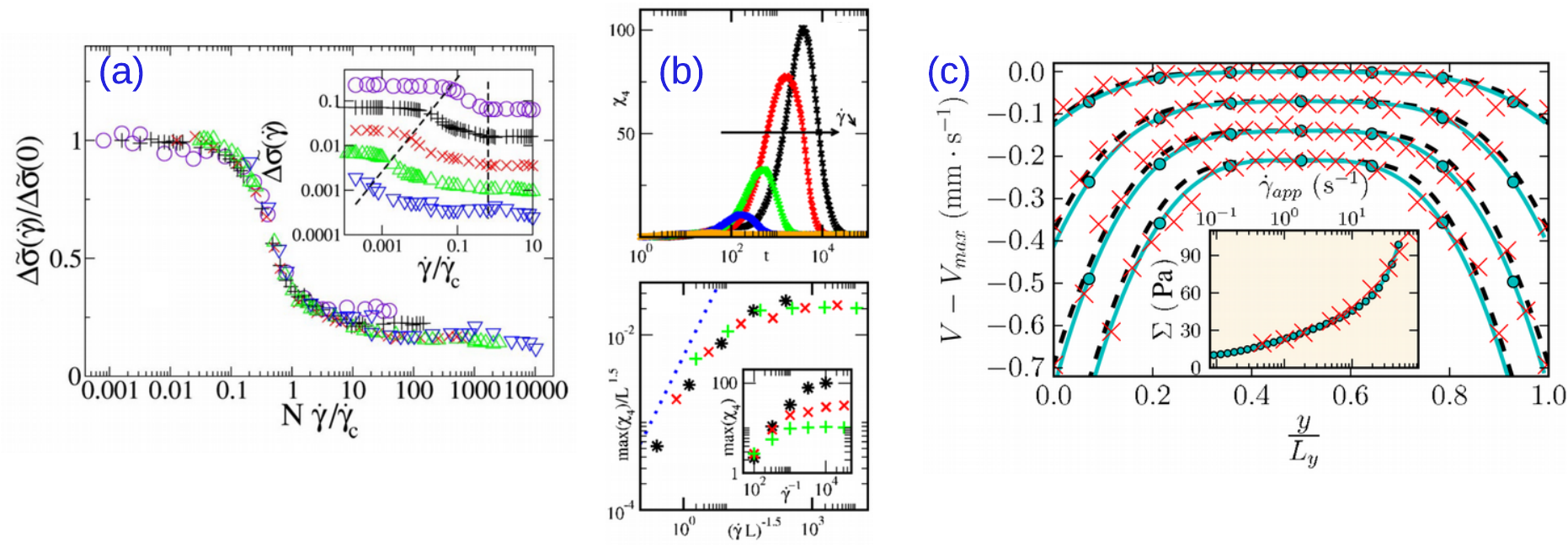}
\par
\end{centering}
\caption{\label{fig:spatialcorrelations} \emph{Evaluation of cooperative effects in EPM.} 
\textbf{(a)} Mean stress drop $\Delta\tilde{\sigma}$ as 
a function of the shear rate $\dot\gamma$. (Both variables were appropriately rescaled and made
dimensionless). From \citep{Picard2005}. 
\textbf{(b)} \emph{Top}: Time evolution
of the dynamical stress susceptibility $\chi_4(t,\dot\gamma)$ at different $\dot\gamma$. \emph{Bottom}:
Finite-size scaling plot of the maxima of $\chi_4$.
Adapted from \citep{Martens2011}. 
\textbf{(c)} Velocity profiles across a
microchannel for different applied pressures. (\emph{Red crosses}) Experimental
data for an oil-in-water emulsion; (\emph{dashed cyan lines}) EPM;
(\emph{solid black lines}) bulk rheology predictions. The profiles
have been shifted vertically for legibility. Adapted from \citep{Nicolas2013a}.}
\end{figure}

\blue{
Correlations in the flow dynamics are found in macroscopically homogeneous flows,
but their impact is most conspicuous when
the loading or the flow is inhomogeneous over the correlation length scale. In these 
situations, marked cooperative effects are generated.
}

This is the case in pressure-driven flows through a
narrow channel, of transverse width $w$ ($w\approx 100\,\mathrm{\mu m}$ for
microchannels). In this geometry, the streamline-averaged shear stress
\blue{
$\Sigma$ varies linearly across the channel, from zero in the center (in 2D, but also in 3D)
to $\frac{w}{2} \nabla p$ at the wall. Therefore, a `plug' of advected, but unsheared
material 
is expected in the central region where $|\Sigma|<\Sigma_{y}$,
for yield-stress fluids. 
}

\blue{
These expectations were reshaped following
seminal experiments on concentrated oil-in-water emulsions by \citet{goyon2008spatial}.
Indeed,  compared to the predictions from the bulk rheology, the observed 2D velocity profiles are more rounded
and, overall, the flow is enhanced.}
Thus, there is no unique
relation between the local strain rate and the local stress \citep{Goyon2010}:
the rheology is nonlocal. Using a similar system, \citet{Jop2012}
 demonstrated the existence of finite strain rate fluctuations
$\delta\dot{\gamma}(\boldsymbol{r})\equiv \sqrt{\langle \dot{\gamma}(\boldsymbol{r})^{2} \rangle - \langle \dot{\gamma}(\boldsymbol{r}) \rangle ^{2}}$
in the plug, which reach their minimum at the channel center. Numerical
simulations of athermal soft disks confirmed the impact of confinement
on the rheology: In 2D periodic Poiseuille flows, 
\blue{which put side by side Poiseuille flows of alternate directions, the
wall stress below which flow stops substantially increases with decreasing channel width $w$ \citep{Chaudhuri2012}.
All these phenomena clearly arise because of the interactions between
}
streamlines subjected to different
stresses, via the elastic fields generated by plastic events.

The remarkable effect of spatial correlations on inhomogeneous
flows is often rationalized by means of a nonlocal term in the equation
controlling the fluidity $f$. This variable, defined here as the inverse viscosity
$\nicefrac{\dot{\gamma}}{\sigma}$,
\blue{is thought of as a proxy for
}
the rate of plastic events. Owing to the symmetry
of the propagator, the leading-order correction to the local fluidity
involves the Laplacian $\nabla^{2}f$. \blue{The} steady-state
fluidity diffusion equation \blue{thus reads}
\begin{equation}
\xi^{2}\nabla^{2}f+\left[f_{b}\left(\Sigma\right)-f\right]=0,\label{eq:fluidity_eq}
\end{equation}
where $\xi$ is a cooperative length and $f_{b}$ is the fluidity
in a bulk flow subject to stress $\Sigma$. The KEP model of \citet{Bocquet2009}
provides a formal justification for Eq.~\eqref{eq:fluidity_eq} to linear order in $f$, with $\xi\sim\left(\Sigma-\Sigma_{y}\right)^{\nicefrac{-1}{2}}$,
by accounting for the mechanical noise generated in the immediate
vicinity of plastic events. In fact, using a \emph{constant }value
of $\xi$ (for each material) in Eq.~\eqref{eq:fluidity_eq} already
provides very good fits of the experimental curves, not only for concentrated
emulsions \citep{goyon2008spatial} and lattice-Boltzmann simulations thereof \citep{Benzi2014}, but also for
polymer microgels (Carbopol) \citep{Geraud2013}. In the case of emulsions,
$\xi$ \blue{vanishes} below the jamming point, and reaches
up to 3 to 5 droplet diameters ($20-30\,\mathrm{\mu m}$) in the
very dense limit \citep{goyon2008spatial}. Similarly, for Carbopol samples,
$\xi$ \blue{measures} 2 to 5 structural sizes (sizes of optical heterogeneities)
 \citep{geraud2017structural}. 

However, the fitting \blue{is highly sensitive to} the adjustment of the fluidity $f_{w}$ at the wall
\citep{Geraud2013}, which limits the accuracy of
the experimental measurement of $\xi$. This difficulty highlights
the value of EPM for testing the validity of theoretical predictions.
In EPM descriptions of channel flow, the driving term $\mu\dot{\gamma}$
is set to zero in Eq.~\eqref{eq:gen_eq_of_motion}; flow arises on
account of the initially imposed transverse stress profile $\Sigma\left(y\right)$.
\blue{Channel walls are} accounted for by a no-slip boundary
condition. This adds a correction to Eq.~\eqref{eq:elastic_propagator_Fourier}
for the elastic propagator, which can be calculated via
a method of images \citep{Nicolas2013b}, and leads to a faster local relaxation for plastic
events near walls  [also see \citep{hassani2018wall}]. Combined with appropriate
dynamical rules, the model semi-quantitatively reproduces the shear
rate fluctuations $\delta\dot{\gamma}(\boldsymbol{r})$ observed by \citet{Jop2012} in the plug as well
as the moderate deviations of the velocity profiles from the bulk
predictions witnessed with smooth walls, provided that the EPM block
size corresponds to around 2 droplet diameters (see Fig.~\ref{fig:spatialcorrelations}c).
The fluidity diffusion equation, Eq.~\eqref{eq:fluidity_eq} either
with $\xi=\mathrm{cst}$ or $\xi\sim\dot{\gamma}^{\nicefrac{-1}{4}}$,
captures the EPM fluidity profiles reasonably well, albeit imperfectly.

\citet{gueudre2016scaling} \blue{took} a closer look at the decay of $\dot{\gamma}(y)$
in a region ($y>0$) subject to $\Sigma<\Sigma_{y}$ contiguous to
a sheared band ($y<0$). \blue{They found} that EPM
results obey a scaling relation involving a length scale $\xi (\Sigma)\sim\left(\Sigma-\Sigma_{y}\right)^{-\nu}$
but the scaling exponents
differ from mean-field predictions and are also inconsistent with
KEP-based Eq.~\eqref{eq:fluidity_eq}. In particular, for $\Sigma\approx\Sigma_{y}$,
$\dot{\gamma}(y)$ is argued to decay algebraically with $y>0$ instead
of exponentially. \citet{gueudre2016scaling} further claim
that pressure-driven flows display larger finite-size effects than
simple shear flows.  \blue{Indeed, the} finite size $L$ of a system
\blue{shifts} the critical stress for flow initiation by $\Delta\Sigma\propto L^{\nicefrac{-1}{\nu}}$
 in a homogeneous setup (so
that $\xi(\Sigma)=L$ at initiation). On the other hand, in a pressure-driven flow, the 
length scale entering the critical stress
\blue{for flow cessation should not be the system size $L$, but}
the (much smaller) width
of the \emph{sheared} band near the wall.

The description of nonlocal effects by Eq.~\eqref{eq:fluidity_eq}
has also been applied to granular matter, which generically display
heterogeneous flow and shear bands \citep{Kamrin2012}. To do so,
the fluidity was redefined as $\nicefrac{\dot{\gamma}}{\mu}$, \blue{because}
the rheology of dry frictional grains is best expressed
\blue{in terms of} the inertial number $\mathcal{I}$
(a rescaled shear rate) and \blue{its dependence on} the friction $\mu\equiv \Sigma / P$
(with $P$ the pressure). The resulting model \blue{successfully} captures
cooperative effects and accounts for the global velocity
profile observed in discrete element simulations of a simple shear
flow with gravity, a gravity-driven flow in a channel \citep{Kamrin2012}
as well as the flow of a granular layer on an inclined plate, which
is sensitive to the thickness of the layer \citep{Kamrin2012}. Nevertheless,
the validity of the definition of a `granular fluidity', which
is not an intrinsic state variable (because of the denominator $\Sigma$
or $\mu$), has been questioned; employing another
variable would also lead to an exponential decay of the flow away
from an actively sheared zone \citep{bouzid2015non}. Other \blue{suggestions}
for the fluidity variable $f$ that should enter a diffusive equation
\blue{include} the ratio between the `static' and the `fluid' part of the
stress tensor \citep{aranson2006patterns} and the inertial number $\mathcal{I}$, which
\citet{bouzid2015microrheology} claim to best match their
discrete-element simulations, in particular regarding the \blue{necessary} continuity of $f=\mathcal{I}$
at the interface between differently-loaded regions.

\subsubsection{Cooperative effects due to boundaries}

Coming back to emulsions, \citet{goyon2008spatial}'s observations indicate
that the flow deviates much more from the bulk predictions, with an
enhanced fluidization, when smooth walls are replaced by rough walls.
Further experimental studies on regularly patterned surfaces show
that the wall fluidization enhancement varies nonmonotonically with
the height of the (steplike) asperities, for asperities smaller than
the droplet diameter, as does the wall slip velocity \citep{Mansard2014}. 

These strong deviations in the presence of rough walls exceed by far
what is found in EPM. This points to another physical origin than
the coupling to regions subject to higher shear stresses. Since wall
slip was experimentally observed, it has been suggested that the `collisions'
of droplets against surface asperities, as they slide along the wall,
are the missing source of plastic activity; adding sources of mechanical
noise along the walls in EPM can indeed capture the experimental features
\citep{Nicolas2013a}. \citet{derzsi2017fluidization} experimentally
confirmed the presence of roughness-induced scrambles at the wall
and, with the help of lattice-Boltzmann simulations, the ensuing increase
in the rate of plastic rearrangements near rough walls.

\blue{
\subsection{Alleged causes of permanent shear localization or fracture\label{sub:VI_Ingredients_shear_loc}}
}

Several EPM have been able to reproduce permanent strain localization
\citep{Bulatov1994b,Jagla2007,Coussot2010,Vandembroucq2011,Martens2012,Li2013,Nicolas2014u}.
In these cases, after a transient, the plastic activity will typically
concentrate in a narrow region of space, generally a band, that may
slowly diffuse over time. \blue{Hints at the ingredients suspected of causing} this phenomenon
{come from its observation in}
certain (but not all) EPM and for a certain range of parameters
only. \blue{Suspicions particularly target}
the rules for yielding or for
elastic recovery, \blue{as we will see.} Of course, relating
these \blue{somewhat abstract} rules to microscopic physical properties \blue{may not be straightfoward.}
\blue{Therefore, the interpretation remains mostly qualitative,} with very few detailed comparisons so far between microscopic calculations and EPM rules.

To start with, one notices that large applied stresses $\Sigma\gg\Sigma_{y}$
are incompatible with localization. Indeed, the applied stress then
exceeds the local yield stresses: Plastic events pervade the system,
which globally flows in a viscous manner. In other words, large loadings
fluidize the material, consistently with experimental observations
\citep{divoux2012fluidization}.

On the other hand, at lower stresses (hence, lower shear rates), plastic
events are sparser and may hit the same regions over and over again,
provided that the latter are strongly or durably weakened by these
events. Meanwhile, in the rest of the material, the driving term in
Eq.~\eqref{eq:gen_eq_of_motion} is compensated by the nonlocal contributions
due to a band of plastic events, \emph{i.e.}, a uniform relaxation \citep{Martens2012}.
The general cause for localization thus evidenced is the insufficient
healing of regions following rearrangements \citep{Nicolas2014u}.
In the following, we look into the distinct possible origins of this
weakening.

\blue{
\subsubsection{Long breakdowns (rearrangements), slow recovery}
}

\citet{Coussot2010} rationalized shear-banding in jammed systems
by considering the formation and breakage of particle clusters. Locally,
these events delimit periods of solid and liquid behavior, \blue{with elastic stresses
$\sigma^\mathrm{el} = \mu\dot{\gamma}t$ and $\sigma^\mathrm{el}=0$,} respectively,
that comes on top of a constant viscous stress $\eta\dot{\gamma}$. Here, $\mu$ is the
shear modulus and $\dot{\gamma}t$ is the local strain.
On the basis of a mean-field argument,  they showed that
if the liquid-like phase lasts longer than $\nicefrac{\eta}{\mu}$,
then the flow curve becomes nonmonotonic, which is the hallmark of
shear-banding. The idea was elaborated by \citet{Martens2012}, who
used a spatially resolved EPM of the Picard type with a variable rearrangement
(`healing') time $\tau_{\mathrm{res}}$ as a parameter, with the
notation of Eq.~\eqref{eq:V_Picard_rates}. Their findings confirmed
the formation of shear bands in space for large $\tau_{\mathrm{res}}$,
associated with the emergence of nonmonotonicity in the macroscopic
flow curve. The banded flow shares many properties with systems at
a first-order transition in which different phases coexist; the shear
rate is well defined (independent of the driving) inside the band
and there is an interface with the nonflowing phase. This spatial organization in the form of a band is
intrinsically related to the \blue{band being a soft mode of the propagator}
(see Sec.~\ref{sub:Pointwise-idealisation-of}); this would not be possible without its
long range and its anisotropy.

Attractive interactions in adhesive colloidal systems \citep{Irani2014}
and directional bonds in molecular systems are tentative candidates
for possible microscopic origins of long rearrangements, \emph{i.e.}, long
time delays before the destabilized region reaches another stable
configuration.
Similarly, \blue{deactivating potential forces for a finite `pinning delay' after yielding} 
enhances strain localization \citep{papanikolaou2016shearing}.

\blue{
\subsubsection{Influence of initial stability (aging) and shear rejuvenation (softening): \label{sub:VI_Structural-softening}}
}

\blue{
In (thermal) amorphous solids, with age comes strength and above all stability (see Sec.~\ref{sub:II_thermal_systems}).
Yielding will then be more abrupt. Indeed, 
}
letting a system age in the absence of strain
favors strain localization, or even fracture,
\blue{as indicated by experimental \citep{rogers2008aging} and numerical \citep{shi2005strain} data.} 
The EPM proposed by
\citet{Vandembroucq2011} and inspired by the weakening mechanism
in \citet{fisher1997statistics}'s model for earthquakes helped interpret
this effect: The distribution $P(\sigma_y)$  used for resetting
the local yield stresses $\sigma_y$ following a plastic event was shifted by an amount
$\delta$ with respect to the initial $P(\sigma_y)$,
to mimic the lower structural temperature 
%(or fictive temperature,in the glass community's terminology) 
of the \blue{pristine} material.
For large enough negative $\delta$ (strain weakening), the first
regions to yield are rejuvenated to a state with lower threshold,
so that the system gets trapped in a banded structure. The bands thus
created are localized and pinned in space if the elementary slip distance
is small; otherwise, larger slip events are created, enhancing nonlocal
effects and making bands less stable and more diffusive.

 \citet{Nicolas2014u}
introduced a healing process in this picture, by allowing the blocks
that have just become elastic again to age and gradually recover higher
energy barriers, viz.,
\[
\dot{E}_{y}(t)=k\frac{E_{y}^{\infty}-E_{y}(t)}{E_{y}^{\infty}-E_{y}^{\mathrm{min}}},
\]
where $k$ is the rate of recovery at which the energy barrier rises
from its post-yielding value $E_{y}^{\mathrm{min}}$ to the asymptotic
value $E_{y}^{\infty}$. For low enough recovery rates $k$,
 shear localization was observed. However,
the localized behavior tends to fade away when $\dot{\gamma}$
reaches very small values. This may be paralleled with the recovery
of a homogeneous flow in the dense colloidal suspension studied by
\citet{Chikkadi2011} for shear rates below a certain value, which
allow the strained system to structurally relax before further deformation.

\textbf{\emph{}}
\begin{figure}
\begin{centering}
\includegraphics[width=14cm]{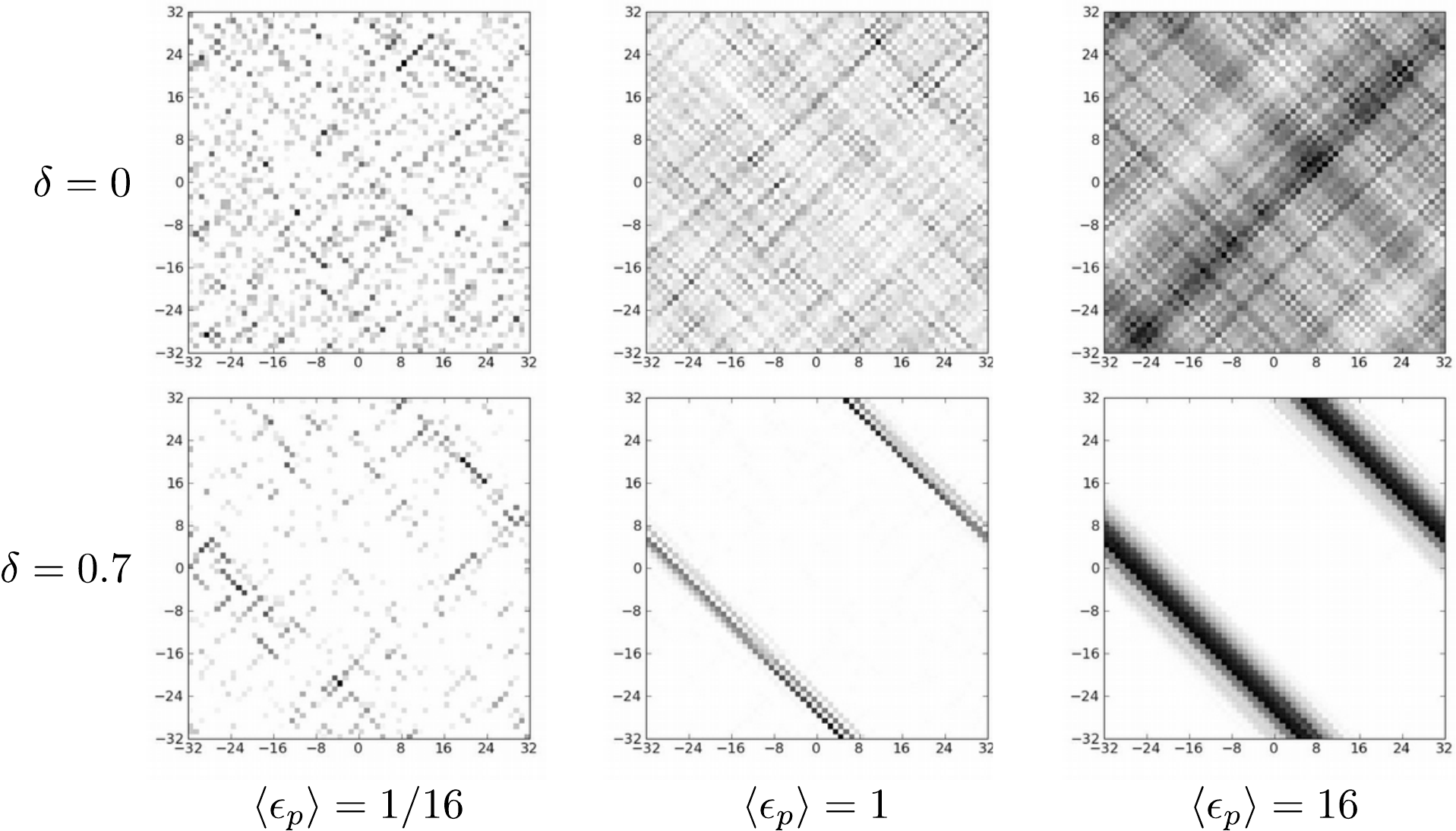}
\par
\end{centering}
\caption{ \blue{\emph{Strain localization in an EPM.}} Maps of the cumulated plastic strain $\epsilon_{p}$
at different rescaled `times' $\left\langle \epsilon_{p}\right\rangle $;
$\delta$ quantifies the weakening of the local yield stresses when they are renewed, as explained in the main text
(the bottom row is strain weakening, while the top row is not).
Darker colors represent larger values of $\epsilon_{p}$. The principal
strain directions are the horizontal and vertical directions. Adapted
from \citep{Vandembroucq2011}. }
\end{figure}

Along similar lines, \citet{Li2013} implemented a process
of free volume creation and annihilation in a finite-element-based
EPM designed to describe the deformation of the metallic glass Vitreloy
1. In their model, free volume is created by the dilation accompanying
a shear transformation (ST) and is annihilated gradually in strictly local
diffusional events. The activation of STs,
in turn, is facilitated by a local excess of free volume. Simulations
relying on a kinetic Monte Carlo scheme for the dynamics showed that at low temperatures
the deformation localizes in bands and that the
variations of free volume are critical for this localization. A parallel
can obviously be drawn between the creation of free volume during
STs and the lowering of yield stresses in other
EPM. There is perhaps an even stronger connection with the plasticity-induced
enhancement of the local effective temperatures $x$ and $\chi$ in variants of the
Soft Glassy Rheology model \citep{Fielding2009} and the Shear Transformation
Zone theory \citep{Manning2007}, respectively.
\blue{In these models, the effective temperature evolves locally in response to three processes:
(i) rises due to plastic
rearrangements, (ii) relaxation to a steady-state value, and (iii) diffusion in the shear gradient direction. For a range of parameters, the homogeneous temperature profile is unstable and a high-temperature shear band emerges in the midst of 
a low-temperature unsheared background.
}

Another approach to account for the competition between local relaxation
and driving-induced plastic events was proposed by \citet{Jagla2007}.
In his continuous model (see Sec.~\ref{sub:Continuous-approaches}),
the system relaxes via a slow drift of the local energy landscape
seen by a given site towards lower energies. Sites whose evolution
towards potential minima is not interrupted by plastic deformations
benefit from this local `structural relaxation'.
Their elastic energy decreases and the local yield stress increases;
while their plastically active counterparts have no time to undergo
structural relaxation, and their yield stress remains consequently
low. Again, this leads to a nonmonotonic flow curve in a mean-field
analysis, and to strain localization at low $\dot\gamma$.

To what extent precisely these strain localization mechanisms are
connected with the weakening-induced runaway (system-spanning) events
observed in \citet{fisher1997statistics}'s model for earthquakes
or \citet{Papanikolaou2012} and \citet{Jagla2014}'s topple-down
oscillations due to viscoelastic relaxation between earthquakes remains
uncertain.

\blue{
\subsubsection{Shear bands like it hot}
}

A temperature rise $\Delta T$ has been experimentally evidenced during the operation
of shear bands in metallic glasses \citep{Lewandowski2006temperature,Zhang2007local}.
The dominant view is that it is however not the initial cause of the
shear banding observed at low strain rates, as $\Delta T$
is small in this case. Still, local heating may result in the recrystallisation
of the material, with associated changes in its mechanical properties
(presumably more brittle behavior). Such effects are obviously not
included in EPM, and probably better described at the level of macroscopic
equations as a thermomechanical instability. The discussion above
is therefore only relevant for the initiation of the instability and
for systems in which thermal effects are weak.

A related mechanism leading to a nonmonotonic flow curve, first identified
in MD simulations \citep{nicolas2016effects} and then also seen in
finite-element-based EPM \citep{karimi2016continuum}, is at play
when \blue{one enters the underdamped regime.} At a given strain rate $\dot\gamma$, long-lived
inertial vibrations \blue{can then be sustained, because of which the yield threshold
may be exceeded earlier than if mechanical equilibrium had been maintained the whole time.}
In other words, the energy
dissipated in the \blue{underdamped} flow remains long enough in the relevant degrees
of freedom to \blue{activate plastic relaxation.}
In MD simulations this
facilitation was shown to be \blue{quantitatively} described by a heating effect, whereby a more strongly
damped system \blue{is heated to a strain-rate-dependent temperature $T(\dot\gamma)$. For strongly underdamped
systems, this leads to a nonmonotonic constitutive curve and to the formation of shear bands, if the system is
large enough\footnote{J.-L. Barrat, K. Martens and V. Vasisht, in preparation.}.
}

\section{Critical Behavior and Avalanches at the Yielding Transition\label{sec:VIII_Avalanches}}

Amorphous solids retain complex solid-like properties
under continuous flow, but the \emph{onset} of flow is of particular interest
from a physical viewpoint owing to the critical behavior that may
come along with \blue{this transition.} Far from being a weakness, the simplified description
provided by EPM (which were originally phenomenological models) represents
an asset for the study of these \blue{critical properties.} In this section
we review the thriving literature about the statistics
of avalanches close to the yielding transition.

\subsection{Short introduction to out-of-equilibrium transitions}

Statistical physics is largely concerned with phase transitions, whereby
some properties of a system abruptly change upon the small variation
of a control parameter. The paradigmatic example of an equilibrium
phase transition is the Ising model, which consists of spins positioned
on a lattice and interacting with their first neighbors. This model
describes the ferromagnetic to paramagnetic transition of a magnet
as the temperature $T$ rises above a critical temperature $T_{c}$;
the transition is marked by the presence of correlated domains of
all \blue{sizes} and the vanishing of the magnetization $m$ (the
``order parameter'') as 
\begin{equation}
m\sim(T_{c}-T)^{\beta}.\label{eq:magnetization}
\end{equation}
Quite interestingly, the \emph{critical exponents}, $\beta$ and its
kin, are shared by many other, a priori unrelated systems: The latter
are said to belong to the same universality class as the Ising model.

These ideas extend beyond equilibrium, but fewer methods are available
to deal with the \emph{dynamical phase transitions} encountered out
of equilibrium. In this respect, it is worth noting that the Herschel-Bulkley
constitutive law can be recast into an expression analogous to Eq.~\eqref{eq:magnetization},
viz.,
\begin{equation}
\dot{\gamma}\sim(\Sigma-\Sigma_{y})^{\beta}.\label{eq:VIII_beta_exponent}
\end{equation}
This yielding transition is receiving more and more attention as an
example of transition in a driven system. 
\blue{
Still, the existence of a critical behavior as $\dot\gamma \to 0$ is not 
unanimously accepted: While some authors reasonably argue that thermal fluctuations
will wash out criticality at any $T>0$ \citep{Hentschel2010}, others claim that 
the material's state should become independent of $\dot\gamma$ once the driving
gets slower than any internal relaxation rate \citep{langer2015shear}, perhaps 
overlooking the possibility that the latter (for instance, mediated by the propagation of shear waves) may
become unboundedly long for increasingly large systems under slow driving.
Besides, among  the defenders of criticality,}
there have
been lively discussions as to whether it belongs to the same
universality class as the depinning transition of driven elastic
lines (Sec.~\ref{sub:X_Depinning-transition}).

\subsubsection{Avalanches in sandpile models\label{sec:avalanches_SOC}}

As models featuring threshold dynamics and a toppling rule, EPM are
also connected to the somewhat simpler sandpile models, introduced
in Sec.~\ref{sub:IV.Sandpile-models}. Let us clarify some concepts
using the latter class of systems. 

Simulations of 2D sandpile models display avalanches of grains 
of duration $T$ (number of iterations to reach stability) and
size $S$ (total number of transferred grains). \blue{Because of the rules governing grain transfer,}
these avalanches are compact structures, unlike those observed
in EPM. % (where the propagator is long-ranged and inhomogeneous).
\blue{In a nutshell,} at vanishing deposition rate, the
cumulative distributions of $S$ and $T$ exhibit power-law scalings,
with a cut-off at large scales due to the finite size of the system, viz., 
\begin{equation}
C_S(S) = S^{1-\tau} f(S/L^{d_{f}}) \text{ and } C_T(T) = T^{1-\tau\prime} g(T/L^{z}), \label{eq:avalanches_powerlaw}
\end{equation}
where $\tau>0$ and $\tau^\prime>0$ are critical exponents, $f$ and $g$ are
fast decaying functions, and the positive exponents $d_{f}$ and $z$ are called
the fractal dimension of the avalanches and the dynamical exponent, respectively.
This means that small avalanches are more frequent than larger ones, but in such
a fashion that no \textit{typical} or characteristic size can be established,
which has been called \emph{self-organized criticality}.
% Ezequiel's paragraph readapted follows -- AN
Let us note that the \emph{extremal dynamics} used to trigger avalanches
can be substituted by a very slow (quasistatic) uniform loading of the columns of sand 
if some randomness is introduced in the stability thresholds.
In this sense, self-organized criticality in the sandpile model
simply exposes the criticality associated with the dynamical
phase transition undergone by the loaded system.
%
% Let us put forward at this point that, from a conceptual point of view, there is basically no difference between
% discussing a yielding transition defined in the quasi-static limit of external driving protocol and discussing it as a SOC
% system, in which, upon stasis, we load uniformly with the minimum amount of stress needed to set a new instability.
% In this sense, self-organized criticality and criticality associated to a dynamical phase transition are identically the
% same phenomenon

\blue{Different regimes of avalanches can be seen when the deposition rate is varied or, somewhat equivalently, 
when one inspects them at different frequencies $\omega$.}
%probes the power spectrum $S_J(\omega)\equiv\int dt\int d\tau e^{-i\omega\tau}J(t)J(t+\tau)$
%of the current $J$, i.e., the number of grains falling out of the pile
%per unit time.
At large frequencies $\omega$, independent, non-overlapping
avalanches are observed. \blue{As $\omega$ decreases,} the avalanches start interacting.
In this regime, their overlaps cut off the correlation lengths of
single avalanches, but due to mass conservation during grain transfer,
the scale-free behavior is preserved.
On long time scales, i.e., for low $\omega$, the observed features are typical
of discharge events, whereby the whole sandpile becomes unstable after having
been loaded~\citep{hwa1992avalanches}.

\subsubsection{Stress drops and avalanches in EPM}

Similarly to the instabilities in sandpile models, the plastic events
occurring in EPM can trigger avalanches of successive ruptures. To
facilitate the comparison with experiments or atomistic simulations,
these avalanches are usually quantified by looking at the time series
of the macroscopic stress \blue{$\sigma(t)$} and, more specifically, at
the stress drops $\Delta\sigma$ associated with plastic relaxation \blue{(in this chapter,
we use a lowercase symbol $\sigma$ for the stress to underscore that it is an intensive variable).}
Close to criticality, the duration $T$ of these drops and their extensive
size $S\equiv\Delta\sigma L^{d}$ \blue{in a system of volume $L^d$ in $d$ dimensions} most often display \blue{statistics}
formally similar to Eq.~\eqref{eq:avalanches_powerlaw}, viz.
\begin{equation}
P(S)\sim S^{-\tau}f(S/S_{\mathrm{cut}}) \text{ and } P(T)\sim T^{-\tau^{\prime}}g(T/T_{\mathrm{cut}}) ,\label{eq:VIII_avalanches_powerlaw_EPM}
\end{equation}
where the upper cut-offs $S_{\mathrm{cut}}$ and $T_{\mathrm{cut}}$
entering the scaling functions $f$ and $g$ will typically depend
on system size, e.g. $S_{\mathrm{cut}}\propto L^{d_f}$.
In the following, we will pay particular attention
to the possible impact of the peculiarities of the quadrupolar stress
redistribution in EPM, notably its fluctuating sign, on the avalanche
statistics.

\subsection{Avalanches in mean-field models \label{sub:VIII_avalanches_MF}}

Shortly after the emergence of the first EPM, mean-field approximations
were exploited to determine the statistics of avalanches.
Most of these approaches assume a uniform redistribution of the stress released by
plastic events, as exposed in Sec.~\ref{sub:uniform_stress_redistrib}.
An exponent $\tau=\nicefrac{3}{2}$ is then consistently found in
the avalanche size scaling of Eq.~\eqref{eq:VIII_avalanches_powerlaw_EPM}. 

For instance, \citet{sornette1992mean} proposed to map the Burridge-Knopoff
model for earthquakes, \blue{introduced in Sec.~\ref{sub:IV.Sandpile-models},}
onto a fiber bundle \blue{which carries a load equally shared among unbroken fibers} (see Sec.~\ref{sub:X_Fibre_bundles}).
At criticality the extremal load \blue{needed to make the weakest surviving fiber break fluctuates; more precisely, it} performs
an unbiased random walk. \blue{An avalanche of failures lasts as long as this extremal load remains below the initial load,
so its size is given by} the
walker's survival time close to an absorbing boundary, whence an exponent
$\tau=\nicefrac{3}{2}$.
If deformation starts farther from the critical point, a larger exponent
is found, $\tau=\nicefrac{5}{2}$.
A posterior, but widely celebrated model %\citep{dahmen1998gutenberg}
for heterogeneous faults in earthquakes was proposed by \citet{fisher1997statistics},
and later applied to the deformation of crystals by \citet{Dahmen2009}
and more recently to the deformation of granular matter \citep{Dahmen2011}
and amorphous solids \citep{Antonaglia2014}.
Here, the problem is directly mapped onto an elastic line depinning problem
(see Sec.~\ref{sub:X_Depinning-transition}).
Once again, above an upper critical dimension that decreases with
the interaction range, the model yields the mean-field exponent $\tau=\nicefrac{3}{2}$.
But if a post-yield weakening mechanism is introduced or if stress
pulses due to inertial effects are present, the power-law regime only
holds for small avalanches, while larger ones trigger runaway events
that span the whole system and result in a bump at a characteristic
size in the avalanche statistics.

Much more recently, there have been endeavors to extend mean-field
approaches in order to account for the non-positiveness of the redistributed
stress, which undermines the mean-field reasoning. \blue{For instance, the H\'ebraud-Lequeux model introduced
in Sec.~\ref{sub:V_Hebraud_Lequeux} features an additional diffusive
term acting on local stresses.}
\citet{JaglaPRE2015} studied avalanches in a discrete variant of this model
and reported on subtleties that are absent from depinning problems.
Indeed, if avalanches are artificially triggered by picking a random block
and destabilizing it, the problem can yet again be mapped onto a
survival problem for an unbiased random walk, similarly to the fiber bundle,
and the mean-field exponent $\tau \simeq\nicefrac{3}{2}$ is obtained. (\blue{In passing, with a
random-kick protocol of the sort,} \citet{Lin2014Density} had arrived at a similar result, in two EPM variants.)
\blue{But now consider} the physically more relevant protocol of quasistatic loading, \blue{where
stresses are uniformly increased} until a block is destabilized. The foregoing result still
holds in the depinning case, because the distribution
of local stresses $\sigma$ \blue{is fairly homogeneous close to the yield point $\sigma_{y}$, viz.
$p(\sigma\approx\sigma_{y}) \simeq\mathrm{cst}$, and thus}
insensitive to the stress shift induced by the uniform loading. By contrast,
stress fluctuations in disordered solids deplete local stresses close to
$\sigma_{y}$, so much so that $p(\sigma_{y}^{-})=0$ \blue{and
$p(\sigma)$ varies substantially close to $\sigma_{y}$.}
Accordingly, significantly smaller exponents $\tau \simeq 1.1-1.2$ are both
predicted and observed numerically in that case~\citep{JaglaPRE2015}.
Furthermore, the power law is cut off at a value $S_{\mathrm{cut}}$
that depends on the distance to criticality and on the system size.
An extension of these results to heavy-tailed
distributions of stress fluctuations \citep{lin2016mean} is still pending.

\subsection{Experimental observations and atomistic simulations of avalanches}

\subsubsection{Experiments}

\begin{figure}
\begin{centering}
\includegraphics[width=1\columnwidth]{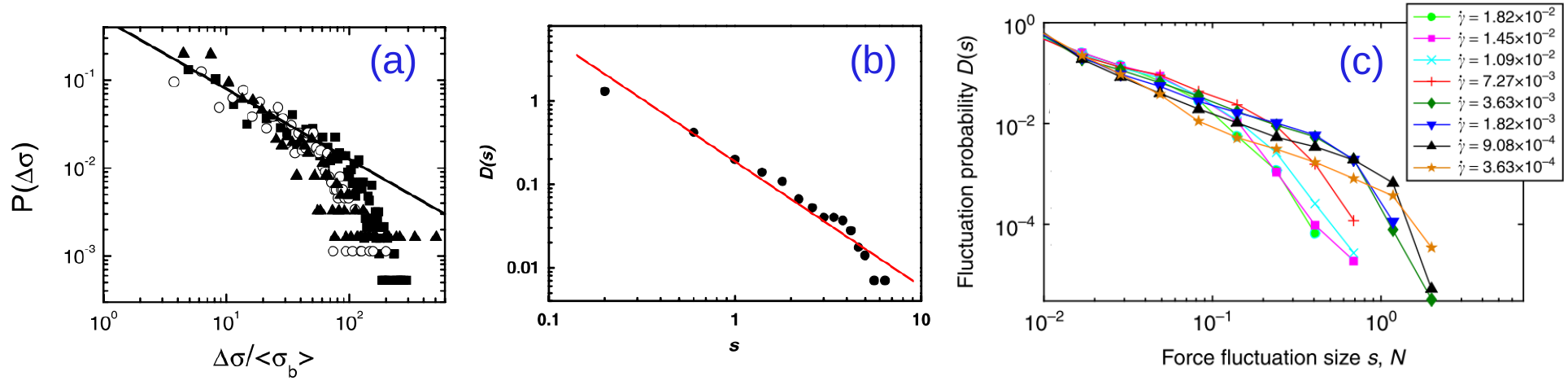} 
\par
\end{centering}

\caption{\label{fig:avalanches_exp} \textit{Distributions of stress drops in the deformation
of amorphous materials}. \textbf{(a)} Distribution of stress drops
$\Delta\sigma$ in a foam that is strained in a Couette cell, for
three different strain rates. From \citep{lauridsen2002shear}
with APS permission. The solid line in this logarithmic plot has a
slope of $-0.8$.\textbf{ (b)} Distribution $D(s)$ of stress drops
of normalized magnitude $s$ in a metallic glass ($\mathrm{Cu}_{47.5}\mathrm{Zr}_{47.5}\mathrm{Al}_{5}$).
Adapted from~\citep{Sun2010plasticity}.
The red line represents a power law with exponent $\tau \simeq 1.49$. \textbf{(c)}
Distribution $D(s)$ of force fluctuation sizes $s$ in a sheared granular
system, for different shear rates and at constant confining pressure
$P=9.6\,\mathrm{kPa}$. Adapted from~\citep{denisov2016universality}
with permission. The data suggest truncated power laws $D(s)\sim s^{-\tau}\exp(-s/\dot{\gamma}^{\mu})$,
with $\tau=1.5$ and $\mu=0.5$. }
\end{figure}

Various experimental settings have been designed to characterize avalanche
statistics in deformed amorphous solids in the last decade, even though
experiments are still trailing behind the theoretical predictions
and numerical computations in this area.
Let us mention examples of such works.

\citet{lauridsen2002shear} sheared a foam in a Couette cell and investigated
its plastic behavior. The distributions $P(S)$ of normalized stress
drops $S$ (plotted in Fig.~\ref{fig:avalanches_exp}a) were shown
to follow a power law at three different shear-rates, with an apparent
exponent $\tau \simeq 0.8$ in Eq.~\eqref{eq:VIII_avalanches_powerlaw_EPM}.
This value was reported to be consistent with the bubble model of \citet{durian1997bubble},
but contrasts with other theoretical predictions, as we will see.
It should however be noted that the power law was fitted over barely
a decade in $S$.

At the other end of the softness spectrum, the compression of millimetric
metallic glass rods was studied by \citet{Sun2010plasticity} and
the stress drops were analyzed. Again, $P(S)$ follows a power law
regime over one decade of experimental measurements, but this time
with exponents in the range $\tau\in\left[1.37,1.49\right]$, as can
be seen in Fig.~\ref{fig:avalanches_exp}b. Among several works that
came in the wake of this seminal paper, \citet{Antonaglia2014}'s
compression experiments of microsamples were argued to be compatible
with the mean-field prediction $P(S)\sim S^{\nicefrac{-3}{2}}$. Following
the same approach, \citet{TongIntJourPlas2016} reported exponents
in the range $\tau\in\left[1.26,1.6\right]$ for four different samples
of a $\mathrm{Cu}_{50}\mathrm{Zr}_{45}\mathrm{Ti}_{5}$ alloy.

\blue{A granular packing} subject to the simultaneous application
of pressure and shear was also shown to display stress drops with
power-law statistics by \citet{denisov2016universality}.
The power-law exponents, which seem to lie in a relatively broad range
in Fig.~\ref{fig:avalanches_exp}c, were not fitted, but, upon rescaling,
were reported to be in good agreement with the mean-field value $\tau=\nicefrac{3}{2}$.
It remains uncertain to what extent the value reported in this work and in the
other ones may have been influenced by the large body of literature
claiming that the deformation of (a large variety of) amorphous materials
belongs to a unique universality class,
the one describing the depinning of an elastic line \citep{Dahmen2009,dahmen2017mean}.
Also note that \blue{the two decades of raw (non-cumulative)
stress drops over which \citet{denisov2016universality} could collect data, at least
for the smaller strain rates, make granular packings particularly promising experimental test systems.
\citet{BaresPRE2017}'s recent study of a
sheared bidisperse mixture of photoelastic particles confirms these promises.
From the gradient of the image intensity,
they quantified the local pressure acting on each
grain, hence the energy stored in it, and tracked the fluctuations of the global energy.
This allowed them to define
avalanches as spontaneous energy drops, with a dissipated
power $E$ related to granular rearrangements.
The researchers reported power-law distributions of avalanches, $P(E) \sim E^{-\tau}$
with $\tau = 1.24 \pm 0.11$, in a range dependent on the threshold
used to filter the signal and spanning over three decades in $E$.
}

\subsubsection{Atomistic simulations}

%\begin{comment}
%[discuss shortly each group of references, without going into any detail of the
%MD simulations, we will keep the focus on EP models and we are only
%interested in the outcome of MD simulations for comparison] 
%\end{comment}

In parallel to experiments, stress drops have
been analyzed in atomistic simulations of the deformation of glassy
materials.  In a 2D packing of soft spheres, \citet{Maloney2004} measured power-law
distributed energy drops with an exponent $\tau=0.5-0.7$ comparable to that obtained in
\citet{durian1997bubble}'s foam experiments. On the contrary, exponential distributions of
 stress drops and energy drops were
then reported in athermal systems of particles interacting with three distinct potentials
in  2D \citep{Maloney2006Amorphous}, but also with a more realistic potential for a metallic
glass in 3D \citep{Bailey2007}. All these studies were however limited
to fairly small system sizes. Using larger systems in 2D and 3D, \citet{Salerno2013}
found power-law distributed energy drops and stress drops, with distinct values for the exponent $\tau$ in
the overdamped regime and the underdamped one, and in 2D and 3D. In the overdamped case, the value is
identical in 2D and 3D, $\tau= 1.3 \pm 0.1$. We also mention that, opposing the rather consensual view of scale-free avalanches and non-trivial spatiotemporal
correlations, \citet{dubey2016statistics} suggested that the characteristics of the stick-slip
behavior stemmed from trivial finite-size effects.

\subsection{Avalanche statistics in EPM}

The large amount of statistics afforded by EPM can enlighten
the debate about the criticality of the yielding transition
and the existence (or not) of a unique class of universality
by overcoming the uncertainty and limitations of some experimental
measurements. In the last years, EPM have tended to challenge the
strict amalgamation of the yielding transition with the depinning
one.

\subsubsection{Avalanche sizes in the quasistatic limit}

\begin{figure}
\begin{centering}
\includegraphics[width=1\columnwidth]{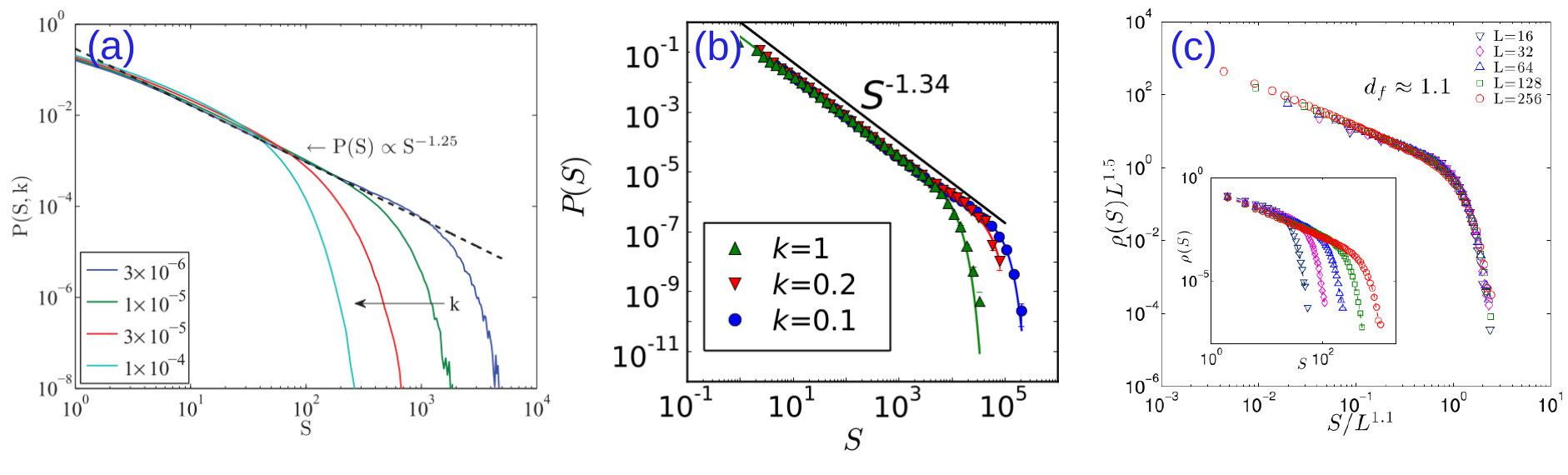} 
\par\end{centering}

\caption{
\blue{ \textit{Distributions $P(S)$ of avalanche sizes obtained with 2D EPM in the quasistatic limit}, in diverse settings:
\textbf{(a)} Under extremal dynamics, where the system (of linear size $L=256$) is driven by spring of variable stiffnesses
$k$, indicated in the legend.} 
The fitted exponent is $\tau \simeq 1.25$. From~\citep{Talamali2011}.
\textbf{(b)} In strain-controlled simulations with the `image sum'
implementation of the elastic propagator kernel (see Sec.~\ref{sub:Pointwise-idealisation-of}.2), with
fits to Eq.~\eqref{eq:PSbudrikisempirical}.
Notice that the driving springs are much stiffer than in panel (a).
From~\citep{Budrikis2013}.
\textbf{(c)} Under extremal dynamics, for systems of different sizes $L$.
The exponent reported for the unscaled curves (inset) is $\tau \simeq 1.2$, while
the rescaled curve shown in the main plot was fitted with $\tau \simeq 1.36$.
From~\citep{lin2014scaling}. }
\label{fig:avalanches_epm1} 
\end{figure}

Avalanches are most easily defined in the limit of quasistatic driving,
in which the external load is kept fixed during avalanches (Sec.~\ref{sub:III_Driving}).
Applying extremal dynamics to a 2D EPM, \citet{Talamali2011} defined
an avalanche size $S$ as the number of algorithmic steps $\Delta t$
during which the external stress $\Sigma$ remains lower than $\Sigma_{\mathrm{start}}-k\Delta t$,
as though the system were driven by a slowly moving spring of stiffness $k$.
\blue{Quite interestingly, driving the system by pulling on it with a moving spring is equivalent to a strain-controlled protocol in the limit
$k\to \infty$, while a stress-controlled protocol is recovered in the opposite limit $k \to 0$ \citep{popovic2018elasto}.} 
\citet{Talamali2011}'s numerical simulations displayed a scale-free distribution
$P(S)\propto S^{-\tau}$ with $\tau = 1.25 \pm 0.05$ cut by a Gaussian tail
(Fig.\ref{fig:avalanches_epm1}a).
It was made explicit that this result is at odds with the mean-field
exponent $\tau=\nicefrac{3}{2}$.
On the other hand, the measured value is similar to that measured
by \citet{durin2000scaling} ($\tau \simeq 1.27$) for one class of Barkhausen
avalanches, due to the motion of ferromagnetic domain walls under
an applied magnetic field, and to that predicted for this effect using
a model of elastic line depinning with anisotropic (dipolar, but positive)
interactions \citep{zapperi1998dynamics}. Of at least equal relevance
is the similarity with the avalanche size exponent $\tau \simeq 1.25$ found
when simulating differential equations \citep{BonamyPRL2008} or cellular
automata \citep{LaursonPRE2010} to describe the interfacial growth
of a crack in a heterogeneous medium. Indeed, the alignment of plastic
events along the Eshelby `easy' axes was seen as an effective dimensional
reduction, leading to avalanches belonging to a quasi 1D problem with
positive interactions decaying as $r^{-2}$, similarly to the interfacial
crack growth model of \citet{BonamyPRL2008}.

A couple of years later, \citet{Budrikis2013} exploited a closely
related EPM, with randomly distributed stress thresholds, to investigate
the effect of two distinct implementations of periodicity for the
long-range elastic propagator $\mathcal{G}$ defined in Sec.~\ref{sub:Pointwise-idealisation-of}.2.
A first series of simulations focuses on the nonstationary plastic
activity below the macroscopic yield stress $\Sigma_{y}$, by adiabatically
increasing the applied stress $\Sigma$. 
Overloaded blocks yield simultaneously;
their strain is increased by $d\gamma=0.1$ and a new local yield
stress is drawn.
For $\Sigma\ll\Sigma_{y}$, avalanche distributions  $P(S)$
are found to decay as exponentials (or compressed exponentials);
but for stresses closer to $\Sigma_{y}$, a short power-law regime appears.
The distributions can be fitted by \citet{LeDoussalPRE2012}'s
first-order correction to the mean-field \blue{prediction} for depinning
(see Eq.~\ref{eq:LeDoussalWiessePS}), but with $\tau \simeq 1.35$ instead.
The tails of $P(S)$ collapse \blue{upon rescaling with}
a cutoff depending on the distance to the critical point
$S_{\mathrm{cut}}\propto(\Sigma_{y}-\Sigma)^{\nicefrac{-1}{\sigma}}$
with $\nicefrac{1}{\sigma} \simeq 2.3$.
In a second series of simulations, apparently inspired by \citet{Talamali2011},
the system was pulled by \blue{a spring of stiffness $k\in[0.1,1]$, moving adiabatically,}
hence an external stress $\Sigma=k(\gamma_{\mathrm{tot}}-\gamma)$,
where $\gamma_{\mathrm{tot}}$ is the position of the spring and $\gamma$
is the plastic strain.
In this case, \blue{the statistics can be improved because the system eventually reaches a critical steady state.}
The avalanches size distributions show a larger power-law regime and 
a little `bump', and the authors fit them with the empirical
shape
\begin{equation}\label{eq:PSbudrikisempirical}
P(S)=c_{1}S^{-\tau}\exp(c_{2}S-c_{3}S^{2}),
\end{equation}
\blue{where all coefficients are free.} Two very close, but not strictly identical, exponents $\tau \simeq 1.35$ (again) were measured for different
implementations of the propagator; the precise values of $\tau$ can be found in Fig.~\ref{fig:avalanches_epm1}b and Table~\ref{tab:VIII_critical_exponents}.
These values somewhat differ from \citet{Talamali2011}'s measurement, presumably because \blue{larger spring constants $k$ were used.}
Still, they definitely deviate from the mean-field
value, too.
The authors also considered the avalanche durations $T$, measured in algorithmic steps
\blue{and fitted them to the} power law $P(T)\propto T^{-\tau'}$,
with $\tau'\simeq 1.5$.
Joining these researchers, \citet{SandfeldJSTAT2015}
tested the robustness of these avalanche statistics to \blue{(i)} variations
of the boundaries, \blue{(ii)} different computations of stress redistribution and \blue{(iii)}
finite-size effects. \blue{To do so,} they used an eigenstrain-based finite element
method with different types of meshgrids, and found that these variations have no influence on the critical exponents.

\citet{Lin2014Density} implemented two slightly different automata
 based on the H\'ebraud-Lequeux model but embedded in finite dimensions.
In stress-controlled simulations at $\Sigma \sim \Sigma_{y}$, in which sites are
randomly `kicked' to trigger an avalanche, \blue{they found an exponent $\tau$
larger than that of the quasistatic simulations described above: They measured $\tau \approx 1.42$ in both model variants.
This may be a consequence of the random-kick protocol [see Sec.~\ref{sub:VIII_avalanches_MF} and \citep{JaglaPRE2015}].}
Yet, later on, \citet{lin2014scaling} reported $\tau \simeq 1.36$ in 2D and $\tau \simeq 1.43$
in 3D, for the same protocol.
Besides,  power-law distributions were reported for the avalanche durations, with exponents
$\tau^\prime \simeq 1.6$ in 2D and $1.9$ in 3D. In parallel, \emph{extremal dynamics} were implemented and
yielded smaller exponents for the same models,
$\tau \simeq 1.2$ in 2D and $\tau \simeq 1.3$ in 3D, closer to previous
quasistatic approaches, even though not devoid of finite-size effects.

\begin{table}[bh]
\caption{\label{tab:VIII_critical_exponents}
List of values measured for the critical exponents characterizing avalanches in EPM.
Only values measured in EPM with \textit{extremal dynamics} (or akin) and a \textit{quadrupolar propagator} are reported. 
Mean field values are added for comparison.
}
\begin{centering}
\begin{tabular}{>{\centering}m{4.5cm}|>{\centering}m{2.5cm}>{\centering}m{2.5cm}>{\centering}m{2.5cm}>{\centering}m{2.5cm}>{\centering}m{2.5cm}}
%\begin{tabular}{c|cccc}
\hline 
\hline 
{\bfseries Exponent} & $\tau$ & $\tau^{\prime}$ & $d_{f}$ & $\theta$ & $\gamma$ \tabularnewline
\hline 
{\bfseries Expression} & $P(S)\sim S^{-\tau}$ & $P(T)\sim T^{-\tau^{\prime}}$ & $S_{\mathrm{cut}}\sim L^{d_{f}}$ & $p(x)=x^{\theta}$ with $x\equiv\sigma_{y}-\sigma$ & $S\sim T^\gamma$ \tabularnewline
\hline 
\multicolumn{6}{c}{\scshape 2D EPM}\tabularnewline
\hline 
\citep{Talamali2011} [spring coupling $k \to 0$] & $1.25 \pm 0.05$ & --- & $\sim 1$ & --- & --- \tabularnewline
%\hline 
\citep{Budrikis2013} [spring coupling $k \gtrsim  0.1$] & $1.364\pm 0.005$ & $1.5 \pm 0.09$ & $\gtrsim 1^\dagger $ & --- & $\sim 1.85$ \tabularnewline  %%$1.342\pm 0.004$ \text{and} 
%\hline 
\citep{lin2014scaling} [extremal] & $\sim 1.2$ & $\sim 1.6$ & $1.10\pm0.04^*$ & $\sim 0.50$ & --- \tabularnewline
%\hline 
\citep{liu2015driving} [$\dot{\gamma}\rightarrow0$] & $1.28\pm0.05$ & $1.41 \pm 0.04$ &  $0.90 \pm 0.07$ & $0.52 \pm 0.03$ & $1.58 \pm 0.07$ \tabularnewline
%\hline 
\citep{budrikis2015universality} [adiabatic loading] & $ 1.280\pm0.003 $ & --- & --- & $0.354 \pm 0.004$ & $1.8 \pm 0.1$ \tabularnewline
\hline 
\multicolumn{6}{c}{\scshape 3D EPM}\tabularnewline
\hline 
\citep{lin2014scaling} [extremal] & $\sim 1.3$ & $\sim 1.9$ & $1.50\pm0.05^*$ & $\sim 0.28$ & --- \tabularnewline
%\hline 
\citep{liu2015driving} [$\dot{\gamma}\rightarrow0$] & $1.25\pm0.05$ & $1.44 \pm 0.04$ & $1.3 \pm 0.1$ & $0.37 \pm 0.05$ & $1.58 \pm 0.05$ \tabularnewline
%\hline 
\citep{budrikis2015universality} [adiabatic loading] & $ 1.280\pm0.003 $ & --- & --- & $0.354 \pm 0.004$ & $1.8 \pm 0.1$\tabularnewline
\hline 
\multicolumn{6}{c}{\scshape Mean field}\tabularnewline
\hline
\citep{fisher1997statistics} [depinning] & $\nicefrac{3}{2}$ & 2 & --- & 0 & 2 \tabularnewline
%\hline 
\citep{JaglaPRE2015} [H\'ebraud-Lequeux like] & $1.1-1.2$ & --- & --- & 1 & --- \tabularnewline
\hline 
\multicolumn{6}{l}{{\footnotesize \emph{Legend -- }: $\dagger$ Estimated from the avalanches shape. $*$ Obtained using the $\tau$ exponents from the random-kick protocol. }}\\
\hline 
\hline 
\end{tabular}
\par\end{centering}
\end{table}

\subsubsection{Connection with other critical exponents}

A discussion on the density of zones close to yielding and its connection
with the critical exponents was opened up by \citet{Lin2014Density}.
Denoting $x\equiv\sigma_{y}-\sigma$ the distance to threshold of
local stresses, a stark contrast was emphasized between depinning-like
models, with only positive stress increments and $p(x)\sim x^{0}$
for small $x$, and EPM, where a pseudo-gap emerges at small $x$,
viz., $p(x)\sim x^{\theta}$ with $\theta>0$.
% In stress-controlled simulations, in which sites are randomly `kicked'
% (i.e., have their local stress altered) to trigger an avalanche,
In \citet{Lin2014Density}'s stress-controlled simulations with randomly `kicked' sites,
identical values of $\theta$ were obtained in two variants of the model
embedded in 2D ($\theta \simeq 0.6$) and 3D ($\theta \simeq 0.4$),
whereas the stress-strain curves differed (see Table~\ref{tab:VIII_critical_exponents}
for the slightly smaller values of $\theta$ measured using extremal dynamics).

Shortly afterwards, \citet{lin2014scaling} proposed to link $p(x)$ with
$P(S)$, in a scaling description of the yielding transition.
Their scaling argument can be summarized as follows.
Starting from Eq.~\eqref{eq:VIII_avalanches_powerlaw_EPM}, one obtains
$\langle \Delta \sigma \rangle \propto L^{d_f (2-\tau)-d}$. 
Now, in a stationary situation, on average this stress drop must balance the stress
increase that is applied to trigger an avalanche.
Among the $L^d$ sites, the one with the smallest
$x$, $x_\mathrm{min}$, will start the avalanche, so $\Delta\sigma \propto x_\mathrm{min}$.
If $p(x)\sim x^{\theta}$, then $x_\mathrm{\min} \propto L^{-\frac{d}{\theta+1}}$.
Identifying the two expressions \blue{of $\Delta \sigma$} leads to
\begin{equation}
\tau=2-\frac{\theta}{\theta+1}\frac{d}{d_{f}},\label{eq:VIII_scaling_relation_tau}
\end{equation}
which is supported by their EPM simulations (notably with the random-kick
protocol).
%$d_{f}$, the  fractal dimension of the avalanches, is obtained from a finite-size
%($L$) analysis of the cutoff in $P(S)$.
%viz., $S_{\mathrm{cut}}\equiv L^{d_{f}}$. 
In this regard, the discrepancy
was once again underscored between the depinning transition (with fractal 
dimensions $d_{f}\geq d$ typically, %due to multiple slips possible per site,
due to the compactness of the avalanches,
and a velocity-force exponent $\beta \leqslant 1$) and the yielding
transition (with typically $d_{f}<d$ and a rheological exponent $\beta \geqslant 1$
in Eq.~\eqref{eq:VIII_beta_exponent}).
Generalized scaling relations encompassing both transitions were put forward
(see \citet{lin2014scaling}-Supporting Information).

\begin{figure}
\begin{centering}
\includegraphics[width=1\columnwidth]{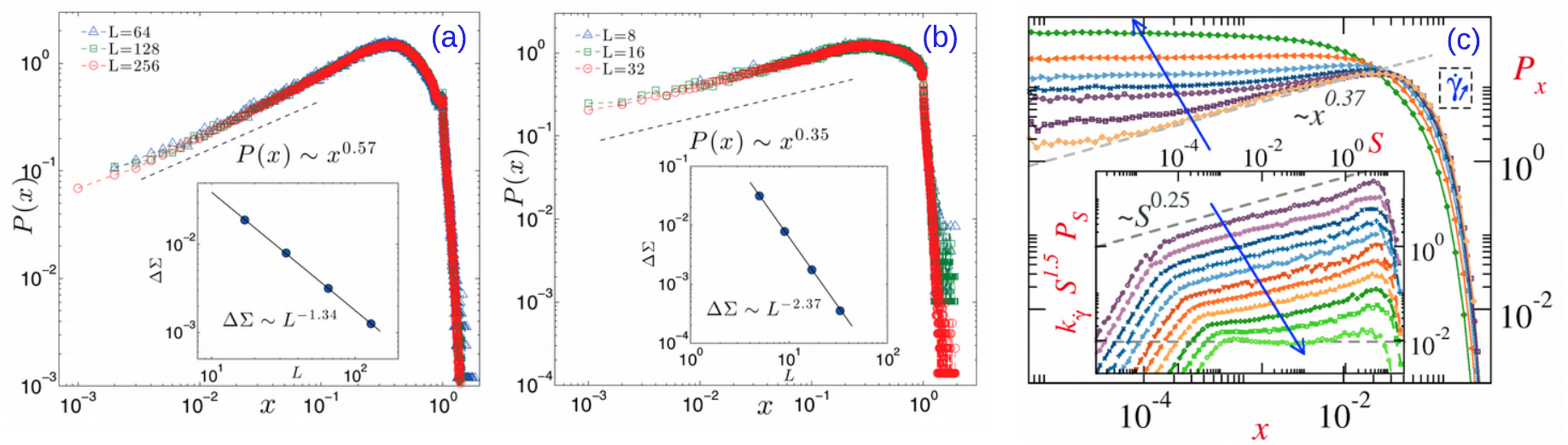} 
\par\end{centering}
\caption{\textit{Probability densities $P(x)$ of the distances $x$ to the yield threshold  in EPM}: \textbf{(a)}
In 2D and \textbf{(b)} in 3D systems, whose size $L$ is varied.
From~\citep{lin2014scaling}. \textbf{(c)} In a 3D system,
as the shear rate is varied. The inset shows the rescaled distribution of avalanche sizes. From~\citep{liu2015driving}. }
\label{fig:pofx-epm} 
\end{figure}

\subsubsection{At finite strain rates \label{sub:VIII_finite_strain_rates}}

Seeking to narrow the gap between experiments and EPM, \citet{liu2015driving}
analyzed the EPM stress signal with methods mimicking the experimental
ones and studied the effect of varying the applied shear rate $\dot{\gamma}$.
At very low $\dot{\gamma}$, avalanches are power-law distributed
with an exponent $\tau \simeq 1.28$ in 2D and $\tau \simeq 1.25$ in 3D, cut off
by finite size effects with $d_{f}=0.90$ and $1.3$, respectively.
These results coincide very well with MD simulations in the quasistatic
limit and support the nascent convergence towards an avalanche size
exponent $\tau \simeq 1.25$ in 2D or 3D EPM, deviating from the (depinning)
mean-field value $\nicefrac{3}{2}$.
Much more tentatively, there \blue{might} be a downward trend of $\tau$ with
increasing dimensions, which would be compatible
with \citet{JaglaPRE2015}'s suggestion $\tau \simeq 1.1-1.2$ above
the upper critical dimension.

Interestingly, \citet{liu2015driving} observe a systematic crossover
towards higher values of $\tau$  when the shear rate is
increased, so that $\tau$ reaches $\tau \simeq 1.5$ at intermediate
$\dot{\gamma}$, before entering the high-$\dot{\gamma}$ regime of
pure viscous flow.
At the same time, the external driving starts to
dominate over the signed stress fluctuations originating from mechanical
noise; this nudges the system into a depinning-like scenario, with
an exponent $\theta$ in $p(x)\sim x^{\theta}$ decreasing towards
zero as $\dot{\gamma}$ reaches finite values both in 2D and 3D. Note that the same effect
occurs in a stress-controlled system, as soon as the imposed stress gets
perceptibly lower than the yield stress \citep{budrikis2015universality}. 
Similarly to pulling the system with a stiff spring (large $k$), 
increasing the shear rate generates simultaneous uncorrelated plastic
activity in the system, which leads to larger $\tau$, closer to $1.5$.
Overall, applying a finite shear rate does not destroy the criticality
of avalanche statistics; but it affects the critical exponents and
eventually produces more trivial effective statistics.

\subsubsection{Insensitivity to EPM simplifications and settings}

\begin{figure}
\begin{centering}
\includegraphics[width=1\columnwidth]{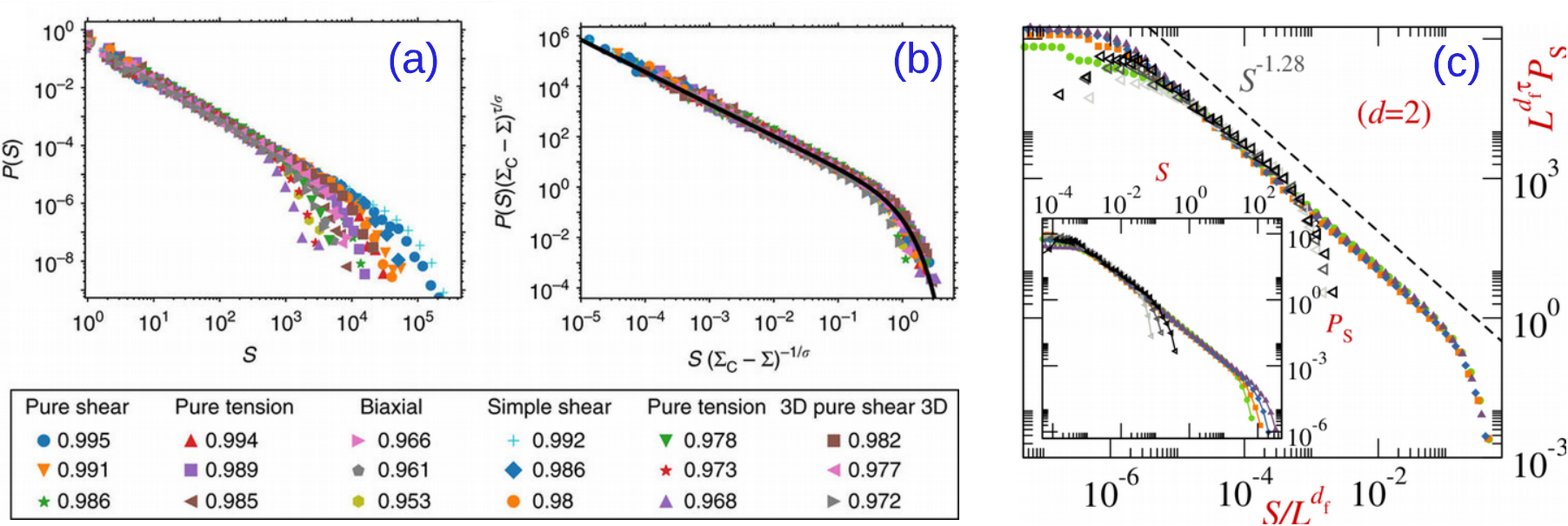} 
\par\end{centering}
\caption{\label{fig:avalanches-fem-df} \textit{Avalanche size distributions
$P(S)$ in EPM}. \textbf{(a)} Dependence on the loading conditions and the external stress $\Sigma$, as computed in
an EPM based on a Finite Element routine. The values in the legend refer to the ratio $\Sigma/\Sigma_{y}$.
In panel \textbf{(b)}, the data are collapsed  using exponents $\tau \simeq 1.280$ and
$1/\sigma \simeq 1.95$ and Eq.~\eqref{eq:LeDoussalWiessePS}.
From~\citep{budrikis2015universality}. \textbf{(c)} Rescaled distributions $L^{d_{f}\tau}P_{S}$ vs. $S/L^{d_{f}}$
in 2D, compared to MD simulations, in the quasistatic limit.
The fitted exponents are $\tau \simeq 1.28$ and $d_{f} \simeq 0.90$.
The inset shows the raw data. From~\citep{liu2015driving}.}
\end{figure}

At present, technical difficulties \blue{still} hamper a clear discrimination between
theoretical predictions \blue{through} experiments.
The simplifications used in the models thus need to be carefully examined.
\citet{budrikis2015universality} investigated the effect of the scalar
approximation of the stress (see Sec.~\ref{sub:III_scalar_vs_tensorial}) by
comparing the results of a scalar model to those of a finite-element-based fully
tensorial model, under different deformation protocols (uniaxial tension,
biaxial deformation, pure shear, simple shear) and in both 2D and 3D.
Irrespective of the dimension, and (most of the) loading and boundary conditions,
a universal scaling function is observed for the avalanche distribution,
shown in Fig.~\ref{fig:avalanches-fem-df} and coinciding with
\citet{LeDoussalPRE2012}'s proposal
\begin{equation}\label{eq:LeDoussalWiessePS}
P(S)=\frac{A}{2\sqrt{\pi}}S^{-\tau}\exp\left(C\sqrt{u}-\frac{B}{4}u^{\delta}\right),
\end{equation}
with an exponent $\tau = 1.280\pm0.003$ [note the perfect agreement
with \citet{liu2015driving}'s result], $u\equiv S/S_{{\tt max}}$
and $S_{{\tt max}}\propto(\Sigma_{c}-\Sigma)^{-1/\sigma}$ (with $1/\sigma \simeq 1.95$).
The constants, $A$, $B$, and $C$ are functions of $\tau$, as is
$\delta=2(1-\tau/3)$. 

Heterogeneous deformations, such as bending and indentation, were
also considered and yielded similar values for $\tau$.
Nevertheless, the cutoff value is different from the homogeneous cases.
\blue{This} is not unexpected: An independent length scale enters
the problem and the yield stress $\Sigma_{y}$ used to measure exponent $\sigma$
is not universal.
Also, while the observed $\tau$ value was nearly identical in the different
(homogeneous) loading cases when treated separately, some range
of variation was observed for exponent $\sigma\in[1.53,\,2.05]$. 
Finally, the average avalanche size was related to its duration $T$
via $S\sim T^{\gamma}$ with $\gamma=1.8\pm0.1$.

A possible explanation for the insensitivity of avalanche statistics
to the aforementioned aspects may lie with the quasi-1D geometry of the avalanches,
resulting from the quadrupolar propagator. Most cooperative phenomena thus appear to be controlled 
by the stress component along one direction, and a scalar description may be sufficient in this respect.
(Scalar models do indeed reproduce the same power-law exponent and evidence a fractal
dimension $d_{f}\approx1$ in 2D and 3D, as shown in Fig.\ref{fig:avalanches-fem-df}c).

\subsubsection{Effects of inertia}

Without the assumption of instantaneous stress redistribution, stress
waves are expected to propagate throughout the system (see Sec.~\ref{sub:IV_Finite_Elements}
and Fig.~\ref{fig:IV_FE_propagation}), in a ballistic way or a diffusive
one depending on the damping. 
This is not described by the traditional elastic propagator $\mathcal{G}$ of
Eq.~\eqref{eq:propagator_real_space}, but finite-element
based EPM have recently made it possible to account for inertial effects
\citep{karimi2016continuum}. \citet{karimi2016inertia} exploited
this type of model to study \citet{Salerno2013}'s claim, based on
extensive atomistic simulations in the quasistatic regime, that inertial
effects drive the system into a new (underdamped) class of universality.
At odds with this claim, but consistently with results \blue{of} sandpile
models \citep{prado1992inertia,KhfifiPRE2008} and seismic fault models
\citep{carlson1989properties}, they found that inertial effects 
\blue{destroy} the universal, scale-free avalanche statistics.
A characteristic hump (or secondary peak) of large events emerges in the avalanche
size distribution $P(S)$, similarly to \citet{fisher1997statistics}'s
findings.
In \citet{karimi2016inertia}'s work, both the relative weight
and the scaling with the system size of this peak are controlled by
the damping coefficient $\Gamma$, a dimensionless parameter that
quantifies the relative impact of dissipation. The effective fractal
dimension $d_{f}^{\prime}(\Gamma)$ of avalanches, \blue{albeit dependent on}
the damping, satisfies a scaling relation with the
exponent $\theta^{\prime}(\Gamma)$ defined by $p(x)\sim x^{\theta^{\prime}(\Gamma)}$.

These results are compatible with \citet{papanikolaou2016shearing}'s
phenomenological description of inertial effects, which are accounted
for by a temporary vanishing of elasticity after local plastic events
(plastic delay): Simulations of the model showed the appearance of
a hump of large events in $P(S)$, an increase of the exponent $\tau$,
as well as the emergence of dynamical oscillations, accompanied with
strain localization.

\subsubsection{Avalanche shapes\label{sec:avalanche_shapes}}

\begin{figure}
\begin{centering}
\includegraphics[width=1\columnwidth]{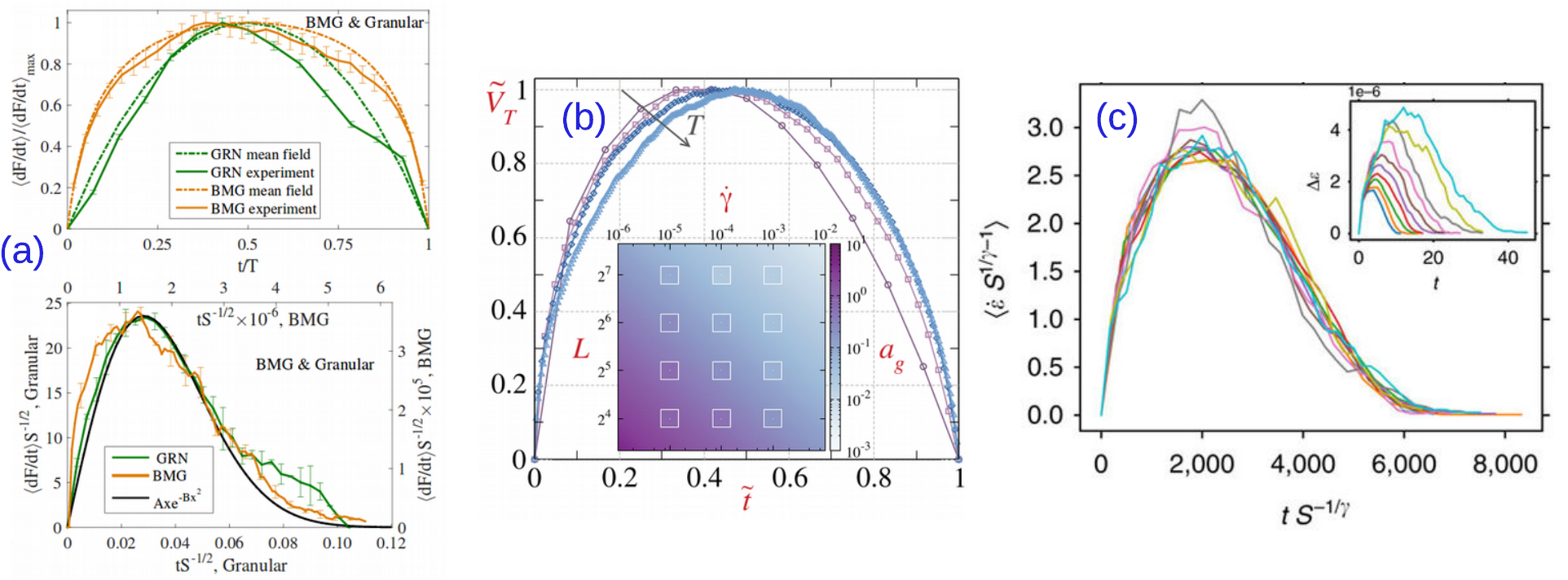} 
\par\end{centering}

\caption{\label{fig:avalanches-shapes}\textit{Avalanche shapes in experiments
and EPM}. \textbf{(a)} Experimental avalanche shapes, for avalanches
of fixed duration (\emph{top}) and fixed sizes (\emph{bottom}) in
a bulk metallic glasses (BMG) and a granular system. From \citep{denisov2016slip}.
\textbf{(b)} Avalanche shapes at different fixed durations in a strain-controlled
EPM simulation at fixed $\dot{\gamma}$. From \citep{liu2015driving}.
\textbf{(c)} Avalanche shapes at fixed sizes. From \citep{budrikis2015universality}.}
 
\end{figure}

In addition to their duration and size, further insight has been gained
into the avalanche dynamics by considering their average temporal
signal, i.e., the `shape' of the bursts. This observable can be determined
experimentally with higher quality \citep{Antonaglia2014,denisov2016slip}.
Avalanche shapes have thus been estimated for various systems displaying
crackling noise; examples include earthquakes \citep{MehtaPRE2006},
plastically deforming crystals \citep{Laurson2013}, and the Barkhausen
noise \citep{MehtaPRE2002,papanikolaou2011universality}. 

In the latter example, the magnetization of a film changes mostly
changes via the motion of domain walls\footnote{Rigorously speaking, this is true in the central part of the
hysteresis loop near the coercive field.}; its rate of change is recorded
as a time series $V(t)$.
When the film thickness, which controls
the long-range dipolar interactions, is such that mean field is valid, the average
shape $V(t|T)$ of avalanches of duration $T$ is well described \blue{in the scaling limit}
by an inverted parabola \citep{papanikolaou2011universality}, viz.,
\begin{equation}
V(t|T)\propto T\tilde{t}(1-\tilde{t})\text{ where }\tilde{t}\equiv\nicefrac{t}{T}.
\end{equation}
Since oftentimes mean field does not hold, a generalized functional
form was proposed by \citet{Laurson2013}:
\begin{equation}
V(t|T)\propto T^{\gamma-1}\left[\tilde{t}\left(1-\tilde{t}\right)\right]^{\gamma-1}\left[1-a\left(\tilde{t}-\frac{1}{2}\right)\right].\label{eq:shapes_Laurson}
\end{equation}
Here, the shape factor $\gamma$ is also the exponent that controls the scaling
between size and duration ($S\sim T^\gamma$), since $S(T)$ is nothing but the integral
$\int_0^T V(t|T) dt$.
$\gamma$ and the parameter $a$ \blue{control} the asymmetry ($a>0$ refers to positive skewness);
the mean-field formula is recovered for $\gamma=2$ and $a=0$.
As the interaction range increases from local to infinite, the university-class parameters
evolve from $\gamma \simeq 1.56$, $a \simeq 0.081$ to $\gamma \simeq 2.0$, $a \simeq 0.01$.
\citet{dobrinevski2015avalanche} provided an analytic formalization
for this expression as a one-loop correction around the upper critical
dimension; these authors also computed the shape of avalanches of
fixed \emph{size} $S$.
The need for this generalization beyond mean field was confirmed by \citet{DurinPRL2016}.

In the deformation of amorphous solids, the inverted-parabola shape
predicted by mean field was shown to provide a satisfactory description
\blue{of some experiments, e.g.} in metallic glasses \citep{Antonaglia2014} and
granular matter \citep{denisov2016slip}, 
\blue{even though the agreement is not perfect} (see Fig.\ref{fig:avalanches-shapes}a). 
\blue{
On the contrary, \citet{BaresPRE2017}'s granular experiments point to a clear asymmetry in the shape
of \emph{long} avalanches.
}

On the EPM side, \citet{liu2015driving} studied the effect of finite
shear rates $\dot{\gamma}$ on the avalanche shape. By sorting the
avalanches according to their duration $T$, at fixed $\dot{\gamma}$,
they found that short avalanches are noticeably more asymmetric and
display faster velocities at earlier times (positive skewness, see
Fig.\ref{fig:avalanches-shapes}b). For larger $T$, it is argued
that avalanches most likely result from the superposition of uncorrelated
activity, which leads to more mean-field like results. This would
explain the gradually more symmetric shapes observed for increasing
$T$ (see the evolution of the asymmetry parameter in the inset of
Fig.~\ref{fig:avalanches-shapes}b). In the quasistatic limit, asymmetric
stress-drop shapes are then expected. Indeed, at low $\dot{\gamma}$,
fits with Eq.~\eqref{eq:shapes_Laurson} give a non-mean-field value
$\gamma \simeq 1.58$ in both 2D and 3D.
This feature gradually disappears at larger $\dot{\gamma}$. \citet{budrikis2015universality}
\blue{extended the study to} different loading conditions, in 2D and
3D, and measured values for $\gamma$
in the range $[1.74,1.87]$.
Contrary to \citet{liu2015driving}'s findings with a scalar EPM, \blue{they saw}
clearly asymmetric avalanches with positive skewness only in
the bending and indentation protocols,
and not (visibly, at least) in the tension and shear simulations.
In addition, the shapes obtained by sorting the avalanches according
to their sizes (see Fig.\ref{fig:avalanches-shapes}c) collapsed well
with the scaling form proposed by \citet{dobrinevski2015avalanche}, with
a shape exponent $\gamma \simeq 1.8$ \blue{(note the difference with respect to
the mean-field value $\gamma=2$).}

\section{Steady-state bulk rheology\label{sec:VII_Bulk_rheology}}

In this Chapter we redirect the focus to materials that flow rather than fail. 
This is the relevant framework for foams, dense emulsions, colloidal suspensions, and various other soft glassy materials exhibiting a yield stress.
\blue{
 Experimentally, in the absence of shear-banding, their flow curve (which characterizes the steady-state
 macroscopic rheology) is generally well described by the Herschel-Bulkley law [Eq.~\eqref{eq:II_Herschel-Bulkley}]
 $\Sigma = \Sigma_y + A\, {\dot\gamma}^n$; the exponent $n$ typically lies in the range 0.2--0.8, often around $0.5$, perhaps closer
 to $0.3$ for foams \citep{Becu2006,Jop2012,bonn2015yield}. Such a nontrivial dependence on the shear rate
 $\dot\gamma$ proves that the rheology of these materials cannot be understood as a mere sequence of $\dot\gamma$-independent elastic loading
 phases interspersed with $\dot\gamma$-independent plastic events \citep{puglisi2005thermodynamics}.
Instead, the violation of the quasistatic conditions of Eq.~\eqref{eq:III_quasistatic} at finite shear rates
implies that the specific dynamical rules implemented in EPM will affect the rheology, at odds with 
the situation observed in the quasistatic limit. In particular, we will see that the local yielding and healing dynamics (notably via rules R2 and R4 in Sec.~\ref{sub:General_philosophy}) play a crucial role in determining the flow properties at finite $\dot\gamma$.
More generally, different flow regimes will be delineated, depending on the material time scales at play.
}

\begin{figure}[t!]
\begin{center}
\includegraphics[width=\columnwidth, clip]{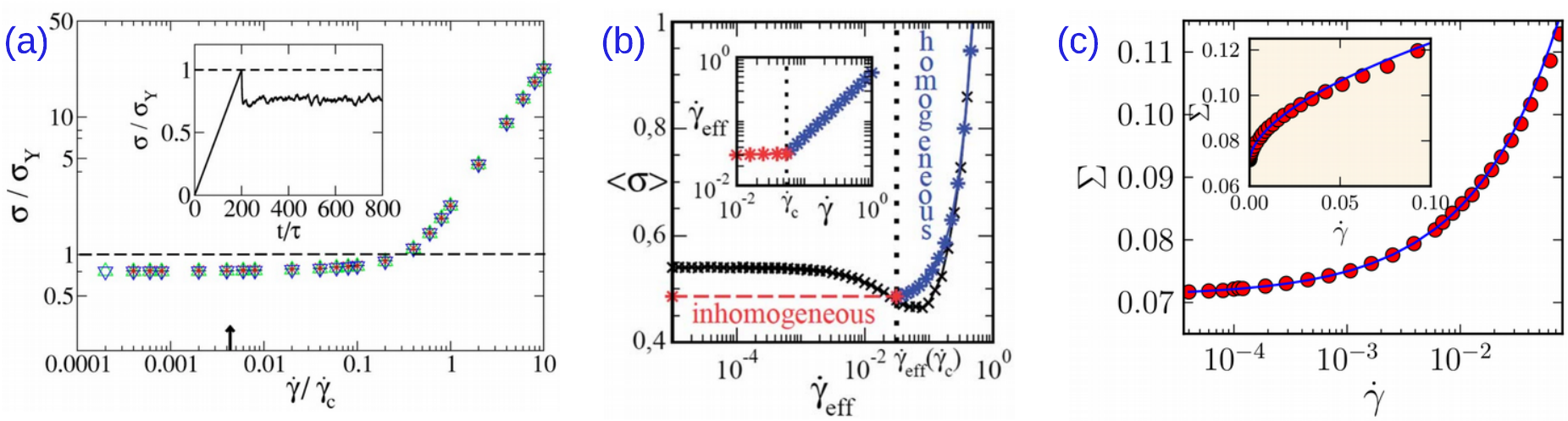}
\end{center}
\caption{ \emph{Steady-state flow curves obtained in variants of Picard's EPM}.
\textbf{(a)} Rescaled flow curves in Picard's original EPM,  for different system sizes, in \blue{logarithmic} representation.
The inset shows a typical stress-strain curve at low shear rate, starting from a stress-free configuration.
From \citep{Picard2005}.
%At large times we see the typical intermittent features in the stress dynamics.
 \textbf{(b)} Non-monotonic flow curve obtained in Picard's model with a long local restructuring time \blue{$\tau_{\mathrm{res}}$, plotted with semi-logarithmic axes.}
 \emph{Inset:} average local shear rate in the flowing regions.  For $\dot\gamma<\dot{\gamma}_c$ a mechanical instability leads to shear-banding,
 with a coexistence of a flowing band and a static region.
 From \citep{Martens2012}.
 \textbf{(c)} Flow curve obtained in a variant of Picard's model.
 The \blue{solid} line is a fit to a Herschel-Bulkley equation, with exponent $n=0.56$. \emph{Inset:} same data, in linear representation.
 From \citep{nicolas2014rheology}.}
\label{fig:flow-curves1}
\end{figure}

\subsection{Activation-based (glassy) rheology v. dissipation-based (jammed) rheology}

\blue{
The rheology of glasses was long thought to be tightly connected with that of jammed systems such as 
foams \citep{liu1998nonlinear}. But the contrast between the role played by thermal activation in the
former and the importance of dissipative processes for the latter has pointed to prominent differences.
The particle-based simulations of \citet{ikeda2012, Ikeda2013}, in particular, contributed 
to disentangling glassy (thermal) and jammed (athermal) rheologies.
% as a function of the P\'eclet number.  -- Peclet number would need to be defined AN
They identified the time scales of the relevant processes
at play, namely Brownian motion and dissipation, in addition to the driving, 
and were able to separate the thermal rheology associated with the former from the
dissipation-based one. As the shear rate $\dot\gamma$ is increased, the driving starts to interrupt
the thermally triggered plastic relaxation and dissipation starts to dominate the rheology.
The idea of a transition from a thermal to an athermal regime was bolstered by experiments on microgel colloidal suspensions,
which are impacted by thermal fluctuations close to the transition to rigidity (glass transition), but obey jamming-like
scalings further from the transition \citep{Basu2014}.
Already in the first EPM, the competition between the driving and the realization of plastic events was
emphasized. Since then, it has been implemented in different ways.
}

For flows dominated by thermal relaxation, it makes sense to consider EPM of the type of
\citet{Bulatov1994}'s and \citet{Homer2009}'s, as well as \citet{sollich1997rheology}'s SGR model
(presented in Sec.~\ref{sub:V_SGR}), in which plastic events are activated at a rate given by an Arrhenius law [Eq.~\eqref{eq:activation_rate_simple}].
For instance, in SGR, where the Arrhenius law is controlled by a fixed effective temperature $x$, 
as $\dot\gamma$ increases, blocks can accumulate more elastic strain before a plastic event is activated.
The macroscopic stress thus increases with $\dot\gamma$.
It does so in a non-universal way: The flow curve at low $\dot\gamma$ follows a Herschel-Bulkley law [Eq.~\eqref{eq:II_Herschel-Bulkley}] with exponent $1-x$ for $x<1$. 

Most other EPM \blue{dedicated to the study of} steady-state rheology consider systems close to the athermal regime,
in particular foams and dense emulsions of large droplets, which in practice undergo negligible thermal fluctuations. \blue{This will be the focus
of the rest of this chapter. These athermal systems will depart from the quasistatic conditions of Eq.~\eqref{eq:III_quasistatic}
because the driving time scale $\gamma_{y} / \dot{\gamma}$ competes with either the time scale $\tau_{\mathrm{pl}}$ of individual rearrangements,
or, at lower shear rates, the duration $\tau_{\mathrm{av}}$ of avalanches. 
}
A rough upper bound for the latter is given by
the propagation time (if applicable) across the system of linear size $L$, viz.,
$\tau_{\mathrm{av}} \sim L^z$,
where the exponent $z$ depends on the damping regime. Concretely, at vanishing shear rate, the flow
\blue{consists of well separated avalanches, some of which span the whole system. These avalanches are gradually perturbed and cut off as $\dot\gamma$ is increased, while higher shear rates add further more local corrections. The external driving thus
hampers the relaxation of the system at increasingly local scales as it gets faster.
Therefore, at least two scaling regimes could be seen as $\dot\gamma$ is varied \citep{bonn2015yield}, 
and, indeed, two regimes were observed by \citet{liu2015driving} in their EPM simulations, as shown in Fig.~\ref{fig:flow-curves2}(a).
We will discuss these regimes separately.}

\subsection{Athermal rheology in the limit of low shear rates \label{sub:VII_flow_curve_mean}}

\blue{
At vanishingly low shear rates $\dot\gamma$ the nonlinear response of athermal materials is anchored in the critical
dynamics discussed in Sec.~\ref{sec:VIII_Avalanches}. Accordingly, there have been many recent endeavors to deduce
the Herschel--Bulkley exponent $n$ (which controls the variations of $\Sigma$ when $\dot\gamma \to 0$) from other critical properties.
}

\begin{figure}[t!]
\begin{center}
\includegraphics[width=\columnwidth, clip]{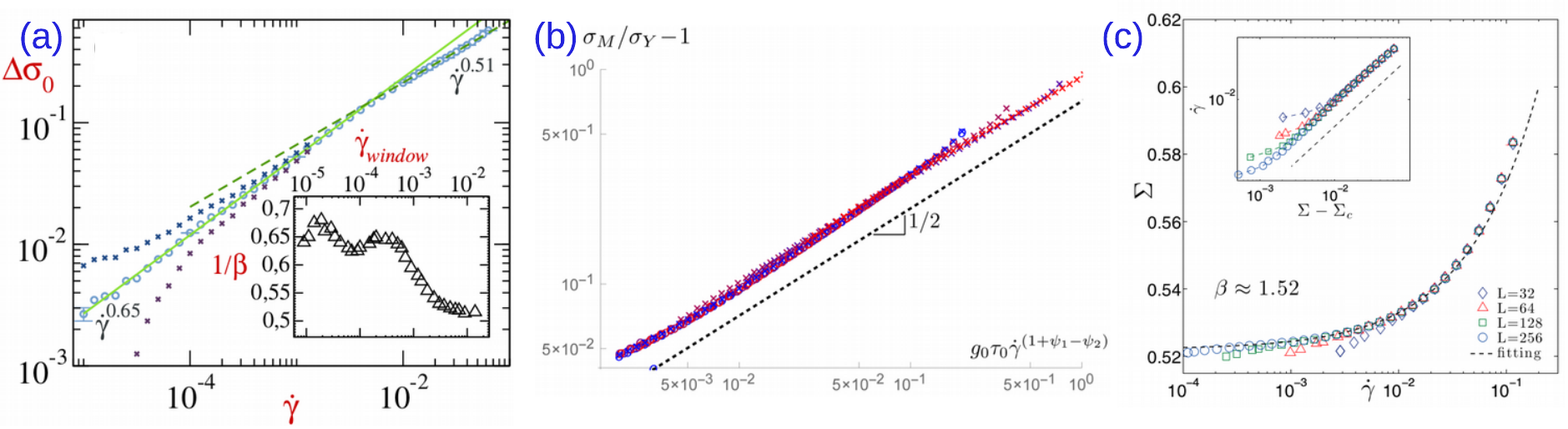}
\end{center}
\caption{ \emph{Dependence of steady-state EPM flow curves on the shear rate $\dot\gamma$.} \textbf{(a)} Difference $\Delta\sigma_0 \equiv \Sigma-\Sigma_y$ (circles) between the steady-state stress $\Sigma$ and $\Sigma_y$ as a function of $\dot\gamma$, in \citet{nicolas2014rheology}'s model. From \citep{liu2015driving}. The data suggest two different scaling regimes: Close to criticality the Herschel-Bulkley exponent is $n=0.65$, whereas at high $\dot\gamma$ $n$ tends towards $1/2$. \textbf{(b)} Flow curves for the same EPM at relatively high $\dot\gamma$ with a shear-rate-dependent local shear modulus $G_0(\dot{\gamma})\sim\dot{\gamma}^{\psi_1}$ and plastic event `duration' $\gamma_c(\dot{\gamma})\sim\dot{\gamma}^{-\psi_2}$.
\blue{These dependences introduce corrections to the Herschel-Bulkley exponent, $n=(1+\psi_1-\psi_2)/2$ (pay attention to the horizontal axis on the plot).
From \citep{agoritsas2016nontrivial}. \textbf{(c)} Flow-curves obtained from stress-imposed EPM simulations in 2D. The dashed line is a fit to a Herschel-Bulkley law with $n\approx0.66$. From \citep{lin2014scaling}.
}
}
\label{fig:flow-curves2}
\end{figure}

\blue{
To start with, the phenomenological mean-field model proposed by \citet{Hebraud1998} and discussed in Sec.~\ref{sub:V_Hebraud_Lequeux} leads to $n=\nicefrac{1}{2}$, a typical experimental
value. Recall that, in this model, the local stresses in the system drift and diffuse due to endogenous Gaussian white noise, and
yield at a finite rate $\tau_c^{-1}$ above a local threshold $\sigma_y$. The growth of the macroscopic stress with $\dot\gamma$
mirrors the associated decrease of the relative yielding rate $(\dot\gamma\tau_c)^{-1}$, which makes the boundary $\sigma=\sigma_y$ more
permeable. The result $n=\nicefrac{1}{2}$ is
robust to several variations of the yielding rules \citep{olivier2010asymptotic, olivier2011glass},
notably the inclusion of a distribution of yield stresses \citep{agoritsas2015relevance},
and tensorial generalizations of the model for multidimensional flows \citep{olivier2013generalization}.
On the other hand, it varies if a shear-rate dependence is introduced in the elastic modulus or the local restructuring time
(Fig.~\ref{fig:flow-curves2}(b) \citep{agoritsas2016nontrivial}). 
}

\blue{
Although this model can fit several aspects of athermal rheology,
the assumption of Gaussian mechanical noise fluctuations has been debated. Indeed, the
distribution of stress releases due to a single plastic event is heavy-tailed, $w_1(\delta \zeta) \sim |\delta \zeta|^{-1-k}$ with $k=1$ for the elastic propagator (see Sec.~\ref{sub:V_Heavy_tailed_fluctuations}). Accordingly,
more cautious approximations have been propounded. Assuming that the stress received by a block is a sum of
random stress increments drawn from $w_1$ in a variant of the H\'ebraud--Lequeux model, \citet{lin2017some} showed 
that the Herschel--Bulkley exponent $n$, while equal to $\nicefrac{1}{2}$ for any $k>2$, rises to $\nicefrac{1}{k}$ for 
mechanical noises characterized by $k\in [1,\,2]$. For the physically relevant value $k=1$, an exponent $n=1$
is thus found, up to logarithmic corrections. To derive these results, the authors perturbed the density of stabilities $P(x)$ around
its critical state at the yield stress ($x=|\sigma|-\sigma_y$ is the local distance to yield).  However,
the mean-field values thus obtained are larger than those measured in
the corresponding spatially-resolved EPM
in 2D ($n\simeq 0.66$) and 3D ($n\simeq 0.72$), although the discrepancy seems to shrink as the dimensionality is increased. 
}

\blue{
Besides, \citet{lin2017some} argued that the flow exponent $\beta=\nicefrac{1}{n}$ should obey the hyperscaling relation
\begin{equation}
\beta=1+ \frac{z}{d-d_f}, \label{eq:VII_beta_scaling}
\end{equation}
where the exponent $z$ relating the duration and the size of avalanches ($T \sim L^z$) is claimed to be close to 1
and the fractal dimension $d_f$ of avalanches can be expressed with Eq.~\eqref{eq:VIII_scaling_relation_tau} 
[also see \citep{lin2014scaling}]. Equation~\eqref{eq:VII_beta_scaling} yields flow exponents $\beta$ somewhat larger
than typical experimental values and
than those actually measured in EPM with instantaneous elastic propagation [Eq.~\eqref{eq:propagator_real_space}]; it was
claimed that the difference originates from a better account of the finite speed of shear waves.
}

\blue{
So far, we have seen that, within mean-field models, the dynamics of shear wave propagation and the (heavy-tailed) statistics of mechanical noise fluctuations may affect the low-shear-rate rheology, and that the finite dimension of space
introduces deviations from mean-field predictions due to correlations in the noise. Another ingredient of the models is
worth studying: the way blocks soften when they are destabilized -- or, in other words, the plastic disorder
potential $V(e_3)$ of Eq.~\eqref{eq:IV_free_energy_disorder}, where $e_3 \sim \gamma$ is the shear strain. We recall that binary EPM rely on linear elasticity
[$V(e_3)\propto e_3^2$] within the elastic regime, which is equivalent to $V$ being a concatenation of parabolas joined by cusps. For the problem of elastic line depinning (Sec.~\ref{sub:X_Depinning-transition}),
it is known that such a cuspy potential will not give the same results as a smooth potential in mean field, because destabilization
is very slow atop a smooth hill (which has a flat crest), whereas it is instantaneous at a cusp. The discrepancy vanishes in
finite dimensions, because sites are destabilized by abrupt `kicks' anyway.
}

\blue{
\citet{aguirre2018critical} suggest that the situation \emph{differs} widely for the yielding transition. 
Within a 2D continuous approach based on a plastic disorder potential [see Sec.~\ref{sub:Continuous-approaches}], they
separate the mean background of the mechanical noise from its zero-average fluctuations $\delta \sigma(t)$ and arrive at an equation of
motion which schematically reads
\begin{equation}
\eta \dot{e}_3= -V^\prime(e_3) + k\,[w(t)-e_3] + \delta \sigma(t),
\label{eq:VII_Prandtl-Tomlinson}
\end{equation}
where $\eta$ is a viscosity and $k>0$ can be interpreted as the constant of a spring connecting the current strain $e_3$ to
a driven `wall' at $w(t)$. This equation can be compared with Eq.~\eqref{eq:gen_eq_of_motion2}, but it should be noted
that the fluctuations $\delta \sigma$ are here cumulative, i.e., integrated since $t=0$.  As mentioned in Sec.~\ref{sub:V_Heavy_tailed_fluctuations}, \citet{aguirre2018critical} argue that spatial correlations (avalanches) affect the distribution of these fluctuations, $W(\delta \sigma)\sim \delta \sigma^{-1-k}$; the value
$k=1$ expected for uncorrelated (and unbounded)
stress releases should thus be substituted by $k=1.5$ if one considers objects as extended
as avalanches. Once again, we should note that this result is heavily impacted by 
the non-positiveness of the elastic propagator, which undermines purely mean-field arguments.
}

\blue{
Equation~\eqref{eq:VII_Prandtl-Tomlinson} describes the motion of a particle 
pulled by a spring on a corrugated potential; it is a stochastic Prandtl--Tomlinson equation.
This model was worked out by \citet{jagla2018prandtl}, who obtained the following scaling relation
\begin{equation}
\beta = k - \frac{1}{\alpha} + 1,
\label{eq:VII_scaling_Jagla}
\end{equation}
where $\alpha=1$ for the cuspy parabolic potentials $V$ and $\alpha=2$ if $V$ is smooth. 
We also recall the speculation 
$k= \theta +1$ from Sec.~\ref{sub:V_Heavy_tailed_fluctuations}.
In the case of parabolic potentials, the proposed scaling relation is nicely obeyed by
the values of $k$, $\theta$, and $\beta\simeq 1.51$ measured in their simulations,
as well as those found by \citet{liu2015driving} in 2D [$\beta \simeq 1.54(2)$] and 3D.
Contrary to $\theta$ or $k$, the flow exponent $\beta$
is thus found to explicitly depend on $\alpha$, i.e., the shape of the potential \citep{jagla2017non}.
The scaling relation \eqref{eq:VII_scaling_Jagla} seems to involve fewer parameters of the problem than Eq.~\eqref{eq:VII_beta_scaling}; one should nonetheless bear in mind that in depinning problems the relevant \emph{effective} potential $V$ entering
mean-field reasoning nontrivially depends on different properties of the system.
}

\blue{
Besides the choice of a specific potential shape, an alternative way to model the different destabilization speeds is to introduce
stress-dependent transition rates \citep{jagla2017elasto-plastic}. This, too, yields diverse exponents $\beta$. In this regard,
note that a dependence of the flow exponent on the specific form of the viscous dissipation was reported in
particle-based simulations \citep{roy2015rheology}.
}

\blue{
Despite the very promising recent works in this direction, no firm theoretical consensus has been reached yet regarding
the flow exponent $\beta$ and how universal it is. This exponent clearly has a value distinct from that encountered
in elastic depinning problem and, as we already discussed at length in Sec.~\ref{sec:VIII_Avalanches}, the mechanical noise fluctuations
induced by the alternate sign of the elastic propagator most probably play a prominent role in these deviations close to
criticality. As one departs from the low-shear-rate limit, among other corrections, the mechanical noise properties are expected to vary. Uncorrelated events will start to occur simultaneously,
presumably leading to a more Gaussian noise distribution (see Table~\ref{tab:fluct_handling}), and scalings different from those obtained at $\dot\gamma \to 0$ may be expected.
}

\blue{
\subsection{Athermal rheology at finite shear rates} 
}

\blue{
Rheology is concerned not only with the onset of flow at $\dot\gamma \to 0$, but with the whole range of $\dot\gamma>0$.
The regime of finite driving rates was already targeted by early EPM, including that of \citet{Picard2005} (see Sec.~\ref{sub:uniform_stress_redistrib}).
}
In this athermal model, elastoplastic blocks can only yield when their stress exceeds a \blue{uniform} local threshold $\sigma_y$.
Yielding is then a stochastic local process and so is the subsequent elastic recovery; these processes have fixed rates
$\tau^{-1}$ and $\tau_\mathrm{res}^{-1}$, respectively [see Eq.~\eqref{eq:V_Picard_rates}]. 
Figure~\ref{fig:flow-curves1}(a) shows the resulting monotonic flow curve
for $\tau=\tau_\mathrm{res}=1$. 
Its shape broadly matches that of many
experimental flow curves, but more quantitatively the simulated rheology does not follow a Herschel-Bulkley law: It crosses over to a Newtonian 
regime at stresses $\Sigma$ only slightly above the yield stress $\Sigma_y$. This is due 
to the postulated elastic stress accumulation above the threshold $\sigma_y$ for a fixed duration $\tau$ on average.

These seemingly oversimplified yielding and healing rules have been refined since then.
To make the picture more realistic, \citet{nicolas2014rheology} opted for an instantaneous triggering
of plastic events at $\sigma_y$; they also introduced a yield stress distribution.
In their model, the event lasts for a fixed local strain `duration'  $\gamma_c$.
Therefore, the local dissipation process can be disrupted by the external driving, which contributes to the local strain.
\blue{This mimicks the fact that, contrary to any problem of depinning on a fixed substrate, a deforming region in a solid
will not wade through the same potential landscape at different driving rates.}
The ensuing flow curves are more compatible with experimental ones and are well described by a Herschel--Bulkley law,
as shown in Fig.~\ref{fig:flow-curves1}(c). \citet{liu2015driving} noticed that this model actually exhibits a transition
from a low-shear-rate regime, characterized by a Herschel--Bulkley exponent $n \simeq 0.65$, to a regime with an exponent 
$n \simeq 0.5$ over the range $\Sigma \in [1.06,\,1.6]$ in units of $\Sigma_y$, beyond which further corrections to scaling set in.

It is worth noting that many experimental soft systems exhibit a qualitative change of their flow behavior when the adhesion
\blue{
properties of their constituents are modified,
with e.~g. higher propensity to shear-banding
when an emulsion is loaded with bentonite, creating attracting links between droplets \citep{ragouilliaux2007transition} 
[also see \citep{Becu2006}].
This discovery prompted the idea that there exist different classes of jammed systems depending on microscopic
interactions.
\citet{Coussot2010} suggested that adhesion results in longer local restructuring events.
\citet{Martens2012}'s EPM-based studies confirmed that long plastic events lead to a nonmonotonic constitutive curve
and the formation of permanent shear bands in the unstable parts of the flow curve,
as discussed in detail in Sec.~\ref{sub:VI_Ingredients_shear_loc} and shown in Fig.~\ref{fig:flow-curves1}(b).
}

\blue{
Thus, EPM and experiments highlight the sensitivity of the finite-shear-rate rheology to the specific microscopic interactions between particles or dynamical rules
at play. On the other hand, quite interestingly, this finite-shear-rate regime appears to be amenable to mean-field approaches oblivious to correlations in the flow.
While the latter are pivotal for avalanches, stronger driving decorrelates plastic events. The mechanical noise felt in a given region, then, 
results from the superposition of a large number of events and its distribution acquires a Gaussian shape \citep{liu2016thesis}. Concomitantly, as we noted
in Sec.~\ref{sub:V_uniform_mean_field}, mechanical noise fluctuations play a less important role. As a result, one should not be particularly surprised that
\citet{Martens2012} succeeded in reproducing the finite-shear-rate rheology of \citet{Picard2005}'s model with a mean-field approach discarding fluctuations.
In the same vein, we remark that overall flow curves are only moderately altered by finite-size effects, whether it be in particle-based \cite{roy2015rheology} or EPM computations [see Fig.~\ref{fig:flow-curves1}(a)].
}

\subsection{Strain-driven vs.~stress driven protocols} 

Most EPM works consider strain-controlled protocols (defined in Sec.~\ref{sub:III_Driving}). Some of the counterexamples are given by
\citet{lin2014scaling} [see Fig.~\ref{fig:flow-curves2}(c)] and the recent work by \citet{jagla2017non}. 
Another example of stress-imposed modeling is the numerical work in \citet{liu2016thesis}'s PhD thesis. 
In a section dedicated to the transient dynamics prior to fluidization, a stress-controlled EPM is introduced. To this end, the internal stress 
resulting from plastic events is separated from the externally applied stress field, which can be chosen arbitrarily.

In this type of protocols, depending on the initial condition, two types of stationary solutions are obtained, namely, steady flow and
a dynamically frozen state.
%(see Fig.~\ref{fig:flow-curves3}(a))
Under athermal conditions the system may always reach a configuration with large local yield stresses, in which the dynamics gets stuck, 
even if the applied stress $\Sigma$  is larger than the dynamical yield stress $\Sigma_y$. The smaller $\Sigma$ and
the smaller the system size, the more likely becomes the visiting of such an absorbing state. But if a flowing stationary state
is reached for a given time and granted that the mechanical properties do not show history dependence \citep{narayanan2017mechanical}, 
strain-controlled and stress-controlled protocols yield identical flow curves \citep{liu2016thesis}.
 %, as shown in Fig.~\ref{fig:flow-curves3}(b) and \ref{fig:flow-curves3}(c)

%\begin{figure}[t!]
%\begin{center}
%\includegraphics[width=\columnwidth, clip]{VII_flow-curves-figure3.pdf}
%\end{center} 
%\caption{ \emph{Stress-controlled simulations using \citet{liu2016thesis}'s EPM variant or a mean-field approach.} 
%(a) Time evolution of the plastic strain rate 
%response to an applied stress step $\Sigma$. From \citep{liu2016thesis}. 
%For small $\Sigma$, the system gets stuck in a stable configuration and $\dot\gamma^\mathrm{pl}\rightarrow 0$.
%For large $\Sigma$, $\dot\gamma^\mathrm{pl}\rightarrow 0$ can reach a finite limit in the long run. 
%(b) Comparison of flow curves obtained in strain-controlled and stress-controlled protocols, showing a perfect overlap. From \citep{liu2016thesis}. 
%(c) Stress-controlled flow curves obtained in a mean-field framework inspired by the H\'ebraud-Lequeux model, for different
%values of the coupling parameter $\alpha$ defined in Sec.~\ref{sub:V_Hebraud_Lequeux}. Points are semi-analytical results; dashed lines
% are analytical calculations for a strain-controlled scenario. From \citep{liu2017mean}.}
%\label{fig:flow-curves3}
%\end{figure}

\section{Relaxation, Aging and Creep phenomena\label{sec:Relaxation} }

So far EPM have mostly been exploited to investigate the macroscopic
flow behavior and flow profiles (Sec.~\ref{sec:VI_Macroscopic_Shear_Deformation}),
characterize stationary flow (Sec.~\ref{sec:VII_Bulk_rheology}),
or study fluctuations and correlations in the steady flow close to
criticality, where one finds scale-free avalanches
(Sec.~\ref{sec:VIII_Avalanches}). Still, some works, however
few, are concerned with relaxation, aging, and creep phenomena. This
section is dedicated to both the dynamics in the temperature
assisted relaxation (aging) of disordered systems and to the transient
dynamics under loading (creep), prior to yielding or complete arrest.
The latter phenomenon can be either an athermal process, provided
that the stress load is above, but close to, the yielding point, or
thermally assisted creep, in response to a load below the dynamical
yield stress.

\subsection{Relaxation and aging\label{sub:IX_relaxation}}

A striking feature in the theory of viscous (glassforming) liquids
is their response to an external perturbation, close to the glass transition:
They do not exhibit an exponential structural relaxation, with a simple
time scale, but a stretched exponential relaxation. More
specifically this means that the temporal behavior of the response
function $R(t)$ (e.g., the response in stress $\Sigma(t)$ to the
application of a strain step at time $t=0$) can often be described
by the so called Kohlrausch-Williams-Watt (KWW) function 
\begin{equation}
R(t)\propto\exp\left[-\left(\frac{t}{\tau}\right)^{b}\right]\text{ with }R(t)\equiv\frac{\Sigma(t)-\Sigma(\infty)}{\Sigma(0)-\Sigma(\infty)}.
\label{eq:IX_response_function}
\end{equation}
In this expression, $b$ typically takes a value between 0 and 1,
which stretches the exponential relaxation. This was ascribed to the
formation of dynamical heterogeneities close to the glass transition,
thus producing separately relaxing domains and leading to a broad
distribution of relaxation times \citep{macedo1967effects}, hence
a stretched exponential relaxation \citep{campbell1988nonexponential,bouchaud2008anomalous}.

With this picture in mind, it came as a surprise \blue{that} a series of
dynamical light-scattering measurements on colloidal gels showed 
the opposite behavior, namely, a compressed exponential structural
relaxation, characterized by an exponent $b>1$ \citep{cipelletti2000universal,cipelletti2003universal,ramos2001ultraslow}.
More recent experiments using X-ray photon correlation spectroscopy
have found that this feature is not specific to gels \citep{orsi2012heterogeneous},
but also arises in supercooled liquids \citep{caronna2008dynamics},
colloidal suspensions \citep{angelini2013dichotomic} and even in
hard amorphous materials like metallic glasses \citep{Ruta2012,ruta2013compressed}.
Although this anomalous relaxation was observed ubiquitously
in experimental systems, it took more than a decade to reproduce dynamics
with compressed exponential decay in molecular-scale
simulations, until \citet{bouzid2016elastically} and \citet{chaudhuri2016ultra}
eventually reported such dynamics in microscopic models for gels.
The main obstacle had been to probe the right parameter range, notably
with respect to temperature and also length scales.

From the outset, \citet{cipelletti2000universal} suggested that the
faster than exponential relaxation stems from the elastic deformation
fields generated by local relaxation events. Shortly afterwards \citet{bouchaud2001anomalous}
put forward a mean-field model based on the assumption of elasticity 
to explain this anomalous relaxation. In case this explanation is correct,
EPM should be the ideal tool to test it \citep{Ferrero2014}. In a quiescent
system, the driving term vanishes in Eq.~\eqref{eq:gen_eq_of_motion},
which turns into

\[
\dot{\sigma}_{i}(t)=\sum_{j}2\mu\mathcal{G}_{ij}\dot{\epsilon}_{j}^{\mathrm{pl}}(t)\,,
\]
where $\epsilon_{j}^{\mathrm{pl}}(t)$ \blue{denotes the local
plastic deformation at site $j$} and the other notations were defined below Eq.~\eqref{eq:gen_eq_of_motion}.
As before, this equation describes the response of the surrounding
medium to local relaxation events. Here, only thermally activated
processes are relevant, and their modeling is inspired by the the
trap model of \citet{denny2003trap} and \citet{sollich1997rheology}'s
Soft Glassy Rheology model [SGR, see Eq.~\eqref{eq:SGR_t_esc}], with an
Arrhenius-like yielding rate for sites below the threshold, viz.,
\begin{equation}
p_{\pm}\sim\exp\left[-\frac{\sigma_{y}^{2}\mp\mathrm{sgn}(\sigma)\sigma^{2}}{2\kappa T}\right],\label{eq:activation_probabilities}
\end{equation}
while sites with $|\sigma|>\sigma_y$ yield instantaneously.
In Eq.~\eqref{eq:activation_probabilities}, the signs correspond to
the direction of the yielding event, $\sigma_{y}$ is a local yield
stress, $\kappa$ is a dimensional prefactor, and $T$ the  ambient
temperature.

\begin{figure}
\begin{centering}
\includegraphics[width=1\textwidth]{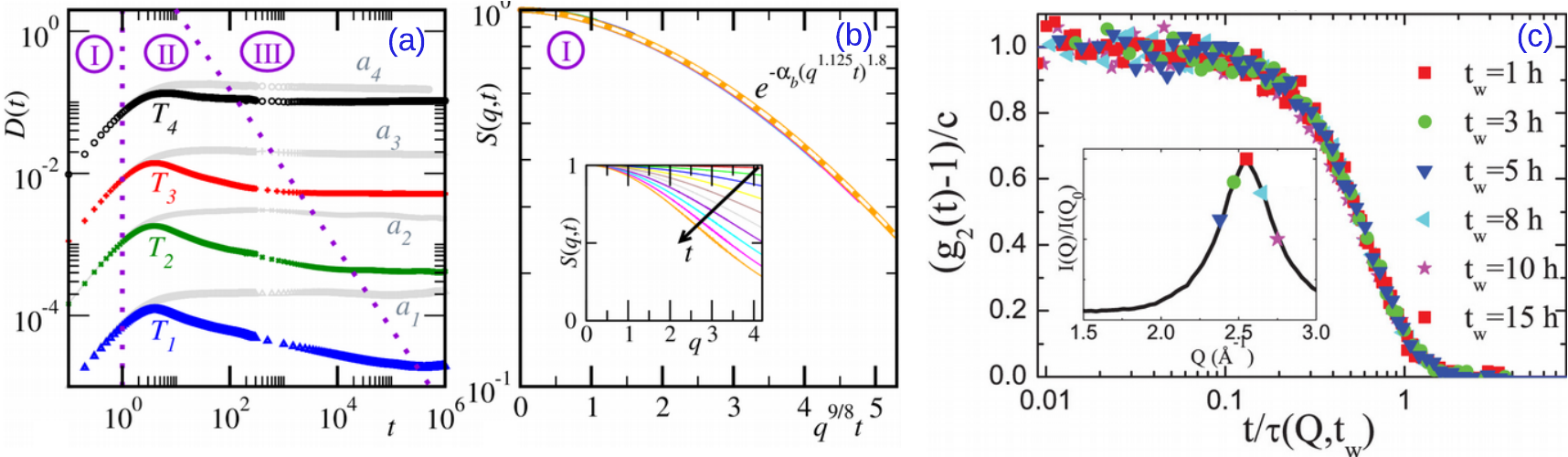}
\end{centering}
\caption{\label{fig:figure-compressed-expo}
% \emph{Motion of particle tracers and
% structural relaxation in a quiescent system obtained in an EPM.} (a)
% Diffusivity $D(t)$ (i.e., mean-square displacement divided by time)
% of tracer particles (\emph{see main text}) for four different temperatures,
% increasing from bottom (blue) to top (black). (b) Rescaled self-part
% of the intermediate scattering function $S(q,t)$ for $t$
% in the first ascending regime of $D(t)$. The motion is close
% to ballistic (linear in time), with $\tau\approx q^{-1}$ in Eq.~\eqref{eq:IX_response_function},
% and the form factor $\beta\simeq2$ (defined in the same equation)
% implies a compressed exponential relaxation. Panels (c) and (d) display
% the same quantities in dynamical regimes II (subdiffusive) and III (diffusive). All insets show
% the data from the main graph without rescaling. From \citep{Ferrero2014}.}
\blue{
\emph{Structural relaxation in quiescent systems.} \textbf{(a)}
Diffusivity $D(t)$ (\emph{i.e.}, mean-square displacement divided by time)
of tracer particles measured in an EPM, at four  temperatures,
increasing from bottom (blue) to top (black).
\textbf{(b)} Rescaled self-part of the intermediate scattering function $S(q,t)$ for $t$
in the first ascending regime of $D(t)$ in panel (a).
The motion is close to ballistic (linear in time), with $\tau\approx q^{-1}$ in Eq.~\eqref{eq:IX_response_function},
and the form factor $\beta\simeq2$, defined in the same equation, implies a compressed
exponential relaxation. From \citep{Ferrero2014}.
\textbf{(c)} Relaxation of thin Zr$_{67}$Ni$_{33}$ metallic glass ribbons with time,
measured by the the decay of the X-ray photon intensity autocorrelation $g_2$ at $T=373K$,
for different waiting times $t_w$ and wavectors $Q$ (shown in the inset).
The characteristic relaxation time $\tau(Q,t_w)$ was determined by fitting $R=g_2 - 1$ to the KWW form of 
Eq.~\eqref{eq:IX_response_function},
which yielded a shape parameter $b\simeq 1.8 \pm 0.08$.
From \citep{ruta2013compressed}.
}
}
\end{figure}

Such models confirm the dependence of the shape parameter $b$
of structural relaxation on the dimensionality of the system, which
\citet{bouchaud2001anomalous}'s mean-field arguments predict to be
$b=\frac{3}{2}$ in 3D and $b=2$ in 2D. Moreover, in EPM insight
into the microscopic dynamics can be gained by following the motion
of tracers advected by the elastic displacement field, as explained in
Sec.~\ref{sub:V_tracers}.
This led \citet{Ferrero2014} to distinguish three dynamical regimes
in 2D, namely (I) ballistic, (II) subdiffusive and (III) diffusive.
In the ballistic regime (see Fig.~\ref{fig:figure-compressed-expo}),
compressed relaxation was found, with a shape parameter $b\approx2$.
The subdiffusive regime was ascribed to correlations in the relaxation
dynamics, a feature that has not been reported in experiments. This
disagreement can either be due to oversimplifications of the model
or to the fact that experiments are usually performed in 3D, and not
2D. Preliminary EPM studies in 3D observed ballistic
motion at short times, with a compressed exponent $b=\frac{3}{2}$,
followed by a diffusive regime \footnote{{Unpublished data of \citet{Ferrero2014}.}}.

There remain many other open questions that could be addressed by
EPM. For instance the $q$-dependence of the experimental intermediate
scattering functions $S(q,t)$ \citep{cipelletti2003universal} cannot be captured
in EPM at present, but could be included by implementing hybrid models
that consider smaller-scale dynamics as well. Besides, the self and
the intermediate part of $S(q,t)$ cannot be distinguished in EPM
yet, because the tracers do not interact, but \blue{the two} may 
differ in reality. Other questions include the 3D dynamics and the
possibility of intermittency in time as well as spatial correlations
of the localized relaxation events.

\subsection{Creep}

Another field that has stimulated much experimental work in the last
years \citep{bonn2015yield} but few rationalization attempts at the mesoscale
is creep. The definition of creep is somewhat ambiguous.
In some contexts it may refer to stationary motion at a vanishingly
small velocity, in particular the creep dynamics of a driven elastic
manifold over a disordered landscape at finite temperature \citep{ferrero2017spatiotemporal},
but also the flow of a granular medium subjected to a constant stress
$\Sigma\ll\Sigma_{y}$ supplemented with an additional small cyclic
stress modulation \citep{pons2016spatial}. But here we will restrict
our attention to the traditional definition in material science, namely,
the slowdown of deformation prior to failure, fluidization or complete
arrest, under load $\Sigma$. This load is usually comparable
to, or smaller, than the material yield stress $\Sigma_{y}$ and creep
can in principle be both of thermal and athermal nature.

For $\Sigma>\Sigma_{y}$ the usual response of most dense soft glassy materials can
be separated into three regimes \blue{(in polymeric systems, five regimes are listed by \citet{medvedev2015stochastic}).}
Primary creep corresponds to a first
slowdown of the dynamics, with a gradual decrease of the (initially
high) strain rate $\dot{\gamma}$. The deformation rate is roughly constant in the
secondary creep regime but abruptly shoots up in the tertiary regime,
which ultimately culminates in macroscopic failure or fluidization.
The measured macroscopic quantities are usually the time-dependent $\dot{\gamma}(t)$ and the fluidization
or failure time $\tau_{f}$ \citep{skrzeszewska2010fracture,Divoux2011}.

Creep is observed in many experimental systems, from crystalline and
amorphous solids to soft materials. In the former materials, a power-law
slowing down of the deformation rate with an exponent close to or slightly less than $\nicefrac{2}{3}$
is often reported \citep{miguel2002dislocation}, viz.,
\[
\dot{\gamma}(t)\sim t^{\nicefrac{-2}{3}}\text{ or, equivalenty, }\gamma(t)\sim t^{\nicefrac{1}{3}}.
\]
This law is commonly called Andrade creep and hints at a possible
universality of the dynamics. However, experiments and simulations
on creep in amorphous systems have found a variety of power-law exponents
for the decay of $\dot{\gamma}(t)$ in primary creep, ranging between
\blue{$\nicefrac{-1}{3}$} \citep{bauer2006collective} and \blue{$-1.0$} (the latter
value corresponding to logarithmic creep $\gamma(t)\sim\ln(t)$),
with a multitude of values in-between \citep{Divoux2011a,Leocmach2014,ballesta2016creep,Chaudhuri2013,sentjabrskaja2015creep,landrum2016delayed}.
\citet{bonn2015yield} extensively reviewed the literature on
the topic. Scaling results for the fluidization (or failure) time
$\tau_{f}$ also vary
and basically fall in two classes. Among other works, \citet{Divoux2011a}
found a power-law scaling of $\tau_{f}$, defined as the time to reach
a homogeneous stationary flow, viz., $\tau_{f}\sim(\Sigma_{y}-\Sigma)^{-\beta}$
, where $\beta$ varies between 4 and 6. On the other hand, other
works defined $\tau_{f}$ as the duration of the rapid increase of
$\dot{\gamma}(t)$ at the end of secondary creep and reported an inverse
exponential dependence $\tau_{f}\sim\exp\left(\frac{\Sigma_{0}}{\Sigma}\right)$,
where a characteristic stress scale $\Sigma_{0}$ has been introduced
\citep{gopalakrishnan2007delayed,gibaud2010heterogeneous,Lindstrom2012}.

Thus, rather than a universal behavior, experiments suggest a multitude
of dependencies, notably on the preparation protocol prior to the
application of the step stress (quench or pre-shear), on temperature,
age and also on the dominant physical process at play during creep.
In some systems the initial creep regime appears to be completely
reversible and one expects the creep to be a result of visco-elasticity.
Accordingly, \citet{jaishankar2013power} were able to reproduce the
experimental power-law creep in Acacia gum solutions using a modified
Maxwell model featuring fractional time derivatives. On the other
hand, on the basis of molecular dynamics simulations, \citet{shrivastav2016yielding}
claim that the power-law creep in a variety of glassy systems can
be related to a percolation dynamics of mobile regions, thus plasticity,
which would render EPM particularly suitable to tackle the open questions
in the field. Among the `hot topic' highlighted by \citet{bonn2015yield},
the detection of precursors that may point to incipient failure stands as
the Atlantis in many disciplines from material science
to engineering and geology. 

\begin{figure}
\begin{centering}
\includegraphics[width=1\textwidth]{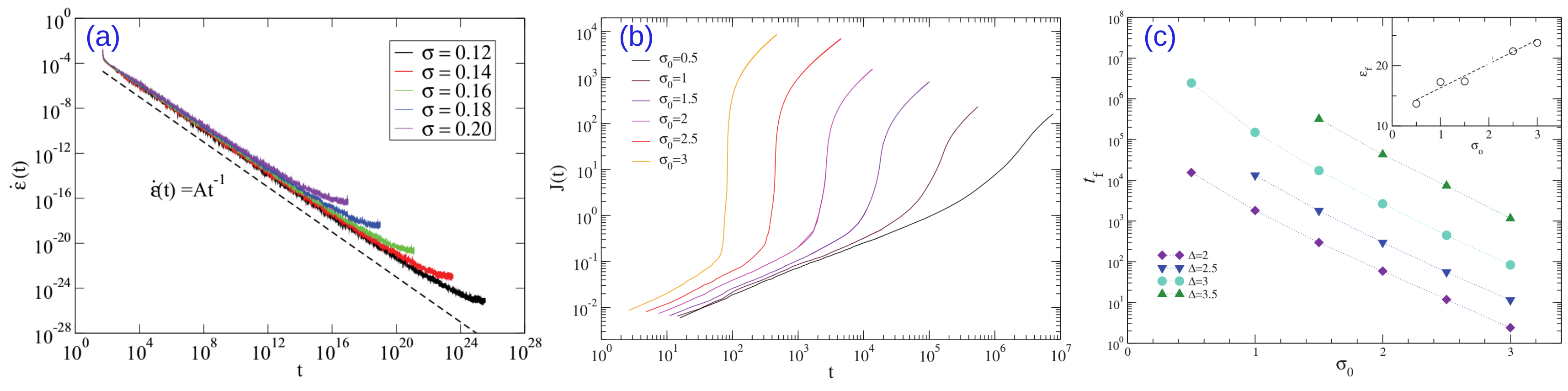}
\par\end{centering}

\caption{\label{fig:figure-meso-creep}\emph{EPM characterization of creep.} (a) Strain
rate $\dot{\epsilon}$ as a function of time $t$ for different applied
stresses $\Sigma$ in the EPM of \citet{bouttes2013creep}. (b) Non-linear
compliance $J\equiv\frac{\epsilon}{\Sigma_{0}}$ as a function of
time for different applied stresses $\Sigma_{0}$, 
obtained with \citet{merabia2016creep}'s mesoscopic model.
(c) Dependence of the fluidization time $t_{f}$ on $\Sigma_{0}$.
Panels (b) and (c) are extracted from \citep{merabia2016creep}.
}
\end{figure}

Using a lattice-based EPM,
\citet{bouttes2013creep} made a first endeavor to address thermal
creep and showed its strong dependence on initial conditions and the
impact of aging on the creep behavior. In the model, each site is assigned an energy barrier
 $E_{0}$ (renewed after every plastic event) in the stress-free configuration, 
 with a uniform distribution of $E_{0}$. The elastic
stress redistributed by plastic events via the usual elastic
 propagator [Eq.~\eqref{eq:elastic_propagator_Fourier}] biases this potential. The
plastic activation probabilities are analogous to Eq.~\eqref{eq:activation_probabilities},
with an Arrhenius-like law, and are resolved with a kinetic Monte-Carlo
algorithm. The resulting creep dynamics $\dot{\gamma}(t)$, studied
in pure shear, depend on the applied stress $\Sigma$ and temperature
$T$ and all display an apparent exponent suggestive of logarithmic
creep (see Fig.~\ref{fig:figure-meso-creep}a). Besides, the fluidization
time $\tau_{f}$ is found to decrease with increasing $\Sigma$ and $T$.

\blue{
\citet{merabia2016creep} explored the transient thermal creep that occurs upon application of
a stress step, prior to steady flow, at relatively high temperatures.
}
Within an EPM,
they also resorted to a kinetic Monte-Carlo scheme and Arrhenius-type plastic
activation rates, but they used a non-uniform distribution of intrinsic
trap depths $\rho(E_{0})$. With an exponential distribution $\rho(E_0) \sim \exp[-\alpha E_0]$ (leaving aside
a lower cutoff), the model is
formally similar to the SGR model (see Sec.~\ref{sub:V_SGR}), 
but here the temperature parameter is interpreted as the room temperature,
instead of an effective noise temperature, and samples are assumed to
be thermally equilibrated before stress is applied ($\alpha k_B T>1$). 
%Furthermore, the energy barriers assigned to site $i$ are
%drawn from a steady distribution (no aging) of the form $\rho(E_{i})\exp(\frac{E_{i}}{k_{b}T})$.
Contrary to \citet{bouttes2013creep}, the simulated creep does not
always slow down logarithmically. Instead, a power-law decay 
$\gamma(t) \sim t^{\alpha-1}$ is observed, for $1<\alpha<2$, in agreement with a mean-field analysis; 
it tends to logarithmic creep as $\alpha \rightarrow 1$. \citet{merabia2016creep} also considered a Gaussian distribution
$\rho(E_{0})$. In that case, the steady-state flow curve grows logarithmically, $\Sigma\sim\ln(\dot{\gamma})$. 
Regarding the creep regime before steady state, the cumulative strain contains a term that grows linearly in time
and the fluidization time $\tau_{f}$ follows the inverse exponential
dependence on $\Sigma$ [i.e., $\tau_{f}\sim\exp\left(\frac{\Sigma_{0}}{\Sigma}\right)$, see
Fig.~\ref{fig:figure-meso-creep}c)] found in experiments on carbopol black
gels by \citet{gibaud2010heterogeneous}.
The latter result is robust to variations of the Gaussian half-peak width. 

The authors also tried different stress propagators of short range
character, besides the quadrupolar (Eshelby-like) one. It turns out
that their mean-field predictions agree best with the simulations
with a short-range propagator and an exponential distribution of energy
barriers, whereas there is a systematic offset in the creep exponent
with respect to the more realistic quadrupolar propagator. This is somewhat
counter-intuitive because increasing the interaction range usually leads to
a more mean-field-like behavior.

\blue{An alternative mean-field approach is based on the
H{\'e}braud-Lequeux model (see Sec.~\ref{sub:V_Hebraud_Lequeux}).
Its initial purpose was not to describe aging, but \citet{sollich2017aging}
have shown that the modeled systems that age under zero stress rapidly
freeze into a preparation-dependent state; the initial stress does not fully relax.
}
Within the same framework, \citet{liu2017mean} studied athermal
creep under a load $\Sigma\equiv \langle \sigma \rangle > \Sigma_{y}$ and they, too,
reported a strong dependence on the preparation. The initial distribution of stresses
$\mathcal{P}(\sigma,t=0)$ was taken as a proxy for the sample age insofar as,
in real systems, aging results in stress relaxation and thus a narrower
distribution $\mathcal{P}(\sigma,t=0)$. For $\Sigma$ slightly above
 the yield stress and long aging, there is first a power-law decay $\dot{\gamma}(t)\sim t^{-\mu}$
($\mu>0$) to a minimal value and then an acceleration up to the steady-state value. This
evolution is consistent with several experimental measurements in
bentonite suspensions and colloidal hard-sphere systems. But, contrary
to expectations, the model exhibits a parameter-dependent (thus, non-universal)
power-law exponent $\mu$. Within the model, the first creep regime
is dominated by the plastic activation of sites that have not yielded
yet, which become rarer and rarer, until the memory of the initial
configuration is lost and steady-state fluidization is achieved. This
occurs at a fluidization time $\tau_{f}$ that decreases as $\Sigma$ increases,
 but in a non-universal way.

In conclusion, these few seminal papers proposing a mesoscopic approach
to creep leave room for further exploration with EPM, for instance
about the universality (or not) of the long-time response in thermal
and athermal systems. It would also be interesting
to determine if precursors can be defined to predict failure and,
once the validity of EPM is established, to upscale the mesoscopic
approach into a valid macroscopic description of the creep response.

\section{Related topics\label{sec:Related_topics}}

Amorphous solids seem to form a specific class of materials. However, the phenomenology 
exposed \blue{in the previous chapters} suggests underlying theoretical connections with other problems. And, indeed,
EPM are related to a spectrum of other models, \blue{notwithstanding physical differences,} in particular
in the interaction kernels. This section  reviews, and attempts to compare to EPM, some of these related
approaches, from mesoscale models for crystalline plasticity and elastic line depinning to fiber bundles, fuse networks
and random spring models.
\blue{The ample connections with seismology, hinted at in Sec.~\ref{sec:Elastic_couplings}, and
tribology \citep{persson1999sliding,lastakowski2015granular,jagla2018prandtl} -- the latter
being plausibly mediated by fracture mechanics \citep{svetlizky2014classical,svetlizky2017brittle} -- will not be discussed here.
}

\subsection{Mesoscale models of crystalline plasticity\label{sub:Mesoscale-models-of}}

\subsubsection{Crystal plasticity}

Like amorphous solids, driven crystalline materials respond elastically
to infinitesimal deformations, via an affine deformation of their
structure, but undergo plastic deformation under higher loading. To
be energetically favorable, plastic deformation increments must somehow
preserve the regular stacking of atoms. The question is whether it 
saves energy to jump to the closest regular structure ('switch
neighbors'), rather than to keep on with the affine deformation of the current
structure. For a perfect crystal, such a criterion would predict an
elastic limit of around 5\%. 

Real crystals actually have a much lower elastic limit because they
harbor structural defects, which were created at the stage of their
preparation and which play a key role in the deformation. These defects
in the regular ordering take the form of dislocations and grain boundaries
separating incompatible crystalline domains. Dislocations are
line defects obtained by making a half-plane cut in a perfect crystal
and mismatching the cut surfaces before stitching them back together.
\blue{Dislocations are similar to creases on a carpet in that they} can
glide across the crystal (and occasionally ``climb'' when they encounter
a defect), thereby generating slip planes, in the same way as creases can be pushed across the rug
to move it gradually without having to lift it as a whole. Grain boundaries also promote
deformation; in these regions, gliding is facilitated by the mismatch-induced
weakness of the local bonds. On the other hand, the presence of impurities,
e.g., solute atoms in the crystal, may pin a dislocation at some location
in space until it is eventually freed by a moving dislocation, which
results in a dent in the stress \emph{vs. }strain curve; this is the
so called Portevin-Le Chatelier effect.

The stress field around a dislocation is well known (it decays inversely
proportionally to the distance to the line) and the attractive or
repulsive interactions between dislocations can also be rigorously
computed. As a matter of fact, the elastic propagator used in EPM
can be regarded as the stress field induced by four edge dislocations
whose Burgers vectors sum to zero \citep{benzion1993earthquake,IspanovityPRL2014,tuzes2016disorder}.
However, owing to the vast lengthscales separating the individual
dislocation from the macroscopic material, it is beneficial to coarse-grain
the description to the mesoscale, by considering the dislocation density
field.

\subsubsection{Models and results}

Mesoscale dislocation models, which exist in several variants (Field
Dislocation Model, Continuum Dislocation Dynamics), bear formal similarities
with EPM.

Noticing that the plastic deformation induced by crystallographic
slip generates an elastic stress field $\tau_{int}\left(\boldsymbol{r}\right)$
(via the very same elastic propagator as in EPM), \citet{zaiser2005fluctuation}
separated this internal stress $\tau_{int}\left(\boldsymbol{r}\right)$
from the aspects more specific to dislocations and crystals and arrived
at the following equation in 2D:
\begin{equation}
\frac{1}{B}\partial_{t}\gamma\left(\boldsymbol{r}\right)=\tau_{ext}+\tau_{int}\left(\boldsymbol{r}\right)+\frac{DG}{\rho}\partial_{x}^{2}\gamma+\delta\tau\left(\boldsymbol{r},\gamma\right),\label{eq:X_crystal_meso_eq_Zaiser}
\end{equation}
where $B$, $D$, and $G$ are material constants, $\tau_{ext}$ is
the externally applied stress, and $\rho$ is the dislocation density.
The last two terms on the rhs have no strict counterparts in EPM;
they account for the mechanisms generated by interactions between
dislocations that alter the stress required to set a dislocation in
motion. The third term is a homogenizing term while the fourth one
is a ($\rho$-dependent) fluctuating term; its dependence on the plastic
strain $\gamma$ may be used to effectively describe strain hardening
effects due to the multiplication of dislocations. In EPM, such effects
would belong to the rules that govern the onset of a plastic event.
Armed with this model, the authors then studied the slip avalanches
in order to explain the experimentally observed deformation patterns
consisting of slip lines and bands, echoing the endeavors in this
direction on the EPM side. They found scaling exponents for such avalanches
that are comparable, but not strictly equal, to the mean-field exponents
for the depinning problem; this difference is not unexpected, owing
to the fluctuating sign of their elastic propagator, which is identical
to the EPM one (see Sec.~\ref{sec:VIII_Avalanches}). Also, large
avalanches are cut off due to strain hardening, which is one possible
explanation for the macroscopic smoothness of the deformation.

Contrasting with this macroscopically smooth situation, the deformation
dynamics may feature strong intermittency, which points to collective
effects. Power-law-distributed fluctuations have recently been evidenced
in the acoustic emissions as well as in the stress \emph{vs.} strain
curves of loaded crystals \citep{weiss2015mild,zhang2017taming}.
These fluctuations may be ``mild'', with bursts superimposed on
a relatively constant, seemingly uncorrelated fluctuation background,
which is the case for many bulk samples, especially those with an
\emph{fcc} (face-centered cubic) structure. On the other hand, intermittency becomes dominant
in \emph{hcp} (hexagonal close-packed) crystals and in smaller samples, where large bursts
dominate the statistics. Samples with fewer defects also tend to have
``wilder'' fluctuations. A mean-field rationalization of these phenomena
considers the density $\rho_{m}$ of mobile dislocations and expresses
its evolution with the strain $\gamma$ as
\[
\frac{d\rho_{m}}{d\gamma}=A-C\rho_{m}+\sqrt{2D}\rho_{m}\xi\left(\gamma\right),
\]
where $A$ is a nucleation rate, $C$ is the rate of annihilation
of dislocation pairs, and $D$ controls the intensity of the white
noise $\xi$ \citep{weiss2015mild}. Notice that the latter is multiplied
by $\rho$, owing to the long-ranged interactions between dislocations;
the presence of multiplicative mechanical noise makes collective cascade
effects possible. Such a model allowed the authors to capture the
distinct types of fluctuations in the dynamics, from mild to wild,
depending on the noise intensity $D$. More recently, \citet{valdenaire2016density}
rigorously coarse-grained a fully discrete 2D dislocation picture
into a continuum model centered on a kinetic equation for the dislocation
density, with superficial similarities with the EPM equation of motion,
Eq.~\eqref{eq:gen_eq_of_motion}.

\subsubsection{Relation to EPM}

Although the microscopic defects and the microscopic deformation mechanisms
differ between crystals and disordered solids, the macroscopic phenomenology
and, to some extent, the mesoscopic one share many similarities: Microscopic
defects interact via long-range interactions and their activity is,
in some conditions, controlled by temperature. Globally, the dynamics
are highly intermittent at low shear rates and involve scale-invariant
avalanches, as indicated, \emph{inter alia}, by acoustic emission
measurements on stressed ice crystals \citep{miguel2001intermittent}.
This intermittency is generically known as crackling noise \citep{Sethna2001}
and does not connect EPM only to crystal plasticity, but also to the
fields of seismology and tribology.

The phenomenological similarity is paralleled by a proximity in the
models. In some EPM, the stress redistributed by a shear transformation
is actually described as the effect of a combination of dislocations
\citep{benzion1993earthquake,IspanovityPRL2014,tuzes2016disorder}.
Conversely, quadrupolar interactions may be directly implemented in
mesoscale models of crystal plasticity, for instance in Eq.~(2) of
\citep{Papanikolaou2012}. More generally, the basic equations of
evolution in the two fields look very much alike, and models sometimes
seem to have bearing on both classes of materials \citep{Shiba2010}.
\citet{Rottler2014} numerically investigated the transition between
the dislocation-mediated plasticity of crystals and the shear-transformation-based
deformation of amorphous solids. They found that the directions of
the \emph{nonlinear} displacements under strain could be well predicted
from the low-frequency \emph{vibrational modes} and that polycrystals
already behave comparably to glasses, despite their regular structure
at the grain scale. 

Nevertheless, the connection between crystals and disordered solids
should not be overstated. Even though flow defects (``soft spots'')
in the latter might to some extent persist over rearrangements \citep{schoenholz2014understanding},
on no account can they be assimilated to well identified structural
defects moving through a crystal. Following from this discrepancy
are the facts that, contrary to plastic rearrangements, dislocations
are strongly dependent on the preparation of the material (which determines
the dislocation density), and may be pinned by defects,  annihilate
through the merger of partials (``opposite'' defects) or multiply.

\subsection{Depinning transition\label{sub:X_Depinning-transition}}

\subsubsection{The classical depinning problem}

In several systems, an interface is driven through a disordered medium by a uniform
external force.
This interface can be a magnetic or ferroelectric domain wall, the water front (contact line)
in a wetting problem, the fracture front, or even charge density waves and
arrays of vortices in superconductors. In all these cases, the interplay between the
quenched disorder (e.g., due to impurities) and the elastic interactions along the interface
is at the root of a common phenomenology and a universal dynamical response.

If the external force is weak, the interface will advance and soon get pinned and unable to advance 
any further. If the force is strong enough, instead, the interface will overcome
even the largest pinning centers, reaching a steady state of constant
velocity. This is the well documented dynamical phase transition known as depinning.
Beyond the transition itself, the literature now also describes the equilibrium configuration
of the elastic line, several variations of the problem (short/long-range
elasticity, different disorder types, etc.), thermally activated dynamical
regimes and, in general, tackles the transport problem and its relation with the geometry of the interface.
The interested reader is referred to one of the following self-contained works or reviews:
\citep{FisherPR1998,ChauvePRB2000,KoltonPRB2009,AgoritsasPB2012,FerreroCRP2013}.

\subsubsection{Models}

The most celebrated model to describe the depinning problem is the
quenched Edwards-Wilkinson (QEW) equation. A $d$=1-dimensional interface without overhangs is driven by an 
external pulling force $f$.   In the overdamped
limit, its local displacement at time $t$, described by a single-valued function, $h(x,t)$,
obeys 
\begin{equation}
\eta\partial_{t}h(x,t)=c\nabla^{2}h(x,t)+f+F_{p}(x,h)+\xi(x,t)\label{X_depinning_short_range}
\end{equation}
where $c\nabla^{2}h(x,t)$ represents the elastic force due to the
surface tension, the (quenched) disorder
induced by impurities is encoded in the pinning force $F_{p}(x,h)$
and thermal fluctuations are included as a Langevin thermal noise
$\xi(x,t)$. In general, two different kinds of disorder are considered:
 \emph{random bond} disorder, in which the pinning potential
is short-range correlated in the direction of motion ($<V(h,i)V(h',j)>=\delta_{ij}\delta_{hh'}$),
and \emph{random-field} disorder, where the pinning \emph{force} is short-range
correlated (thus generating correlations of the potential in the direction
of motion, $<V(h,i)V(h',j)>=\delta_{ij}\min(h,h')$ ). 

Of course, the QEW model just mentioned is minimal. Some of its variants
take into account additional ingredients. For example,
charge density waves and vortices involve a periodic elastic structure,
in fracture and wetting the elastic interactions are long-ranged,
and anharmonic corrections to elasticity or anisotropies could
also be relevant. These features would call for a rewriting of Eq.~\eqref{X_depinning_short_range} into a more
general form involving an elastic interaction energy $\mathcal{H}^\mathrm{el}$
\begin{equation}
\eta\partial_{t}h(x,t)= \frac{-\delta\mathcal{H}^\mathrm{el}[h]} {\delta h(x)}
+f+F_{p}(x,h)+\xi(x,t)\label{eq:X_depinning_general}
\end{equation}
 Remarkably, all these different problems, grouped
in a few distinct universality classes, share the same basic
physics, discussed in the following.

\subsubsection{Phenomenology}

\begin{figure}
\begin{centering}
\includegraphics[width=1\columnwidth]{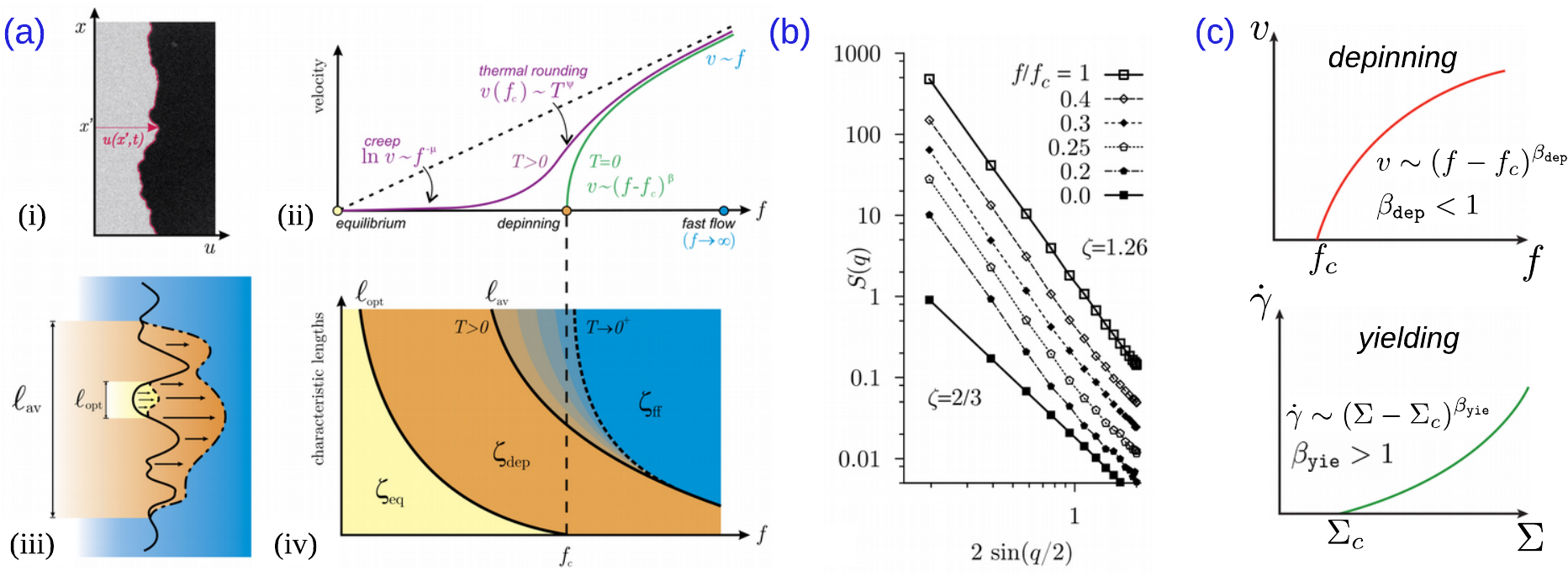} 
\par\end{centering}

\caption{\textit{The depinning picture}. \textbf{(a)} Connection between transport and
geometry in depinning. From~\citep{FerreroCRP2013}.
(i) Snapshot of a domain wall in a 2D ferromagnet. (ii)
Typical velocity-force characteristics. (iii) Crossover lengths $\ell_{opt}$
and $\ell_{av}$ representing the optimal excitation and the deterministic
avalanches, respectively. (iv) Geometric crossover diagram. \textbf{(b)}
Steady-state structure factor $S(q)$ of the line in the
limit of vanishing temperature for different forces (curves are shifted for clarity). Adapted
from~\citep{KoltonPRL2006}. \textbf{(c)} Comparison of the depinning
and yielding critical transitions in a correspondence $(v\leftrightarrow\dot{\gamma}),(f\leftrightarrow\Sigma)$. }
\label{fig:depinning} 
\end{figure}

The velocity-force characteristics $<\dot{h}>=v(f)$ is well known for the depinning problem
(see Fig.~\ref{fig:depinning}a); the information conveyed by this ``equation of state''
is enriched by a vast analytical and numerical knowledge of universal
properties at three special points: (i) equilibrium, i.e., $f=0$; (ii) depinning, i.e., $f=f_{c}$ at $T=0$; and (iii)
fast-flow $f\gg f_{c}$. Around these points, at vanishing temperature, the steady-state interface $h(x)$
displays a self-affine geometry (in the sense that it is invariant under dimensional rescaling, viz.,
$h(ax)\sim a^{\zeta}h(x)$) above
a microscopic length scale, with characteristic roughness exponents: (i) $\zeta_{eq}$, (ii)
$\zeta_{dep}$, and (iii) $\zeta_{ff}$.

Turning to transport properties, at \textit{equilibrium},
the mean velocity is zero and the dynamics is glassy. When the applied
force approaches zero, macroscopic movement can be observed only
at finite temperatures and at very long times. Collective
rearrangements on a scale of size $\ell_{opt}$ ($\ell_{opt} \rightarrow \infty$ as $f\rightarrow 0$) are needed
in order to overcome barriers $E_{b}(\ell_{opt})$ growing
as $E_{b}\sim\ell_{opt}^{\theta}$, with $\theta > 0$ a
universal exponent related to the roughness by $\theta=d-2+2\zeta_{eq}$.
This is the \textit{creep} regime. At the zero temperature
\textit{depinning transition} the velocity vanishes as $v(f,T=0)\sim(f-f_{c})^{\beta}$
for $f>f_{c}$ while $v=0$ for $f<f_{c}$. Approaching $f_{c}$ from
above the motion is very jerky and involves collective rearrangements
of a typical longitudinal size $\ell_{av}$ that diverges at $f_{c}$.
The avalanche size $S$, defined as the area covered by the moving interface,
has power-law statistics, viz.,
\begin{equation}
P(S)\sim S^{-\tau_{dep}}\text{, with }\tau_{dep}=2-\frac{2}{d+\zeta_{dep}}.
\end{equation}
 At finite temperature, the sharp depinning transition is rounded,
the velocity behaves as $v(f_{c},T)\sim T^{\psi}$ and the size $\ell_{av}$
is finite at the transition.
In the \textit{fast-flow} regime $f\gg f_{c}$, the response is linear, viz.,
$v\sim f$. Here impurities generate an effective
thermal noise on the interface. Therefore, the fast-flow roughness
corresponds to the Edwards-Wilkinson roughness $\zeta_{ff}=(2-d)/2$.

One of the remarkable lessons learned from this simple model is the
possibility to relate transport and geometry. If the applied force $f$ lies
in between two of the above mentioned reference points, the interface geometry
[in particular the roughness exponent, see Fig.~\ref{fig:depinning}(b)] depends on the observation scale
and its relative position compared to the characteristic
lengths ($\ell_{opt}$,$\ell_{av}$,\ldots{}). Granted that one knows the functional dependencies
of these characteristic lengths with $f$ and the velocity-force
characteristics for a given system,  transport properties (which intrinsically pertain to the dynamics) 
can be deduced from the static interface geometry, and vice-versa.

\subsubsection{Similarities and differences with EPM}

The manifest qualitative similarity between the yielding transition
and the depinning one has enticed many researchers to look for a unification
of these theories. The analogy has promoted the vision of yielding
as a critical phenomenon and has given rise to interesting advances, but, in our opinion, the \blue{(misguided)} belief in
a strict equivalence of the problems has been deceptive in some regards.

To stay on firm ground, a formal approach consists in finding an EPM analog to the depinning equation, Eq.~\eqref{eq:X_depinning_general}. In the stress-controlled situation (with applied stress $\Sigma^\mathrm{ext}$),
\citet{Weiss2014} (Eq.~S3 of the Supplemental Information) and \citet{tyukodi2015depinning}
thus proposed to substitute the EPM equation of motion [Eq.~\eqref{eq:gen_eq_of_motion}] with

\begin{equation}
\eta\partial_{t}\epsilon^\mathrm{pl}(r,t)= \mathcal{P} \Big[ \frac{-\delta U^\mathrm{el}[\epsilon^\mathrm{pl}]} {\delta \epsilon^\mathrm{pl}(r)}
+\Sigma^\mathrm{ext}-F_{p}(r,\epsilon^\mathrm{pl}) \Big],\label{X_depinninglike_EPM_eq_of_motion}
\end{equation}
where $U^\mathrm{el}[\epsilon^\mathrm{pl}] \equiv - \frac{1}{2} \iint dr dr^\prime \epsilon^\mathrm{pl}(r) \mathcal{G}(r-r^\prime) \epsilon^\mathrm{pl}(r^\prime) $, with $\mathcal{G}$ the elastic propagator, and $\mathcal{P}(x)$ denotes the positive part of $x$ ($x$ if $x>0$, 0 otherwise). In so doing, the deformation of an amorphous solid is mapped
onto a problem of motion through an (abstract) disordered space for the 
$\epsilon^\mathrm{pl}$-manifold pulled by the `force' $\Sigma^\mathrm{ext}$. The positive part $\mathcal{P}$ in Eq.~\eqref{X_depinninglike_EPM_eq_of_motion} creates genuine threshold dynamics; it has no direct counterpart in the depinning equation
but was argued by \citet{tyukodi2015depinning} not to be a core dissimilarity between yielding and depinning.

This formal similarity between the two classes of phenomena seems to buttress
the application of results from the depinning problem (hence mean field, owing to the long range of the elastic propagator) to the question of, e.g., avalanche statistics in disordered solids (see Sec.~\ref{sec:VIII_Avalanches}).
However, the following differences must be borne in mind.

First, and perhaps foremost, as often mentioned along the present review, the interaction kernel in depinning
problems is positive, whereas the quadrupolar elastic propagator $\mathcal{G}$ used in EPM has positive and negative bits.
This has profound consequences on the critical
behavior at the yielding transition observed in EPM, in particular with respect to the possibility of strain
localization and the avalanche statistics. Furthermore, while in depinning $v$ vanishes at
 $f_{c}$ as $v~\sim(f-f_{c})^{\beta}$ with \blue{typically} $\beta<1$, the strain rate
$\dot{\gamma}$ does so at the yielding transition as $\dot{\gamma}~\sim(\Sigma-\Sigma_{c})^{\beta}$
with $\beta>1$, as schematically shown in Fig.~\ref{fig:depinning}(c). Note that, if the systems were at equilibrium,
this difference in the value of $\beta$ would imply a change in the order of the continuous phase transition.
Other consequences can be deduced from the general scaling relations proposed by \citet{lin2014scaling} [Supplementary Information], which are claimed to encompass the depinning and the yielding cases \blue{(not all these relations are strictly obeyed in finite-dimensional EPM)}:
\begin{align}
\beta & =\nu\left(d-d_{f}+z\right)\label{beta}\\
\nu & =\frac{1}{d-d_{f}+\alpha_{k}}\label{nu}\\
\tau & =2-\frac{d_{f}-d+1/\nu}{d_{f}}-\frac{\theta}{\theta+1}\frac{d}{d_{f}}\label{tau}
\end{align}
Here, $d_{f}$ is the fractal dimension of the avalanches, $z$ is
the dynamical exponent, $\nu$ is the exponent controlling the divergence
of the correlating length at the transition and $\alpha_{k}$ is the
dimension of the elastic interaction kernel. In EPM $\alpha_{k}=0$
and $d_{f}<d$ so that $\beta>1$. In depinning, $\alpha_{k}=2$ for short-ranged elasticity and $\alpha_{k}=1$
for long-ranged elasticity, $\theta=0$ and $d_{f}\ge d$.

Secondly comes the question of the nature of the disorder in the pinning force $F_p$.
In elastic depinning models, regardless of how realistic the chosen correlations of $F_p$ are, the origin of the disorder is generally extrinsic. More precisely, it reflects the disorder of the substrate on
which the elastic manifold advances, hence $F_p=F_p(h)$. On the other hand, in the yielding phenomenon, as stated by \citet{papanikolaou2016shearing},
`the pinning disorder for every particle originates in the actual interface that
attempts to depin (other nearby particles); a disordered
solid pins itself during deformation'.  Therefore, it is inaccurate to consider
that $F_p$  only depends on the local value of $\epsilon_{pl}$. In particular, a \emph{given}
system will \emph{not} encounter the same pinning forces $F_p$ along its deformation between, say,
$\epsilon_{pl}=0$ and $\epsilon_{pl}=1$ if it is sheared slowly and if it is sheared fast. Typically,
at high shear rates, the potential energies of the inherent structures of the material are higher (as evidenced by
the variations of potential energies of the inherent structures with the shear rate, in atomistic simulations). This dependence
should impact the $\dot\gamma = f(\Sigma)$ curve \blue{at finite shear rates.}

Lastly, the EPM equation of motion [Eq.~\eqref{eq:gen_eq_of_motion}] cannot always be
reduced to an expression akin to Eq.~\eqref{X_depinninglike_EPM_eq_of_motion}, because of the memory
effects contained in the plastic activity variable $n$.

Let us now mention a subclass of problems that may be more closely related to EPM:
 the so called ``plastic depinning''. This phenomenon is observed for
example in particle assemblies driven over random substrates whenever
irreversible plastic deformations actually occur, or in charge density
wave problems. Unfortunately, this comparison has been much less
exploited by the amorphous solids community, even though the connection was
very recently pointed out in \citet{reichhardt2016depinning}'s review.

To conclude on the topic, there undoubtedly remains much to be learned from
the $30^{+}$ years of studies on depinning phenomena. Some intriguing open questions
left from this comparison are the following:
Are the transport properties of driven amorphous
solids related to geometrical properties, as they are in elastic manifolds?
Is it possible, for example, to infer from a picture at which strain-rate
a dense emulsion is being sheared?

\subsection{Fiber bundle, fuse networks and continuum models for the study of
cracks and fracture\label{sub:X_Fibre_bundles}}

\subsubsection{Brief introduction to cracks and fracture}

In partial overlap with the scope of EPM, the question of the failure
of hard solids under loading, e.g. in tension, has attracted much
attention over the last centuries. Pioneering in this respect, as
recalled by \citet{alava2006statistical}, is Leonardo da Vinci's
observation that\blue{, if one loads a metal wire in tension
with a weight, it will fail more readily if it is longer, for the same cross section;} this runs counter
to basic continuum mechanics predictions for a uniform medium. In
fact, the failure of brittle solids, in particular rocks, is ascribed
to the growth and propagation of pre-existing cracks (at the scale
of the crystalline grains constituting the material) or, more generally,
defects.

If one considers an individual crack in a homogeneous medium, according
to \citet{griffith1921phenomena}'s criterion, its growth hinges on
a competition between a surface energy term averse to the opening
of solid-air interfaces and an elastic energy term favoring its growth
and thereby reducing the elastic energy stored in the bulk. For example,
for a single elongated elliptic crack of length $a$ in a 2D
medium, the sum of these competing terms reads 
\[
E_{T}=\underset{\text{elastic energy}}{\underbrace{\frac{-\pi\Sigma^{2}a^{2}}{2E}}}+\underset{\text{surface energy}}{\underbrace{2\gamma a}},
\]
where $E$ is the Young modulus of the material, $\gamma$ is the
interfacial energy, and $\Sigma$ is the applied stress. Thus, the
evolution of the crack depends on the sign of the derivative $\frac{dE_{T}}{da}$
\citep{alava2006statistical}. However, cracks very seldom have so
simple a geometric shape. Roughly speaking, owing to the presence
of heterogeneities, the crack will zigzag around hard spots. This
will result in undulations and protrusions in the \emph{post-mortem}
fracture surface, which exhibits a self-similar (fractal) pattern:
If the surface height at a point $(x,z)$ is denoted by $h(x,z)$,
the root mean square fluctuation $w(l)$ of the height in a region of size $\Delta x\approx\Delta z\approx l$
obeys
\[
w(l)\equiv\sqrt{\left\langle h(x,z)^{2}\right\rangle -\left\langle h(x,z)\right\rangle ^{2}}\sim l^{\zeta_{\perp}},
\]
where $\zeta_{\perp}$ is the (out-of-plane) Hurst exponent, or roughness
exponent. Interestingly, this exponent seems to be weakly sensitive
to the material or the loading, with values centered around $\zeta_{\perp}\simeq0.8$
and early claims of universality \citep{bouchaud1990fractal}. The
fractal dimension $d_{f}$ of the surface is then related to $\zeta_{\perp}$
via $d_{f}=3-\zeta_{\perp}$ for 3D fracture. While the material is
being fractured, the crack propagates along a rough, scale-invariant
frontline (see Fig.~\ref{fig:X_cracks_and_fuses}a), characterized
by the in-plane roughness exponent $\zeta_{\parallel}$. Roughness
bears practical importance, since it modifies the scaling of the surface
energy term. 

Let us mention two subtleties. First, the exponents $\zeta_{\parallel}$
and $\zeta_{\perp}$ are not independent \citep{ertacs1994anisotropic}.
Second, $\zeta_{\perp}$ might in fact mix two distinct exponents,
insofar as \citet{ponson2006two}'s fracture experiments on silica
and aluminium alloys hint at anisotropic height variations in the
fracture plane, with distinct behaviours along the front line and
along the crack propagation direction.

In addition to being spatially nontrivial, the propagation of the
crack front also displays marked variations in time. The associated
dynamics is highly intermittent and involves avalanches of events
which span a broad range of energies. Indeed, the crackling noise
emitted during these events has a power-law power spectrum, for instance
in composite materials \citep{garcimartin1997statistical}. For instance,
the crack produced when tearing apart two sandblasted Plexiglas sheets
stuck together through annealing undergoes a stick-slip motion at
small scales that is reminiscent of dry solid friction \citep{maaloy2001dynamical},
which in turn may tell us about earthquake dynamics \citep{svetlizky2014classical}. 

At this stage, a discrepancy with respect to soft solids ought to
be mentioned: In (rock) fracture, the microruptures very generally
do not have time to heal on the time scale of the deformation; without
recovery process, the material is thus permanently damaged. However,
the crack velocity may still have an influence on the dynamics of
the process owing to the finite duration of the avalanches.

\subsubsection{Fiber bundles \label{sub:X_fiber_bundles}}

Arguably, the simplest way to model fracture is to consider two blocks
bound by $N$ aligned fibers. These fibers share the global load and
break irreversibly when their elongation $x$ exceeds a randomly distributed
threshold; this is the basis of fiber-bundle models \citep{herrmann2014statistical}.
In \emph{democratic} fiber bundles, the load of broken fibers is redistributed
equally to all survivors. Analytical progress is possible in this
intrinsically mean-field model. In particular, it is easy to show
that, \emph{on average}, when the bundle is stretched by $x$ (with
$x=0$ the reference configuration), a fraction $C(x)$ of fibers
have broken, where $C(x)$ is the cumulative distribution of thresholds,
and the total load (normalized by the initial number of fibers, of stiffness $\kappa$ each) reads $\bar{f}(x)=\kappa x\left[1-C(x)\right]$.
It follows that the maximum strength per fiber of the bundle is, on
average,
\[
f_{c}=\max_{x}\,\kappa x\left[1-C(x)\right].
\]
If one pulls on a \emph{given} bundle, however, the load $f$ will not evolve along the smooth \emph{average} profile $\bar{f}(x)$,
but along a rugged profile $\left\{ f(x_{k})\equiv\kappa x_{k}\left[1-C(x_{k})\right],\,k=1\ldots N\right\} $
due to the randomness of the thresholds $x_{1}\leqslant x_{2}\leqslant\ldots\leqslant x_{N}$,
sorted according to the order of failure. The $f(x_{k})$ thus perform
a random walk in ``time'' $k$ with a time-dependent bias $\left\langle f(x_{k+1})-f(x_{k})\right\rangle $
\citep{sornette1992mean}. If, starting from a stable situation, the
rupture of the $k$-th bond leads to $S$ additional failures, viz.,
\begin{equation}
f\left(x_{k+i}\right)<f(x_{k})\text{ for }i\text{ between 1 and }S\text{ (but
not for }i=S+1),
\end{equation}
an avalanche of size $S$ will occur under fixed
load. Noting that (i) this is a problem of first return for the walker
$f(x_{k})$ {[}or, equivalently, of survival close to the absorbing
boundary $f=f(x_{k})^{-}${]}, and that (ii) close to global failure
$f\approx f_{c}$ the random walk is unbiased, i.e., $\left\langle f(x_{k+1})-f(x_{k})\right\rangle =0$,
\citet{sornette1992mean} showed that the distribution of avalanche
sizes $s$ obeys 
\begin{equation}
p(S)\sim S^{-\tau}\text{, where }\tau=\nicefrac{3}{2}.\label{eq:X_FBM_scaling_3/2}
\end{equation}
More precisely, for a uniform distribution of thresholds between $x_l$
and 1, the distribution reads
\[
p(S)\sim S^{\nicefrac{-5}{2}}\left(1-e^{\frac{-S}{S_\mathrm{cut}}}\right),
\]
where the cutoff size $S_\mathrm{cut}\equiv\frac{1}{ 2 (1-2x_l)^{2} }$
diverges at the critical point $x_l=\nicefrac{1}{2}$ \citep{pradhan2005crossover}.
\blue{For $x_l < \nicefrac{1}{2}$, the fiber fails gradually as the loading is increased,
whereas for $x_l > \nicefrac{1}{2}$ failure occurs all at once. 
A parallel may here be drawn with the discussion about the brittle-to-ductile transition in
amorphous solids in Sec.~\ref{sec:VI_Macroscopic_Shear_Deformation}.
For $x_l \leqslant \nicefrac{1}{2}$,}
the power law with exponent $\tau=\nicefrac{3}{2}$ of Eq.~\eqref{eq:X_FBM_scaling_3/2}
is recovered for $S\ll S_\mathrm{cut}$, whereas for $S\gg S_\mathrm{cut}$ the random
walk of the $f(x_{k})$ is biased upward and a steeper power law is
obtained, with an exponent $\nicefrac{5}{2}$. The scaling $p(S)\sim S^{\nicefrac{-5}{2}}$
is found generically if all avalanches since the start of the deformation
($x=0$) are taken into account \citep{hansen1992distribution}. The
gradual shift to an exponent $\tau=\nicefrac{3}{2}$ then signals
imminent failure. Interestingly, the power-law behavior fades out
in favor of a much faster decay of $p(S)$ if the load released by
broken fibers is redistributed \emph{locally} to the first neighbors
only, instead of being shared by all intact fibers \citep{kloster1997burst}.

\subsubsection{Fuse networks}

\begin{figure}
\begin{centering}
\includegraphics[width=1\columnwidth]{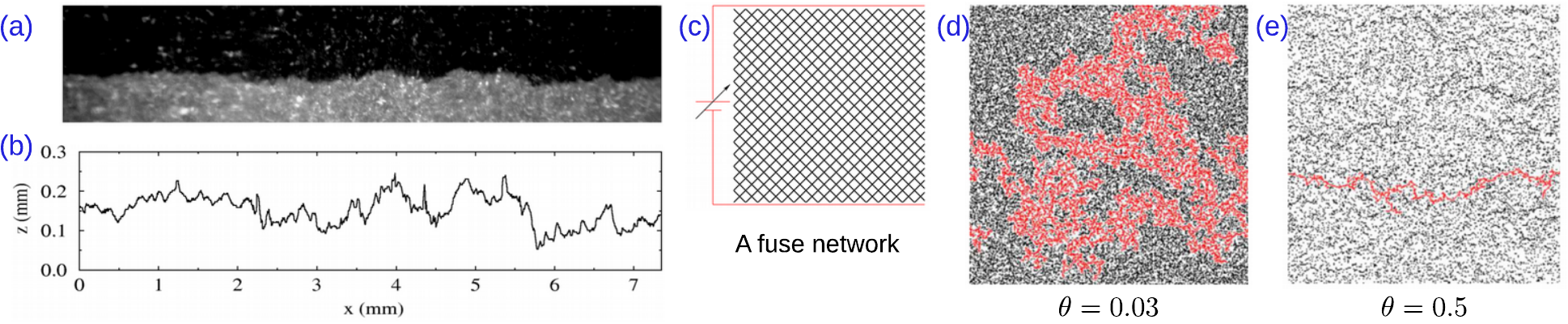}
\par\end{centering}

\caption{\label{fig:X_cracks_and_fuses}Observation and modeling of crack propagation.
(a) Raw image of the front of an in-plane crack propagating between
Plexiglas plates. The intact region appears in black and the image-processed
front line is shown in (b). The roughness along the propagation (\emph{z})
direction has a power-law spectrum characterized by the roughness
exponent $\zeta_{\parallel}$. From \citep{schmittbuhl1997direct}.
(c) Sketch of the random fuse network and (d-e) failure process for
distinct probability density functions for the thresholds, $p(x)\sim x^{\theta}$.
The crack has been colored in red. From \citep{Shekhawat2012}.}

\end{figure}

Unfortunately, the picture promoted by mean-field or 1D fiber bundles
is incapable of describing the heterogeneous and anisotropic propagation
of cracks. Extending the approach to higher dimensions, fuse networks
connect lattice nodes (say, nodes $i$ and $j$) by fuses of conductance
$K_{ij}$ that break past a threshold $x\in[0,1]$, thereby burning
the fuse ($K_{ij}\rightarrow0$). To take an example, the distribution
$p$ of the thresholds can be set as a power law, $p(x)\sim x^{\theta}$
with $\theta>0$. The voltages $V_{i}$ are imposed at two opposite
edges of the system, as depicted in Fig.~\ref{fig:X_cracks_and_fuses}c.
The Hamiltonian of the system reads

\begin{equation}
\mathcal{H}_{nc}=\frac{1}{2}\sum_{\left\langle i,j\right\rangle }K_{ij}\left(V_{i}-V_{j}\right)^{2},\label{eq:hamiltonian_non_central}
\end{equation}
where the sum runs over \blue{all adjacent nodes $(i,j)$.} Note that, if 
\blue{the $K_{ij}$ are constant, then
}
the model can be viewed as
a discretization of Poisson's equation in the vacuum, $\nabla^{2}V$=0.
Fuse networks are thus closer to EPM than fiber bundles, insofar as
the stress redistribution when one fuse burns (in the pristine network)
is strongly anisotropic, with a shielding of the current fore and
aft and an enhancement sideways \citep{barthelemy2002random,rathore2016planar}.
It can then be understood that failure occurs along a line of burnt
fuses, the ``crack'' line, provided that there is finite disorder
($\theta>0$) and the network is large \citep{Shekhawat2012}. Besides,
in a 2D fuse network, \citet{hansen1991roughness} computed a roughness
exponent $\zeta$ approximately equal to $0.7$ for weak disorder,
not far from experimental values for fractured surfaces $\zeta_{\perp}\approx0.8$
(note that $\zeta=\zeta_{\perp}$ in 2D).

The expression of the Hamiltonian in Eq.~\eqref{eq:hamiltonian_non_central2}
evokes a random bond Ising model; the equivalence is formally exact
if the voltages are restricted to the values $\pm1$, and the thresholds
are infinite, thus making bonds unbreakable (\emph{perfect ductility}).
These differences are not negligible in any way. Indeed, the interactions
between nodes are thereby much reduced, in spatial extent and magnitude;
by contrast, in random fiber or fuse models, the impact of breaking
a bond is magnified close to failure, owing to the small number of
intact bonds which will share the load. Nevertheless, the process
of fracture can be mimicked in the random Ising models by imposing
spin +1 (-1) on the left (right) edges of the sample and monitoring
the interface line between the +1 and -1 domains. \citet{rosti2001pinning}
studied the probability that this interface passes through an artificial
``notch'', i.e., a segment in which the bond strengths $K_{ij}$
have been set to zero, and observed a transition from low to high
probabilities as the notch length was increased above a disorder-dependent
threshold value. Similar results were obtained in experiments in which
sheets of papers with pre-cut notches were torn.

\subsubsection{Spring models}

From a mechanical perspective, should one replace the voltage
$V_{i}$ in Eq.~\eqref{eq:hamiltonian_non_central} with the displacement
$\boldsymbol{u_{i}}$ at node \emph{i}, \emph{viz.},
\begin{equation}
\mathcal{H}_{nc}^{\prime}=\frac{1}{2}\sum_{\left\langle i,j\right\rangle }K_{ij}\left(\boldsymbol{u_{i}}-\boldsymbol{u_{j}}\right)^{2},\label{eq:hamiltonian_non_central2}
\end{equation}
the interpretation of the Hamiltonian as the energy of a network of
random springs of stiffness $K_{ij}$ will become apparent.
The \emph{x}, \emph{y}, and \emph{z} components of the dispacements
in $\mathcal{H}_{nc}^{\prime}$ decouple, so that model is actually
scalar \citep{de1976relation}. However, it features noncentral forces:
the force exerted by $j$ on $i$ is not aligned
with $\boldsymbol{e_{ij}}$. A more consistent description of a network
of nodes connected by harmonic springs relies (to leading order) on the Hamiltonian

\begin{equation}
\mathcal{H}_{c}=\frac{1}{2}\sum_{\left\langle i,j\right\rangle }K_{ij}\left[\left(\boldsymbol{u_{i}}-\boldsymbol{u_{j}}\right)\cdot\boldsymbol{e_{ij}}\right]^{2}.\label{eq:hamiltonian_central}
\end{equation}
On a triangular lattice, with bonds of uniform strength $K_{ij}=1$,
the continuum limit of this Hamiltonian represents an isotropic elastic
medium with a Poisson ratio of $\nicefrac{1}{3}$ in 2D and $\nicefrac{1}{4}$
in 3D \citep{monette1994elastic}. As bonds are gradually removed
in a random fashion, the initially rigid system transitions to a non-solid
state with vanishing elastic moduli at a critical bond fraction $p_{c}$.
Such a transition is also observed with the models based on the scalar
Hamiltonian $\mathcal{H}_{nc}$ or the noncentral Hamiltonian $\mathcal{H}_{nc}^{\prime}$,
although at a distinct fraction $p_{c}$. Somewhat surprisingly, the
scalings of the shear and bulk moduli with the fraction of bonds $p$
around $p_{c}$ differ between the $\mathcal{H}_{c}$ and $\mathcal{H}_{nc}$-based
models; the discrepancy stems from the distinct symmetries, in the
same way as the Heisenberg model differs from the Ising model \citep{feng1984percolation}.
The distinction subtly differs from \blue{the dichotomy between} scalar and tensorial
EPM, in that the EPM propagator is always derived from
the same constitutive model (\emph{tensorial} continuum elasticity); the
scalar description simply discards some tensorial components at the end
of the day. 

%Coming back to spring models, a\textbf{ }linear combination of $\mathcal{H}_{nc}^{\prime}$
%and $\mathcal{H}_{c}$, with uniform spring constants $K_{ij}$, leads
%to the Born model:

%\begin{equation}
%\mathcal{H}_{Born}=\frac{1}{2}\sum_{\left\langle i,j\right\rangle }\alpha\left(\boldsymbol{u_{i}}-\boldsymbol{u_{j}}\right)_{\parallel}^{2}+\beta\left(\boldsymbol{u_{i}}-\boldsymbol{u_{j}}\right)_{\perp}^{2},\label{eq:hamiltonian_Born}
%\end{equation}
%where the subscripts $\parallel$ and $\perp$ denote the parallel
%and orthogonal parts of the displacements relative to the bond direction,
%respectively.

Regarding the avalanches of ruptures close to the point of global
failure, i.e., under loading $f\approx f_{c}$, \citet{zapperi1999avalanches}
claimed that both the random fuse network of Eq.~\eqref{eq:hamiltonian_non_central}
and the central-force spring model of Eq.~\eqref{eq:hamiltonian_central}
(supplemented with bond-bending forces) fall in the universality
class of spinodal nucleation, in that the avalanche sizes $S$ are
distributed according to 
\[
p(S)\sim S^{-\tau}\Phi\left[S\,\left(f_{c}-f\right)\right],\text{ where }\tau=\nicefrac{3}{2}
\]
and $\Phi$ is a scaling function. \citet{zaiser2015crack} also found that
fuse networks yielded results similar to spring models regarding the
initiation of failure, with localized correlations in the damage patterns.

We conclude this section on spring models with a historical note referring
to the fact that such models had in fact been pioneered by \citet{de1976relation}
to tackle the ``converse problem'', namely, gel formation (e.g.,
through cross-linking): Instead of gradually destroying bonds, he
cranked up the fraction $p$ of bonds by randomly connecting pairs
of neighbours until bonds percolated throughout the system; this occurred
at a critical fraction $p_{c}$, supposedly corresponding to gel formation.
In any event, the nature of the transition associated with the random
depletion (or creation) of bonds, which pertains to percolation, is
distinct from what is observed in \emph{random} fuse or spring networks.
In the latter models, the disorder in the yield thresholds bestows
critical importance to the spatial redistribution of stresses following
ruptures. This distinction is at the origin of different scaling relations,
e.g., between the failure force and the system size \citep{hansen1989rupture}.

\subsubsection{Beyond random spring models}

Refinements have been suggested to bring random fuse (or spring) networks
closer to models of material deformation and fracture. First, the
irreversible breakage of the fuses past a threshold mirrors perfectly
brittle fracture. At the opposite end, perfect plasticity is mirrored
by the saturation of the fuse intensity past a threshold. But a continuum
of possibilities can be explored between these extreme cases, whereby
the conductivity of the fuse is decreased to mimic partial weakening,
similarly to what can be done in EPM.

Another limitation of the models stems directly from the description
of the bonds on a regular lattice: let alone the presence of soft modes
in several cases, the ($\mathcal{H}_{c}$-based) central-force model,
discretized on a triangular lattice, displays an anisotropic tensile
failure surface (despite an isotropic linear response), with an anisotropy
ratio of 50\% \citep{monette1994elastic}. These deficiencies can
be remedied in part by complementing the spring-stretching energies
in $\mathcal{H}_{c}$ with bond-bending energies.
% contributions proportional to $\left(\cos\theta-\cos\theta_{0}\right)^{2}$, 
% where $\theta$ is the angle between two adjacent bonds sharing a vertex and $\theta_{0}$
% is the equilibrium value of this angle. 
This refinement leads to an
isotropic elastic medium with adjustable Poisson coefficient and a
more isotropic failure surface.

As with EPM, the following step in the endeavor to refine the description
led to the introduction of a finite-element approach, which relies
on a continuum description down to the scale of one mesh element.
The equations of inhomogeneous elasticity are solved and a damage
(of magnitude $D$) is introduced by reducing the local elastic constant
$E\rightarrow(1-D)E$ whenever the local stress exceeds a threshold
value. The process can evolve into avalanches, and eventually to a
vanishing of the elastic resistance through the propagation of a
fracture through the system. Incidentally, this mechanism had first been implemented
by \citet{zapperi1997plasticity} using a fuse network with damage
operating on the fuse resistances; the model displayed scale-free behavior
with power-law distribution of event sizes, $P(S)\sim S^{-\tau}$
with $\tau\simeq1.2$.

\citet{amitrano1999diffuse} refined the modeling approach
by using a pressure-modified (Mohr-Coulomb) criterion for the onset
of plasticity, viz.,
\[
C+\sigma_{n}\tan\phi-\sigma<0,
\]
where $C$ represents the cohesion of the material, $\sigma_{n}$
and $\sigma$ are the normal and shear stresses, respectively. A transition
from brittle failure with very localized damage (at low internal friction
angle $\phi$, i.e., little sensitivity of the yield criterion to
pressure) to ductile with diffuse damage (at large $\phi$) was observed.
At low $\phi$ the damage around a single event is similar to the
stress redistribution considered in EPM, while for large $\phi$ it
becomes much more directional. The transition from ductile to brittle
shares qualitatively similarities with the strain localization transition,
but the control parameter is different from those discussed in Sec.~\ref{sub:VI_Ingredients_shear_loc}.

In the case of large $\phi$ and brittle failure, a description of
compressive failure under uniaxial stress as a critical phenomenon
analogous to depinning was proposed by \citet{Girard2010}
and elaborated by \citet{Weiss2014}. The interpretation in terms
of a criticality notably affords a detailed description of size effects
on the critical stress \citep{girard2012damage}.\emph{ }

\section{Outlook}

In the last ten years, EPM have become an essential theoretical tool
to understand the flow of solids. 
Starting from elementary models intended to reproduce earthquake dynamics,
they have blossomed into more refined approaches that have helped rationalize
many experimentally observed features, at least at a qualitative level,
and unveil new facets of the rheology of these materials.
\blue{Future developments in the field can be expected in a number of directions,
following current experimental and theoretical interests.
}

\blue{
In  rheology, considerable attention has recently been devoted to
the study of transient regimes. For instance, one can mention the study of the load curves at fixed
shear rate $\dot\gamma$, that can exhibit stress overshoots depending on the initial preparation and non-trivial 
scalings of the time to reach the stationary state with $\dot\gamma$.
Other examples include creep under imposed stress, the dynamics of relaxation and the residual 
stresses after sudden cessation of the driving, and oscillatory regimes.}
In the latter category, the Large Amplitude Oscillatory Strain (LAOS) protocol
probes the nonlinear behavior and the frequency dependent one at the same time,
and therefore involves a complex interplay between plastic deformation and internal relaxation.
Reproducing the complex response of particular systems under such
protocols is particularly challenging for simple models.
Several issues could be investigated within the framework of EPM, such as the
onset of tracer diffusion as the amplitude of the oscillatory strain is
increased, or the fatigue behavior leading to failure.
Recently, it was suggested that the LAOS protocol could induce strain localization
in systems with a monotonic flow curve, based on a study of a spatially
resolved version of the soft glassy rheology model, presented in Sec.~\ref{sub:V_SGR}
\citep{Radhakrishnan2016shear,radhakrishnan2017shear}.
\blue{Creep (see Sec.~\ref{sec:Relaxation}) is an equally challenging phenomenon;
a recent mean-field EPM illustrated its very strong
sensitivity to the initial conditions \citep{liu2017mean}.}

A more unexpected emerging avenue is the study of systems with internal activity,
such as living tissues or dense cell assemblies.
\blue{
The general ideas exploited for the description of amorphous systems can indeed be expanded
to incorporate new types of events, such as cell division
(assimilated to a local anisotropic dilation) and cell death (local isotropic contraction).
At the mean-field level, \citet{matoz2017nonlinear} and others conducted a first analysis along these lines.
For further exploration of the collective behavior resulting from the interplay between 
cell division, apoptosis, locomotion, and contractility,
as well as the mechanosensitivity of these processes, an EPM describing all these ingredients at the same time would be invaluable.}
At present, new experimental tools are providing information on the
statistical fluctuations in such systems, which will allow to calibrate
these models. 

% and compare them to others \citep{BiPRX2016}.

From the viewpoint of statistical physics, the yielding transition
described by EPM stands as a new type of dynamical phase transition,
with specificities that are still to be understood in extent.
Considerable efforts have been devoted to the theoretical study of the related
problem of the depinning transition (Sec.~\ref{sub:X_Depinning-transition}).
In the latter case, (mostly) exact exponents, scaling functions, and avalanche shapes
were derived using scaling analysis and renormalization techniques.
For the yielding transition, the (slow) process of consensus building
has not converged yet, but there are reasons to believe that
the results on avalanche statistics obtained in the depinning problem
cannot be directly transposed to this field, because
the propagator controlling stress redistribution is
partly negative, which affects the density of sites close to yielding.
\blue{
Whether this features only induces an effective dimensional reduction,
leaving us in a well known universality class but for $d<D$, or whether it exhibits a
completely distinct set of critical exponents, still needs to be clarified.
}
Scaling relations between critical exponents have been proposed
\citep{lin2014scaling,aguirre2018critical} and tested in diverse EPM, but analytical calculations beyond
mean field are scant.
\blue{
Recent efforts to relate EPM to (better known) problems
of motion through a disordered landscape
open new vistas for the understanding of yielding and transport properties under slow driving \citep{jagla2017elasto-plastic},
but there is still no consensual theory explaining the flow exponents (the low-shear-rate rheology).
}
The situation is somewhat similar on the experimental side:
The depinning phenomenon has benefited from a very detailed experimental
characterization in various systems (magnetic domain walls, contact lines,
vortices), including avalanche statistics and shapes, which has permitted
comparison to the theory.
Amorphous plasticity is not on quite so good a footing, with only
a few attempts to characterize the distribution of stress drops in
deformed systems.
The situation is however improving, thanks to several recent efforts,
e.g. those combining mechanical deformation and confocal microscopy
in colloidal glasses.

The foregoing discussion is related to the critical aspects of the
yielding phenomenon, discussed in Sec.~\ref{sec:VIII_Avalanches} and \ref{sec:VII_Bulk_rheology}.
In a number of real systems \citep{bonn2015yield}, the
\blue{onset of} flow is in fact discontinuous and implies a
coexistence between flowing and immobile states.
EPM and other theoretical studies have proposed
possible mechanisms that may influence the continuous or discontinuous
character of the transition (see Sec.~\ref{sec:VI_Macroscopic_Shear_Deformation}).
Nevertheless, it turns out to be experimentally difficult to control the transition
in a systematic way by changing some experimental parameter.
\citet{wortel2016criticality}'s work on weakly vibrated granular media represents
a notable exception, insofar as the intensity of external shaking could be used to
continuously tweak the flow curve towards nonmonotonicity. 
\blue{
The ensuing emergence of a critical point at a finite driving rate has scarcely been 
addressed in the literature and could be analyzed with the EPM approach.
}
Similar systems of vibrated grains have also permitted the experimental realization
of a Gardner transition \citep{SeguinPRL2016}, a transition which may be important
for the theory of glasses and which has been associated with shear yielding \citep{UrbaniPRL2017}.
\blue{
On a related note, cutting-edge atomistic simulations suggest that the ductility or brittleness
of the yielding phenomenon hinges on the initial preparation of the glass, rather than the microscopic
interactions between particles or the dynamics \citep{ozawa2018random}; this puts EPM in the 
forefront for the study of these questions, but requires them to establish adequate proxys for the
initial stability of the glass.
}

These prospective lines of research have hardly been explored using EPM.
So, for all our efforts to articulate a comprehensive view of the state of the art here,
we can only wish that this review will soon need to be updated with
insightful results in these new avenues.

\begin{acknowledgments}
The authors acknowledge financial support from ERC grant ADG20110209 (GLASSDEF).
E.E.F acknowledges financial support from ERC Grant No.~ADG291002 (SIZEFFECTS).
K.M. acknowledges financial support from ANR Grant No.~ANR-14-CE32-0005 (FAPRES) and CEFIPRA Grant No.~5604-1 (AMORPHOUS-MULTISCALE).
J.-L.B. is supported by IUF.
We thank M.~V. Duprez for professional help with the graphics.
We would like to warmly acknowledge E.~Bertin, Z.~Budrikis, A.~Kolton, C.~Liu, C.~Maloney, S.~Merabia, S.~Papanikolaou, A.~Rosso, J.~Rottler,
P.~Sollich, L.~Trukinovsky, D.~Vandembroucq, J.~Weiss and M.~Wyart for fruitful interactions and useful comments on a first draft
of this review.
\end{acknowledgments}

\newpage{}

\bibliography{ReviewBibClean}

\end{document}